\documentclass[11pt]{article}
\usepackage{a4wide}

\usepackage[dvipdfmx,hiresbb]{graphicx} %   usepackage with driver option
\usepackage{mediabb}                     %   [For PDF graphics] Automatic PDF-bounding box
\usepackage{array}
\usepackage{slashed}

\usepackage{color}
\definecolor{refkey}{rgb}{1,0,1}
\definecolor{labelkey}{rgb}{1,0,1}
\newcommand{\red}{\color{black}}
\newcommand{\blue}{\color{black}}

\newcommand{\al}[1]{\begin{align}#1\end{align}}
\newcommand{\als}[1]{\begin{align*}#1\end{align*}}
\newcommand{\ab}[1]{\left|#1\right|}
\newcommand{\paren}[1]{\left(#1\right)}
\newcommand{\fn}[1]{\!\left(#1\right)}
\newcommand{\sqbr}[1]{\left[#1\right]}
\newcommand{\br}[1]{\left\{#1\right\}}

\usepackage{amsmath,amssymb,amsthm}

\DeclareMathOperator{\Tr}{Tr}
\newcommand{\df}{\text{d}}
\newcommand{\ol}{\overline}
\newcommand{\modeq}{\stackrel{1}{=}}

\usepackage{url}

\usepackage{epsf}

\usepackage{braket}

\newcommand{\GeV}{\ensuremath{\,\text{GeV} }}
\newcommand{\meV}{\ensuremath{\,\text{meV} }}

\newcommand{\nn}{\nonumber\\}

\newcommand{\bmat}[1]{\begin{bmatrix}#1\end{bmatrix}}

\newcommand{\Zhat}{\hat Z}

\newcommand{\p}{\partial}

\newcommand{\bs}{\boldsymbol}

\newcommand{\E}{\mathcal E}
\newcommand{\I}{\mathcal I}

\begin{document}

\title{\vbox{
\baselineskip 14pt
\hfill \hbox{\normalsize KUNS-2538, OU-HET/850
}
} \vskip 1cm
\bf Eternal Higgs inflation and\\
cosmological constant problem\vskip 0.5cm
}
\author{
Yuta~Hamada,\thanks{E-mail: \tt hamada@gauge.scphys.kyoto-u.ac.jp}~
Hikaru~Kawai,\thanks{E-mail: \tt hkawai@gauge.scphys.kyoto-u.ac.jp}~ 
and Kin-ya~Oda\thanks{E-mail: \tt odakin@phys.sci.osaka-u.ac.jp}\bigskip\\
%\\*[20pt]
{\it \normalsize
${}^{*\dagger}$Department of Physics, Kyoto University, Kyoto 606-8502, Japan}\smallskip\\
\it \normalsize
${}^\ddag$Department of Physics, Osaka University, Osaka 560-0043, Japan
}
\date{\today}

%\newpage

%\begin{titlepage}
\maketitle
%\thispagestyle{empty}
%\clearpage
%\tableofcontents
%\thispagestyle{empty}
%\end{titlepage}

\abstract{\noindent
We investigate the Higgs potential beyond the Planck scale in the superstring theory, under the assumption that the supersymmetry is broken at the string scale. We identify the Higgs field as a massless state of the string, which is indicated by the fact that the bare Higgs mass can be zero around the string scale. 
We find that, in the large field region, the Higgs potential is connected to a runaway vacuum
with vanishing energy, which corresponds to opening up an extra dimension.
We verify that such universal behavior indeed follows from the toroidal compactification of the non-supersymmetric $SO(16)\times SO(16)$ heterotic string theory. 
We show that this behavior fits in the picture that the Higgs field is the source of the eternal inflation. 
The observed small value of the cosmological constant of our universe may be understood as the degeneracy with this runaway vacuum, which has vanishing energy, as is suggested by the multiple point criticality principle.
}

%\newpage

%\renewcommand{\thefootnote}{\arabic{footnote}}
%\setcounter{footnote}{0}

\normalsize
\newpage

\tableofcontents

\newpage

\section{Introduction}
The Higgs boson discovered at the LHC~\cite{Aad:2012tfa,Chatrchyan:2012ufa} beautifully fits into the Standard Model (SM) predictions so far~\cite{CMS:yva}. The determination of its mass~\cite{Agashe:2014kda}
\al{
M_H
	&=	125.7\pm0.4\GeV
}
completes the list of the SM parameters, among which the ones in the Higgs potential,
\al{
V	&=	m^2\ab{H}^2+\lambda\ab{H}^4,
}
have turned out to be $m^2\sim-\paren{90\GeV}^2$ and $\lambda\simeq0.13$, depending on the precise values of the top and Higgs masses; see e.g.\ Ref.~\cite{Buttazzo:2013uya}.

We have not seen any hint of a new physics beyond the SM at the LHC, and it is important to guess at what scale it appears, as we know for sure that it must be somewhere in order to account for {\blue the tiny neutrino masses, dark matter, baryogenesis, inflation, etc.} In this work, we assume that the Higgs sector is not altered up to a very high scale,\footnote{
See e.g.\ Refs.~\cite{Haba:2013lga,Haba:2014zda,Hamada:2014xka,Haba:2014zja,Haba:2014sia,Bhattacharya:2014gva,Kawana:2014zxa} for a possible minimal extension of the SM with {\blue the dark matter and right-handed neutrinos}. 
}
in accordance with the following indications:
The renormalization group (RG) running of the quartic coupling $\lambda$ revealed that it takes the minimum value at around the Planck scale $\sim10^{18}\GeV$ and that the minimum value can be zero depending on the precise value of the top quark mass~\cite{Holthausen:2011aa,Bezrukov:2012sa,Degrassi:2012ry,Alekhin:2012py,Masina:2012tz,Hamada:2012bp,Jegerlehner:2013cta,Jegerlehner:2013nna,Hamada:2013cta,Buttazzo:2013uya,Branchina:2013jra,Kobakhidze:2014xda,Spencer-Smith:2014woa,Branchina:2014usa,Branchina:2014rva,Jegerlehner:2014xxa}.
We have also found that the bare Higgs mass can vanish at the Planck scale as well~\cite{Hamada:2012bp,Jegerlehner:2013cta,Jegerlehner:2013nna,Bian:2013xra,Hamada:2013cta,Masina:2013wja,Alsarhi:1991ji,Jones:2013aua}.\footnote{
See also Refs.~\cite{Jegerlehner:2014mua,Cherchiglia:2014gna,Bai:2014lea} for discussion of quadratic divergences.
}
That is, the Veltman condition~\cite{Veltman:1980mj} can be met at the Planck scale. In fact he speculates, {\it``This mass-relation, implying a certain cancellation between bosonic and fermionic effects, would in this view be due to an underlying supersymmetry.''} To summarize, it turned out that there is a triple coincidence: $\lambda$, its running, and the bare Higgs mass can all be accidentally small at around the Planck scale.

\begin{figure}[t]
\begin{center}
\hfill
\includegraphics[width=.5\textwidth]{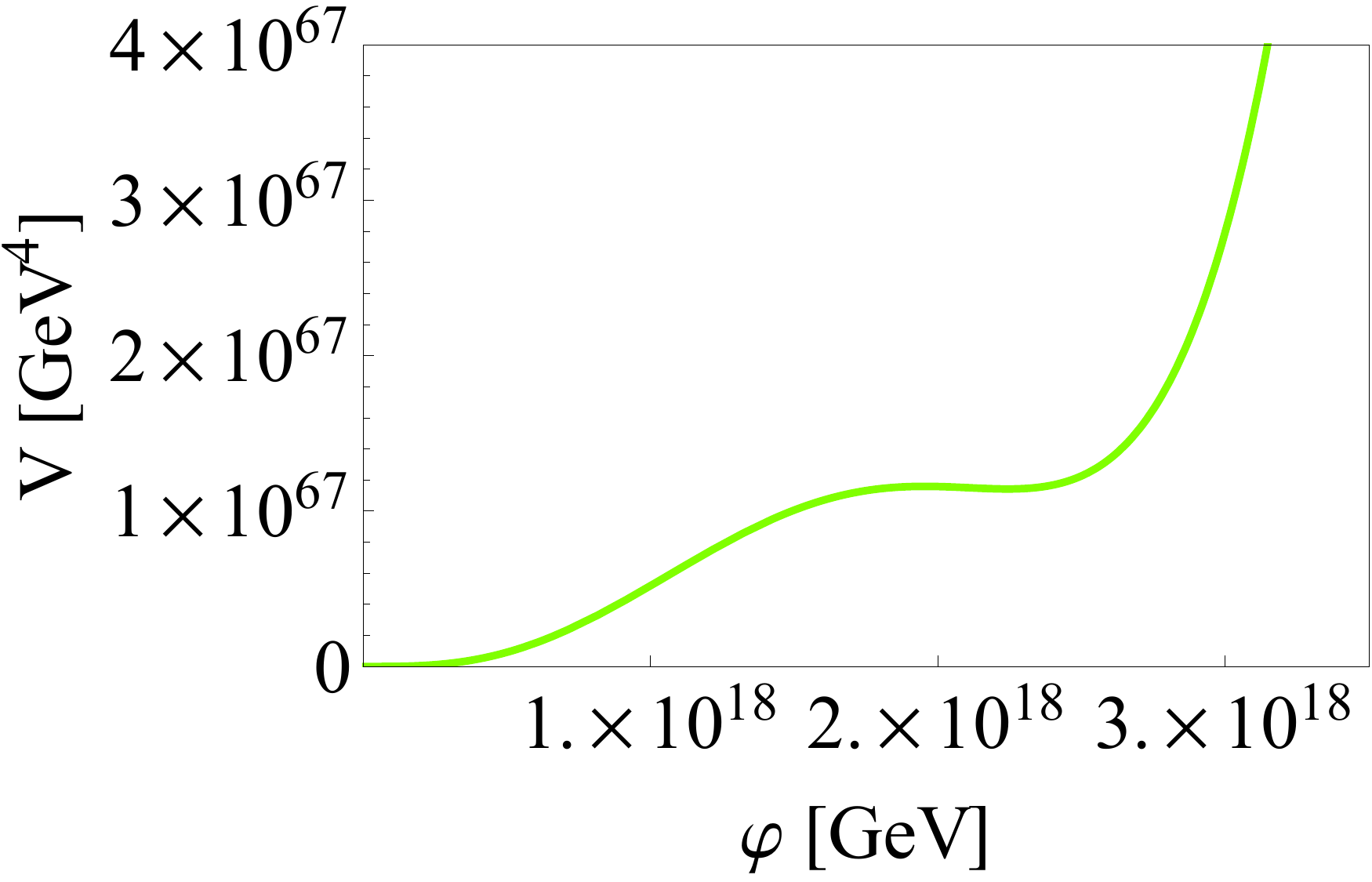}
\hfill\mbox{}
\caption{
{\blue
The SM Higgs potential $V$ as a function of the Higgs field $\varphi$.
Here we take $M_H=125\GeV$ and tune the top mass in such a way that the potential becomes flat; see e.g.\ Ref.~\cite{Hamada:2014wna}.
}
}\label{phenomenological Higgs potential}
\end{center}
\end{figure}

This is a direct hint for Planck scale physics in the context of  superstring theory.
The vanishing bare Higgs mass implies that the supersymmetry is restored at the Planck scale and that the Higgs field resides in a massless string state.
{\blue The} smallness of {\blue both} $\lambda$ and its beta function %suggests that %our electroweak vacuum is quasi-degenerate with the Planck scale one, 
{\blue is consistent with the Higgs potential being very flat around the string scale; see Fig.~\ref{phenomenological Higgs potential}.}\footnote{
\blue This is indeed suggested
%as is predicted 
by the multiple point criticality principle (MPP)~\cite{Froggatt:1995rt,Froggatt:2001pa,Nielsen:2012pu}, the classical conformality~\cite{Meissner:2007xv,Foot:2007iy,Meissner:2007xv,Iso:2009ss,Iso:2009nw,Hur:2011sv,Iso:2012jn,Chankowski:2014fva,Kobakhidze:2014afa,Gorsky:2014una,Kubo:2014ova,Foot:2014ifa,Kawana:2015tka}, the asymptotic safety~\cite{Shaposhnikov:2009pv}, the hidden duality and symmetry~\cite{Kawamura:2013kua,Kawamura:2013xwa}, and the maximum entropy principle~\cite{Kawai:2011qb,Kawai:2013wwa,Hamada:2014ofa,Kawana:2014vra,Hamada:2014xra}.
}
Such a flat potential opens up the possibility that the Higgs field plays the role of inflaton {\blue in the} early universe~\cite{Bezrukov:2007ep,Hamada:2013mya,Hamada:2014iga,Bezrukov:2014bra,Ko:2014eia,Xianyu:2014eba,Hamada:2014wna,Ibanez:2014swa,Bezrukov:2014ipa}.\footnote{
There are different models of the Higgs inflation involving higher dimensional operators~\cite{Germani:2010gm,Kamada:2010qe,Kamada:2012se,Nakayama:2014koa,Lee:2014spa}.
}
{\blue 
To understand the whole structure of the potential, it is crucial to investigate its behavior beyond the Planck scale.
The calculation based on field theory cannot be trusted in this region.
Although it is hard to reproduce the SM completely as a low energy effective theory of superstring, we can explore generic trans-Planckian structure of the Higgs field, under the assumption that the SM is close to a non-supersymmetric perturbative vacuum of superstring theory.
%It is interesting to explore the potential beyond the Planck scale. 
%However, the calculation based on field theory cannot be trusted in this region.
%Therefore, we use string theory to understand the whole structure of the Higgs potential.
}

%例えば, bare Higgs massが消えることはstring scaleでHiggsがstringのmassless stateであることとsupersymmetryがPlanck scaleで回復することを表す. さらに, $\lambda$とそのbeta function $\beta_\lambda$が$0$になることはHiggs potentialがhigh scaleで非常にflatになることを意味し, Higgsがinflatonであるscenarioを導く~\cite{Bezrukov:2007ep,Hamada:2013mya,Hamada:2014iga,Bezrukov:2014bra,Hamada:2014wna}. 

{\red In four dimensions, string theory has {\blue many more} tachyon-free non-super\-sym\-metric vacua than the supersymmetric ones. 
The latest LHC results suggest the possibility of the absence of the low energy supersymmetry, and the research based on the non-supersymmetric vacua is becoming more and more important~\cite{Blaszczyk:2014qoa,Angelantonj:2014dia,Hamada:2014eia,Nibbelink:2015ena,Abel:2015oxa}.}

{\blue 
% A virtue of such non-supersymmetric vacua is that 
In such non-supersymmetric vacua,}
{\red almost all the moduli are lifted up perturbatively,
contrary to the supersymmetric ones which typically possess tens or even hundreds of flat directions that cannot be raised purturbatively.}
{\blue
However, there remains a problem of instability in the non-supersymmetric models: The perturbative corrections generate tadpoles for the dilaton and other moduli such as the radii of toroidal compactifications.}
{\red
The dilaton can be stabilized within the perturbation series when $g_s\sim 1$~\cite{Dine:1985he}, or else by the balance between the one-loop and the non-perturbative potentials when $g_s$ is small~\cite{Abel:2015oxa}. In this paper, we assume that the dilaton is already stabilized.
}{\blue
We will discuss other instabilities than the dilaton direction in Sections~2 and 3.
}

%{\blue The Higgs-like field in this kind of toy model would share essential features with the Higgs field in the following more realistic models.}
We start from the tachyon-free non-supersymmetric vacua of the heterotic string theory.
We assume that the Higgs comes from a closed string and that its emission vertex at the zero momentum can be decomposed into a product of operators whose conformal dimensions are $(1,0)$ and $(0,1)$.
This is realized in the following cases for example:
\begin{itemize}
\item The Higgs comes from an extra dimensional component of a gauge field~\cite{Manton:1979kb,Fairlie:1979at,Salam:1981xd,RandjbarDaemi:1982hi,Hosotani:1983xw,Hosotani:1983vn,Hosotani:1988bm}.
\item The Higgs is the only one doublet in generic fermionic constructions~\cite{Kawai:1986va,Kawai:1986ah,Lerche:1986cx,Antoniadis:1986rn}.
\item The Higgs comes from an untwisted sector in the orbifold construction~\cite{Dixon:1985jw,Dixon:1986jc}; see e.g.\ Ref.~\cite{Blaszczyk:2014qoa} {\blue for a recent model-building example}.\footnote{
In Ref.~\cite{Blaszczyk:2014qoa} the SM-like one Higgs doublet model is constructed,
in which Higgs is realized as an extra dimensional gauge field.
For example, the model under the $\mathbb{Z}_{6-I}$ orbifold compactification of $SO(16)\times SO(16)$ heterotic string with the shift vector
\als{
V=\paren{-1/2,\,-1/2,\,1/6,\,1/2,\,-2/3,\,-1/2,\,0,\,1/6}\paren{-2/3,\,-1/2,\,0,\,-1/2,\,-1/2,\,0,\,1/6,\,2/3}
}
and Wilson lines
\als{
A_5
	&=	\paren{1/2,\,1/2,\,-1/2,\,5/6,\,-1/6,\,1/2,\,1/6,\,-1/2}\paren{1/2,\,-1/6,\,-5/6,\,7/6,\,1/6,\,5/6,\,1/2,\,-1/6} \\
A_6	&=	\paren{1/2,\,1/2,\,-1/2,\,5/6,\,-1/6,\,1/2,\,1/6,\,-1/2}\paren{1/2,\,-1/6,\,-5/6,\,7/6,\,1/6,\,5/6,\,1/2,\,-1/6}
}
fits in all the three criteria.
We thank the authors of Ref.~\cite{Blaszczyk:2014qoa} on this point.
}
\end{itemize}
Then we consider multiple insertions of such emission vertices to evaluate the effective potential.
It is very important to understand the whole shape of the Higgs potential in order to discuss the initial condition of the Higgs inflation, as well as to examine whether the MPP is realized or not. 
%We will show that a certain large field limit of the Higgs field with a fixed compactification radius fits into  three categories: runaway, periodic, and chaotic. 
%In all cases, 
We will show that, in the large field region, the Higgs potential is connected to a runaway vacuum
%This leads to a runaway vacuum 
with vanishing energy, which corresponds to opening up an extra dimension.
%We will argue that the Higgs field can be a source of an eternal inflation.
We find that such potential can realize an eternal inflation.

%We verify that such universal behavior indeed follows from the toroidal compactification of the non-supersymmetric $SO(16)\times SO(16)$ heterotic string theory. 
%We show that this behavior fits in the picture that the Higgs field is the source of the eternal inflation. 
%The observed small value of the cosmological constant of our universe may be understood as the degeneracy with this runaway vacuum, which has vanishing energy, as is suggested by the multiple point criticality principle.

This paper is organized as follows.
In Sec.~\ref{classification}, we show that the potential in the large field limit with fixed radius can be classified into the above three categories. 
In Sec.~\ref{SO(16)}, we compute the one-loop partition function as a function of a background field in $SO(16)\times SO(16)$ non-supersymmetric heterotic string on $\mathbb{R}^{1,8}\times S^1$, as a concrete toy model~\cite{Dixon:1986iz,AlvarezGaume:1986jb,Ginsparg:1986wr,Itoyama:1986ei,Itoyama:1987rc}. We explicitly check that the limiting behavior of the potential fits into the three categories mentioned above. We argue that physically this corresponds to opening up a multi degrees of freedom space above the Planck scale and that the runaway vacuum is a direction in this space.
In Sec.~\ref{eternal section}, we point out a possibility that the Higgs inflation is preceded by an eternal inflation, which occurs either in a domain wall or in a false vacuum.
In Sec.~\ref{solution to cosmological constant}, we show a possible explanation for the vanishing cosmological constant in terms of the MPP, and consider a possible mechanism to yield the observed value of the order of $\paren{\text{meV}}^4$.
In Sec.~\ref{summary}, we summarize our results. 
In Appendix~\ref{notation}, we summarize our notation for several mathematical functions. In Appendix~\ref{fermionic construction}, we review the fermionic construction that we use for the heterotic superstring theory. The computation of the partition function is also outlined. In Appendix~\ref{T-duality section}, we review the T-duality that we use in this work. In Appendix~\ref{MPP review}, we review the MPP.

%%%%%%%%%%%%%%%%%%%%%%%%%%%%%%%%%%%%%%%%%%%%%%%%%%%%%%%%%%%%

\section{Higgs potential in string theory}\label{classification}
In this section, we show how to treat the large constant background of a massless mode in closed string theory. In general, we start from a worldsheet action, say,
\al{
S_0={1\over 2\pi\alpha'}\int \df^2z\,G_{MN}\,\partial X^M  \bar{\partial} X^N+\cdots,
}
where $G_{MN}$ is the target space metric, $M,N,\dots$ run from $0$ to $D-1${\blue , and $\alpha'$ is the string tension}.
In general, a genetic massless string state has the emission vertex
\al{
\mathcal{O}\fn{z,\bar{z}}e^{ik\cdot X},
}
where $k^2=0$ and $\mathcal{O}\fn{z,\bar{z}}$ has conformal dimensions $(1,1)$ to preserve the conformal symmetry on the worldsheet.

\begin{figure}[tn]
\begin{center}
\hfill
\includegraphics[width=.9\textwidth]{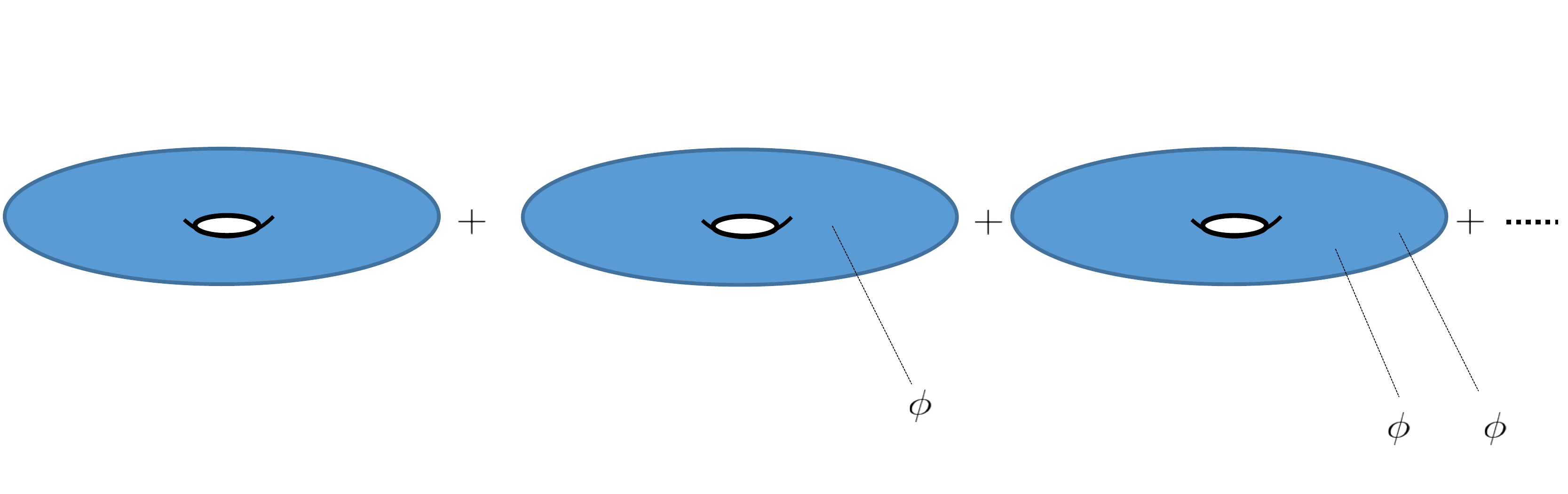}
\hfill\mbox{}
\caption{
Partition function under the presence of the background $\phi$. Summing up all the possible insertions of $\phi$, it exponentiates to yield Eq.~\eqref{shifted action}. This picture shows the one-loop case.
}\label{string diagram}
\end{center}
\end{figure}

As said in Introduction, we assume in this paper that the emission vertex at the zero momentum of the physical Higgs can be decomposed into a product of the $(1,0)$ operator $\mathcal O_L\fn{z}$ and the $(0,1)$ operator $\mathcal O_R\fn{\bar z}$:
\al{
\mathcal O\fn{z,\bar z}
	&=	\mathcal O_L\fn{z}\,\mathcal O_R\fn{\bar z}.
}
An operator of this form is exactly marginal: Insertions of the operator $\phi\,\mathcal O\fn{z,\bar z}$ can be exponentiated without renormalization, and hence the deformation of the worldsheet action
\al{
S	&=	S_0
		+\phi\int \df^2 z\,\mathcal{O} (z,\bar{z})
		\label{shifted action}
}
keeps the theory conformally invariant; see Fig.~\ref{string diagram}.

We want to know the effective potential for the background: $V\fn{\phi}$.
At the tree-level, the potential vanishes
\al{
V_\text{tree}(\phi)=0.
}
This is because the one-point function of any emission vertex, especially that of the graviton, vanishes on the sphere as it has non-zero conformal dimension. At the one-loop level and higher, we have non-zero effective potential.\footnote{
On the whole plane that is mapped from the sphere, an operator $\mathcal O$ with the scale dimension $d_s$ satisfies $\langle\mathcal O\fn{\lambda z}\rangle=\langle\mathcal O\fn{z}\rangle\lambda^{-d_s}$ and the translational invariance reads $\langle\mathcal O\fn{\lambda z}\rangle=\langle\mathcal O\fn{z}\rangle$. Hence we get $\langle\mathcal O\fn{z}\rangle=0$ for $d_s\neq0$. On the other hand, for torus and surfaces with higher genera, we cannot define the scale transformation, unlike the plane.
}

The $D$-dimensional energy density is given by
\al{
V_\text{$g$-loop}
	&=	-{Z_{g}\over \mathcal V_D},
		\label{energy density}
}
where $\mathcal V_D$ is the volume of $D$-dimensional spacetime and $Z_{g}$ is the partition function on the worldsheet with genus $g$ after moduli integration.
We note that the potential~\eqref{energy density} is given in the Jordan frame that does not yet make the gravitational action canonical; we will come back to this point in Secs.~\ref{graviton background} and \ref{S1 with Wilson}.

We emphasize that in string theory, the partition function $Z_g$ can be obtained even for the field value larger than the Planck scale, unlike the ordinary quantum field theory where infinite number of Planck-suppressed operators become relevant and uncontrollable.

Before generalizing to arbitrary compactification, we first analyze two simple examples to build intuition:
In Sec.~\ref{graviton background}, we study the large field limit of the radion, namely an extra dimensional component of the graviton under the toroidal compactification. This limit corresponds to the large radius limit of the compactified dimension.
In Sec.~\ref{momentum boost}, we further turn on the Wilson line and the anti-symmetric tensor field. We can analyze this setup by considering the corresponding boost in the momentum space~\cite{Narain:1985jj,Narain:1986am}.
From the analysis of the spectrum of these modes, we argue that the effective potential in the large field limit can be classified into three categories, namely, runaway, periodic, and chaotic.
(In Sec.~\ref{SO(16)}, we will confirm it by a concrete computation for the toroidal compactification of the $SO(16)\times SO(16)$ heterotic string theory.)

In Sec.~\ref{generalization}, we discuss more general compactifications, and show that the same classification holds.

\subsection{Radion potential}\label{graviton background}
As said above, we start from the toroidal compactification of the $(D-1)$th direction: $X^{D-1}\sim X^{D-1}+2\pi R$.
The emission vertex of the radion, $G_{D-1\,D-1}$, is
\al{
\partial X^{D-1}\bar{\partial} X^{D-1}\,e^{i k\cdot X}.
}
Its constant background is given by setting the momentum $k=0$.

We want the partition function with the radion background $\phi$:
\al{
S_\text{worldsheet}={1\over 2\pi\alpha'}\int \df^2z \paren{1+\phi}\partial X^{D-1}  \bar{\partial} X^{D-1}+\cdots.
}
In this case, we can transform the action into the original form with $\phi=0$ by the field redefinition
\al{
X'^{D-1}=\sqrt{1+\phi}\,X^{D-1},
}
which however changes the periodicity as
\al{
X'^{D-1}\sim X'^{D-1}+2\pi\sqrt{1+\phi}\,R.
}
That is, the radion background changes the radius of $S^1$ to
\al{
R'	&:=	\sqrt{1+\phi}\,R.
}
Therefore if the compactification radius $R'$ is large, the effective action is proportional to it, and the $(D-1)$-dimensional effective action for large $\phi$ becomes
\al{
S_\text{eff}
	&\sim
		\int \df^{D-1}x \sqrt{-g}\,R'\left(\mathcal{R}-C-{2\over R'^2}\paren{\p R'}^2\right)\nn
	&=	\int \df^{D-1}x \sqrt{-g}\sqrt{1+\phi}\,R\left(\mathcal{R}-C-{1\over2\paren{1+\phi}^2}\paren{\p\phi}^2\right)
		\label{potential in lower dimensions}
}
up to an overall numerical coefficient, where we have taken the $\alpha'=1$ units, $\mathcal R$ is the Ricci scalar in $(D-1)$-dimensions, and $C$ is a $\phi$-independent constant that is generated from loop corrections in the non-supersymmetric string theory.
$C$ can be viewed as the $D$-dimensional cosmological constant.

This can be confirmed at the one-loop level as follows. The radius dependent part of the one-loop partition function before the moduli integration is
\al{\label{radius dependence}
\sum_{n,w=-\infty}^\infty \exp\left[
2\pi i \tau_1 n w-\pi \tau_2 \alpha'\left(\left({n\over R'}\right)^2+\left({R' w\over \alpha'}\right)^2\right)
\right],
}
where $n$ and $w$ are the Kaluza-Klein (KK) and winding numbers, respectively, and $\tau=\tau_1+i\tau_2$ is the moduli of the worldsheet torus.
In the large radius limit $R'\gg\sqrt{\alpha'}$, we can rewrite Eq.~\eqref{radius dependence} by the Poisson resummation formula:
\al{
{R'\over\sqrt{\pi\tau_2\alpha'}}\sum_{m,w}
\exp\left[
-{\pi R'^2\over \alpha'\tau_2}|m-w\tau|^2
\right].
	\label{prop to R after Poisson}
}
We see that the partition function becomes indeed proportional to $R'$ in the large $R'$ limit. Note that in the large $R'$ limit, only the $w=0$ modes contribute, and hence that the winding modes are not important here.

We then rewrite the action~\eqref{potential in lower dimensions} in the Einstein frame.
In $(D-1)$-dimensions, the field redefinition by the Weyl transformation, $g_{\mu\nu}^\text{E}=e^{2\omega} g_{\mu\nu}$, gives us the volume element and the Ricci scalar in the Einstein frame as
\al{
\sqrt{-g^\text{E}}
	&=	e^{\paren{D-1}\omega}\sqrt{-g}, \\
\mathcal{R}^\text{E}
	&=	e^{-2\omega}\sqbr{
			\mathcal{R}
			-2\paren{D-2}\nabla^2\omega
			-\paren{D-3}\paren{D-2}g^{\mu\nu}\partial_\mu\omega\partial_\nu\omega
			},\label{curvature}
}
respectively. %, and the potential~\eqref{potential in lower dimensions} becomes
%\al{
%&\quad
%	\int \df^{D-1}x \sqrt{-g^\text{E}}\,e^{-\paren{D-1}\omega}R\paren{
%	e^{2\omega}\mathcal{R}^\text{E}
%	+\paren{D-3}\paren{D-2}e^{2\omega}g^{\text{E}\mu\nu}\partial_\mu\omega\partial_\nu\omega
%	-C
%	}\nn
%&=	
%S_\text{eff}
%	&=	\int \df^{D-1}x \sqrt{-g^\text{E}}\,e^{-\paren{D-3}\omega}R\paren{
%		\mathcal{R}^\text{E}
%		+\paren{D-3}\paren{D-2}g^{\text{E}\mu\nu}\partial_\mu\omega\partial_\nu\omega
%		-e^{-2\omega}C
%		}.
%}
By choosing $e^{\paren{D-3}\omega}=R'$, we get the Einstein frame action:
\al{
S_\text{eff}
	&=	\int \df^{D-1}x \sqrt{-g^\text{E}}\paren{
			\mathcal{R}^\text{E}
			+\paren{D-3}\paren{D-2}g^{\text{E}\mu\nu}\,\partial_\mu\omega\,\partial_\nu\omega
			-e^{-2\omega}C
			-{2\over R'^2}\paren{\p R'}^2
			}\nn
	&=	\int \df^{D-1}x \sqrt{-g^\text{E}}\paren{
			\mathcal{R}^\text{E}
			-{D-4\over D-3}{g^{\text{E}\mu\nu}\over R'^2}\partial_\mu R'\,\partial_\nu R'
			-{C\over R'^{2/\paren{D-3}}}
			}\nn
	&=	\int \df^{D-1}x \sqrt{-g^\text{E}}\paren{
			\mathcal{R}^\text{E}
			-{g^{\text{E}\mu\nu}\over2}\partial_\mu \chi\,\partial_\nu \chi
			-C\exp\fn{-\sqrt{2}\chi\over\sqrt{\paren{D-3}\paren{D-4}}}
			},
			\label{to Einstein}
}
%\al{
%\int \df^{D-1}x\,\sqrt{-g^\text{E}} 
%\left(
%\mathcal{R}^\text{E}+{3\over 2}{1\over R^2}g^{E\mu\nu}\partial_\mu R\partial_\nu R
%-{C\over R}
%\right),
%}
where the second term in Eq.~\eqref{curvature} has become a total derivative and we have defined $R'=:\exp\fn{{\chi\over\sqrt{2}}\sqrt{D-3\over D-4}}$. When $D>4$ and $C>0$, we see that the last term, the potential, becomes runaway for large $R'$ or $\chi$.\footnote{
The small radius limit $R'\ll\sqrt{\alpha'}$ is the same {\red as the large radius limit} due to the T-duality: {\red $R'\longleftrightarrow \alpha'/R'$}~\cite{Kikkawa:1984cp,Sakai:1985cs,Maharana:1992my}.
}

To summarize, the large field limit of the radion $\phi$, the extra dimensional component of the graviton, leads to the decompactification of the corresponding dimension.
{\red This decompactified vacuum corresponds to the runaway potential if the cosmological constant is positive~\cite{Appelquist:1982zs,Appelquist:1983vs}.}
{\red Since the large radius limit is equivalent to the weak coupling limit, the runaway vacuum corresponds to a free theory. Therefore this runaway nature is not altered by the higher order corrections. We will see in Section~\ref{summary} that this argument also applies to the dilaton background.}

\subsection{Boost on momentum lattice}\label{momentum boost}
As the second example, we turn on the backgrounds for graviton, gauge, and anti-symmetric tensor fields.
Let $p$ and $q$ be the numbers of the compactified dimensions in the left and right moving sectors of the closed string, other than our four dimensions. We take $p\geq q$ without loss of generality.
The spectrum of $(p+q)$-dimensional momenta $(\vec k_\text{L},\vec k_\text{R})$ of the non-oscillatory mode %under the compactification
is restricted to form an (even self-dual) momentum lattice, due to the modular invariance~\cite{Narain:1985jj,Narain:1986am}; see Appendix~\ref{bosonic part}. Different lattices that are related by the $SO(p,q)$ rotation of $(\vec k_\text{L},\vec k_\text{R})$ correspond to different compactifications, up to the $SO(p)\times SO(q)$ rotation that leaves $\vec k_\text{L}^2$ and $\vec k_\text{R}^2$ invariant. Therefore the compactifications are classified by the transformation
\al{\label{Narain}
{SO(p,q)\over SO(p)\times SO(q)}.
}
This is the moduli space of the theory at the tree level, which is lifted up by the loop corrections in non-supersymmetric string theory.

%The partition function (effective potential) contains a factor
%\al{
%\exp\fn{
%\pi i\tau_1 {\alpha'\over2}(k_\text{L}^2-k_\text{R}^2)-\pi \tau_2{\alpha'\over2}(k_\text{L}^2+k_\text{R}^2)
%}
%}
%in its integrand.
The boost in the momentum space corresponds to putting constant backgrounds for the degrees of freedom that are massless at the tree-level~\cite{Narain:1985jj,Narain:1986am}:
\al{
C_{ij}\,\p X^i_\text{L}\,\bar\p X_\text{R}^{\bar j},
	\label{C_ij}
}
where $i$ and $\bar j$ run for $1,\dots,p$ and $1,\dots,q$, respectively.
In terms of $q$-dimensional fields, they can be interpreted as the symmetric tensor (metric), antisymmetric tensor, and ${U(1)}^{p-q}$ gauge fields (Wilson lines), whose total number is
\al{
{q\paren{q+1}\over 2}+{q\paren{q-1}\over2}+q\paren{p-q}=pq.
}
Indeed, this agrees with the number of degrees of freedom of the coset space~\eqref{Narain}:
\al{
{\paren{p+q}\paren{p+q-1}\over 2}-{p\paren{p-1}\over2}-{q\paren{q-1}\over2}=pq.
}

\begin{figure}[tn]
\begin{center}
\hfill
\includegraphics[width=.44\textwidth]{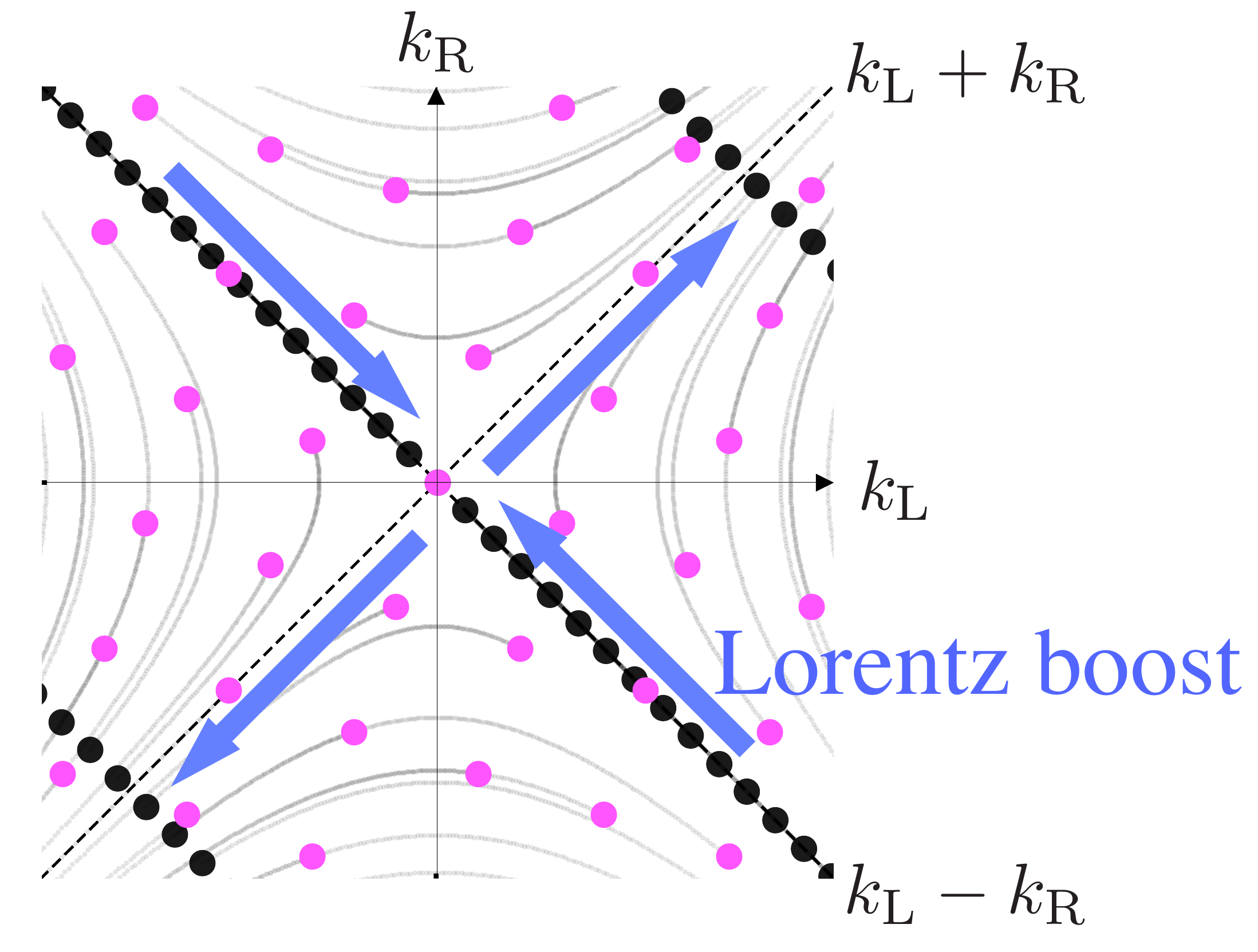}
%\hfill\mbox{}
%\\
\hfill
\includegraphics[width=.45\textwidth]{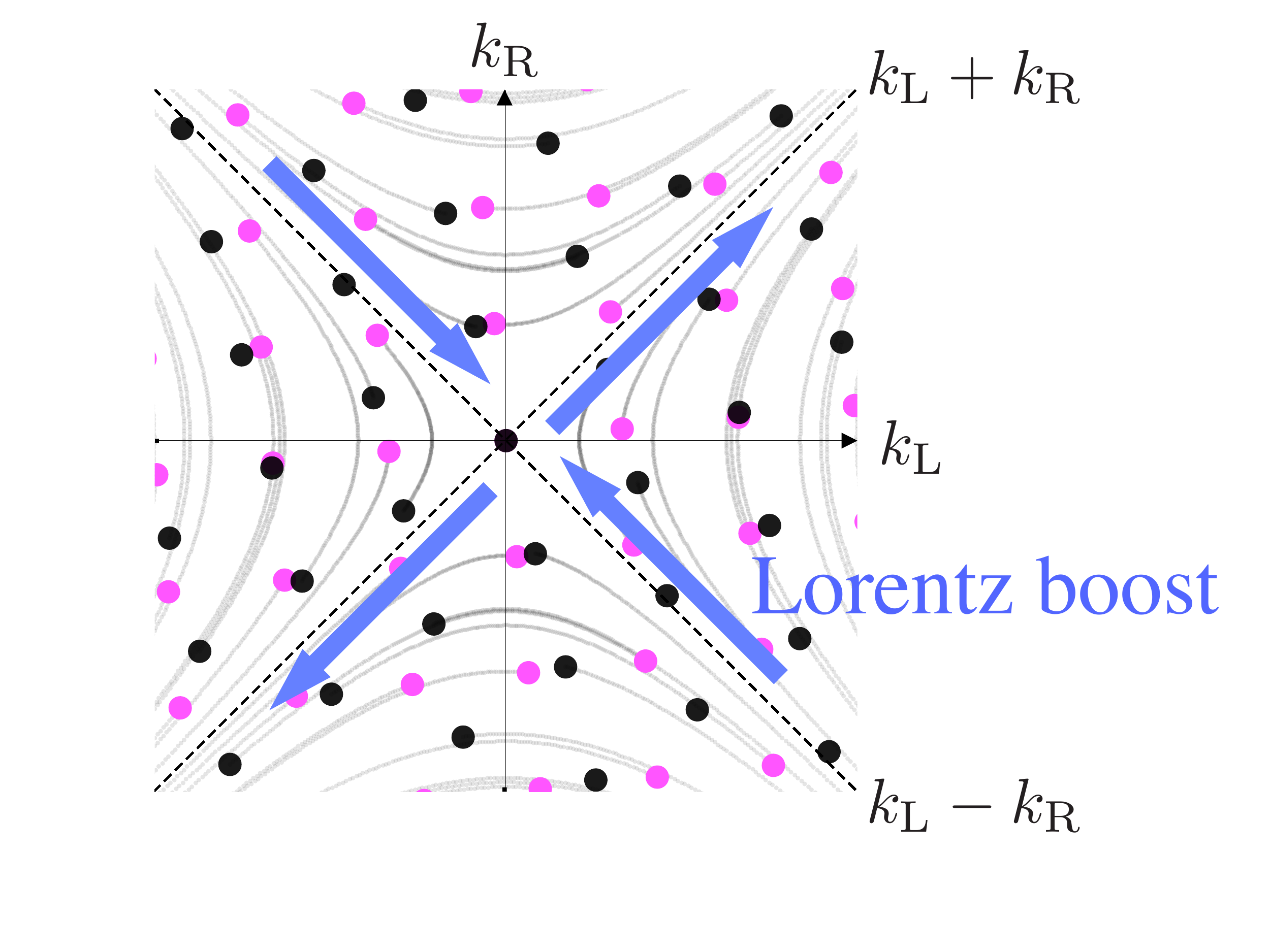}
\hfill\mbox{}
\caption{Schematic picture of the momentum boost in the $k_\text{R}$ vs $k_\text{L}$ plane. The light cone in the momentum space is depicted by the dashed diagonal lines. The sets of lighter (magenta) and black dots represent the initial momentum lattice and the one after the boost, respectively. Left: There exists a point of the initial lattice on the light cone. Then there exist infinite amount of its integer multiplications on the light cone. In the infinite boost limit, they are contracted to form a decompactified dimension, which is represented by the black dots. Right: There is no initial point on the light cone, and such a decompactification does not occur.
}\label{momentum lattice}
\end{center}
\end{figure}
We are interested in switching on the background of a single field. If the emission vertex of the field is given by $c_{i\bar j}\,\p X^i_\text{L}\,\bar\p X_\text{R}^{\bar j}$, this corresponds to adding 
\al{
\lambda\,c_{ij}\,\p X^i_\text{L}\,\bar\p X_\text{R}^{\bar j}
}
to the worldsheet action, where $\lambda$ represents the strength of the background.
In general, the $SO(p)\times SO(q)$ rotation can make $c_{i\bar j}$ into the diagonal form
%\al{
%C_{ij}	&\to
%	\bmat{
%		*	&	0	& \cdots	& 0 \\
%		0	&	*	&	\ddots	& \vdots \\
%		\vdots	&	\ddots	&\ddots &0 \\
%		0	&	\cdots &0 &* \\
%		0	&	\cdots & \cdots &0\\
%		\vdots & \vdots & \vdots & \vdots\\
%		0	&	\cdots & \cdots &0
%		}	
%}
\al{
c_{i\bar j}	&\to
	\bmat{
		*	&		& 	&  \\
			&	*	&		&  \\
			&		&\ddots &  \\
			&	 & &* \\
			&	 &  &\\
			&  &  & 
		},\label{c_ij}
}
where the blank slots stand for zero.
This background corresponds to the combination of $q$ boosts in the $1$-$\bar 1$, \dots, $q$-$\bar q$ planes.
That is, the $(p+q)$-dimensional vector
\al{
k=\paren{k_\text{L}^1,\dots,k_\text{L}^p;k_\text{R}^{\bar 1},\dots,k_\text{R}^{\bar q}}
}
is transformed by
\al{
\bmat{k_\text{L}'^i\\ k_\text{R}'^{\bar i}}
	&=	\bmat{\cosh\eta_i&\sinh\eta_i\\ \sinh\eta_i&\cosh\eta_i}
			\bmat{k_\text{L}^i\\ k_\text{R}^{\bar i}},	\nn
k_\text{L}'^j
	&=	k_\text{L}^j,
}
for $i=1,\dots,q$ and $j=q+1,\,\dots,\,p$.

Let us first consider the effect of a boost in a single plane:
\al{
\bmat{k_\text{L}'\\ k_\text{R}'}
	&=	\bmat{\cosh\eta&\sinh\eta\\ \sinh\eta&\cosh\eta}
			\bmat{k_\text{L}\\ k_\text{R}}.
}
Then one of $k_\text{L}\pm k_\text{R}$ is contracted and the other expanded:
\al{
k'_\text{L}+k'_\text{R}&=e^\eta\paren{k_\text{L}+k_\text{R}},\nn
k'_\text{L}-k'_\text{R}&=e^{-\eta}\paren{k_\text{L}-k_\text{R}}.
}
The effective potential in the large $\eta$ limit depends on whether or not there exists a lattice point on the light cone in this plane, as is illustrated schematically in Fig.~\ref{momentum lattice}.
There are two possibilities in the infinite boost limit:
\begin{itemize}
\item If a point in the initial momentum lattice sits on the light cone as in the left panel in Fig.~\ref{momentum lattice}, infinite amount of its integer multiplications on the light cone are contracted to form a continuous spectrum.
This behavior is the same as that of the KK momenta in the large radius limit discussed in Sec.~\ref{graviton background}.
The resultant partition function becomes proportional to the radius $R$.
The same argument as Sec.~\ref{graviton background} gives us the runaway potential.

\item If no point sits on the light cone in the initial momentum lattice, as in the right panel in Fig.~\ref{momentum lattice}, then the continuum is not formed by the infinite boost.
For a given amount of boost, the closest point to the origin contributes the most to the partition function. Then the potential becomes either periodic or chaotic for larger and larger boost.
\end{itemize}
The fate of the large field limit depends on whether or not a lattice point sits on the light cone of the boost plane in the momentum space.

In the case of the multiple boosts~\eqref{c_ij}, the boost in each plane is independent from the others. However, if there are several degenerate massless states as in Eq.~\eqref{C_ij}, we should better consider all of them simultaneously. As we will see in Sec.~\ref{SO(16)} in a concrete model, the asymptotic behavior of the potential remains essentially the same.

%%%%%%%%%%%%%%%%%%%%%%%%%%%%%%%%%%%%%%%%%%%%%%%%%%%%%%%%%%%%
\subsection{General compactifications}\label{generalization}
We discuss the large field limit in more general setup including compactification on a curved space, possibly involving orbifolding etc., or even the case without having a geometrical interpretation. 
We will show that the classification still holds: runaway, periodic, and chaotic.

\begin{figure}
\begin{center}
\includegraphics[width=.7\textwidth]{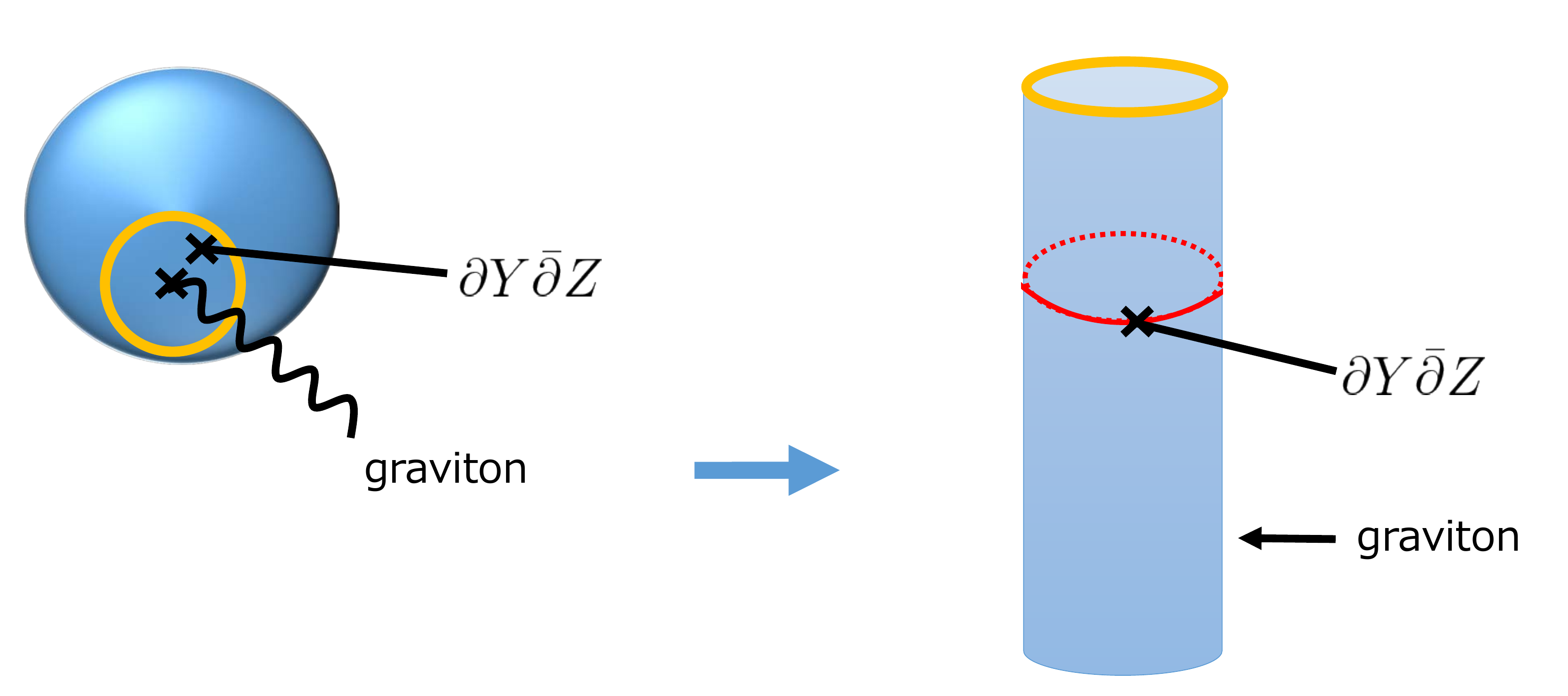}
\caption{Left: The Higgs emission vertex $\p Y\,\bar\p Z$ is single-valued around the graviton emission vertex because $Y$ and $Z$ are independent of the spacetime coordinate $X^\mu$. Right: Exponential mapping around the graviton emission vertex. $\p Y$ and $\bar\p Z$ are periodic around the cylinder, e.g.\ around the (red) circle.}\label{graviton}
\end{center}
\end{figure}
As said above, the emission vertex of a massless field must be written in terms of a $(1,1)$ operator, and we assume that this operator separates into the holomorphic and anti-holomorphic parts,
\al{
\mathcal{O}_{(1,1)}=\mathcal{O}_{(1,0)}\times \mathcal{O}_{(0,1)},
}
on the worldsheet. %{\bf(Kawai-check needed for this statement.)}
Then we can write at least locally, %after a possible bosonization procedure,
\al{
%\mathcal{O}_{(1,1)}
%	&=	\partial Y\,\bar{\partial} Z,
&\mathcal{O}_{(1,0)}=\partial Y,
&\mathcal{O}_{(0,1)}=\bar{\partial} Z,
}
%where $\p Y$ and $\bar\p Z$ are holomorphic and anti-holomorphic parts, respectively.
%Note that $\mathcal O_{(1,0)}$ has an operator product expansion (OPE)
%\al{
%:\partial Y(z)\,\partial Y(w):\,
%	&=	{1\over (z-w)^2}+\text{(regular)},
%}
%and hence 
where $Y$ and $Z$ are free worldsheet scalars.
%\footnote{
%The term $1/(z-w)$ does not appear because of the symmetry $z\leftrightarrow w$.
%}
If we further assume that the Higgs field is uniquely identified, i.e., that it does not mix with other massless states at the tree level, then it suffices to consider a single background as in Eq.~\eqref{shifted action}. In this case we may not need to consider the multi-field potential discussed above.
%\footnote{
%At higher field values than the Planck scale, there may be a mixing with the other string states. Even so, the asymptotic behavior will be essentially the same as the above single boost case.
%}

We can show that $\partial Y$ and $\bar\p Z$ are periodic at least in one sector:
In fact, if we insert the graviton emission vertex $\p X^\mu\,\bar\p X^\nu e^{ik\cdot X}$ near the Higgs emission vertex, the latter is single valued in the neighborhood of the former. This is because $Y$ and $Z$ are independent of the spacetime coordinates $X^\mu$. Therefore, $\partial Y$ and $\bar\p Z$ are periodic in the graviton sector; see Fig.~\ref{graviton}.

In such a sector, we can mode-expand $\partial Y$ and $\bar\p Z$.
Let us consider the simultaneous eigenvalues $\paren{p_Y,p_Z}$ of the constant modes of $\p Y$ and $\bar\p Z$.
The set of the pairs of eigenvalues form a momentum lattice $\Gamma_P=\Set{\paren{p_Y,p_Z}}$: If there exist states $s_1$ and $s_2$ with momenta $\paren{p_{Y1},p_{Z1}}$ and $\paren{p_{Y2},p_{Z2}}$, respectively, there is a state with the momentum $\paren{p_{Y1}+p_{Y2},\,p_{Z1}+p_{Z2}}$; such a state appears when $s_1$ and $s_2$ merge.  
If $\Gamma_P$ contains a non-zero vector, it forms the momentum lattice.
Then the same argument applies as in Sec.~\ref{momentum boost}.
Putting a constant background for $\mathcal{O}_{(1,1)}$ corresponds to the momentum boost.
If there is a point on the light cone with $p_Y/p_Z$ being a rational number, then a runaway direction emerges in the infinite boost limit. If not, namely if there is no such point, then the potential becomes chaotic.

%%%%%%%%%%%%%%%%%%%%%%%%%%%%%%%%%%%%%%%%%%%%%%%%%%%%%%%%%%%%
\section{$SO(16)\times SO(16)$ heterotic string}\label{SO(16)}
We verify the argument in the previous section in the concrete model: the $SO(16)\times SO(16)$ heterotic string theory~\cite{Dixon:1986iz,AlvarezGaume:1986jb}. This model breaks supersymmetry at the string scale but, unlike the bosonic string theory in 26 dimensions, the tachyonic modes are projected out as in the ordinary heterotic superstring theories.
In the fermionic construction, the modular invariance of the partition function restricts the allowed set of the fermion numbers in Neveu-Schwarz (NS) and Ramond (R) sectors.
The classification of the ten dimensional string theories is completed in Ref.~\cite{Kawai:1986vd}.
The $SO(16)\times SO(16)$ model~\cite{Dixon:1986iz,AlvarezGaume:1986jb} is the only one that has neither a tachyon nor a supersymmetry in ten dimensions.

We write the uncompactified dimensions $X^\mu$ ($\mu=0,\dots,9$), and the compactified ones $X_L^I$ ($I=1,\dots,16$) for the left movers.
We then compactify this model on $S^1$~\cite{Ginsparg:1986wr}:
\al{
X^9	&\sim X^9+2\pi R.
}
We further turn on a Wilson line for the gauge field $A_{\mu=9}^{I=1}$, and compute the one-loop partition function.

In Appendix~\ref{fermionic construction}, we spell out the construction of the model and the computation of the partition function; the notations for the theta functions are put in Appendix~\ref{notation}. 
In Sec.~\ref{SO(16)xSO(16) partition function}, we review the partition function in the $SO(16)\times SO(16)$ heterotic string theory in 10 dimensions. In Sec.~\ref{S1 with Wilson} we compute the one-loop partition function of this model for the case described above.
%%%%%%%%%%%%%%%%%%%%%%%%%%%%%%%%%%%%%%%%%%%%%%%%%%%%%%%%%%%%

%%%%%%%%%%%%%%%%%%%%%%%%%%%%%%%%%%%%%%%%%%%%%%%%%%%%%%%%%%%%%%%%%%%%%%
\subsection{Partition function of $SO(16)\times SO(16)$ string}
\label{SO(16)xSO(16) partition function}
We first review the computation of the partition function in the $SO(16)\times SO(16)$ non-supersymmetric heterotic string~\cite{Dixon:1986iz,AlvarezGaume:1986jb}. Here we have chosen a non-super\-sym\-metric string as a toy model because, as discussed in Introduction, the low energy data at the electroweak scale suggests via the Veltman condition that the supersymmetry is broken at the Planck scale. In such a non-supersymmetric theory, the flat direction of the effective potential is raised perturbatively. Detailed procedure of the fermionic construction of the model is explained in Appendix~\ref{formalism section} and \ref{concrete heterotic models}.

Let us write down the contribution from the momentum lattice after the bosonization in each $\alpha\vec w$ sector:
\al{
\Zhat_{T^2,\alpha\vec w}
	&=	\Tr_\text{$\alpha\vec w$} e^{2\pi i\tau_1\paren{L_0-\bar L_0}-2\pi\tau_2\paren{L_0+\bar L_0}}\bigg|_\text{momentum lattice}.
}
In our case, they are
\al{
\Zhat_{T^2,\vec{0}}&=\phantom{-}{1\over8}\paren{\paren{\bar{\vartheta}_{00}}^4-\paren{\bar{\vartheta}_{01}}^4}\paren{\paren{\vartheta_{00}}^8+\paren{\vartheta_{01}}^8}^2,\nn
\Zhat_{T^2,\vec w_0}&=-{1\over8}\paren{\bar{\vartheta}_{10}}^4\paren{\vartheta_{10}}^{16},\nn
\Zhat_{T^2,\vec w_1}&=\phantom{-}{1\over8}\paren{\paren{\bar{\vartheta}_{00}}^4-\paren{\bar{\vartheta}_{01}}^4}\paren{\vartheta_{10}}^{16},\nn
\Zhat_{T^2,\vec w_2}&=\phantom{-}{1\over8}\paren{\paren{\bar{\vartheta}_{00}}^4+\paren{\bar{\vartheta}_{01}}^4}\paren{\vartheta_{10}}^8\paren{\paren{\vartheta_{00}}^8-\paren{\vartheta_{01}}^8},\nn
\Zhat_{T^2,\vec w_0+\vec w_1}&=-{1\over8}\paren{\bar{\vartheta}_{10}}^4\paren{\paren{\vartheta_{00}}^8-\paren{\vartheta_{01}}^8}^2,\nn
\Zhat_{T^2,\vec w_0+\vec w_2}&=-{1\over8}\paren{\bar{\vartheta}_{10}}^4\paren{\paren{\vartheta_{00}}^8+\paren{\vartheta_{01}}^8}\paren{\vartheta_{10}}^8,\nn
\Zhat_{T^2,\vec w_1+\vec w_2}&=\phantom{-}{1\over8}\paren{\paren{\bar{\vartheta}_{00}}^4+\paren{\bar{\vartheta}_{01}}^4}\paren{\vartheta_{10}}^8\paren{\paren{\vartheta_{00}}^8-\paren{\vartheta_{01}}^8},\nn
\Zhat_{T^2,\vec w_0+\vec w_1+\vec w_2}&=-{1\over8}\paren{\bar{\vartheta}_{10}}^4\paren{\paren{\vartheta_{00}}^8+\paren{\vartheta_{01}}^8}\paren{\vartheta_{10}}^8,
		\label{SO(16) partition function}
}
where $\vec 0$ and $\vec{w}_i$ are basis vectors for the boundary conditions on the fermions; see Appendix~\ref{fermion one-loop} for details.

Let us sum up all the above contributions, multiplied by those from the oscillator modes in the bosonization.
Including also the spacetime momentum and oscillator modes from the bosonic $X^m$ ($m=2,\dots,9$), we get the one-loop vacuum amplitude~\cite{Dixon:1986iz,AlvarezGaume:1986jb}:
\al{
Z_{T^2}&={V_{10}\over\alpha'^5}{1\over2\paren{2\pi}^{10}}\int_F {\df\tau_1\,\df\tau_2\over \tau_2^6}{1\over\ab{\eta(\tau)}^{16}{\eta(\tau)}^{16}\,{\bar{\eta}(\bar{\tau})}^4}
\sum_\text{sector $\alpha\vec w$}\hat Z_{T^2,\alpha\vec w}\nn
&=
{V_{10}\over\alpha'^5}{1\over4 (2\pi)^{10}}\int_F {\df\tau_1\,\df\tau_2\over \tau_2^6}{1\over\ab{\eta(\tau)}^{16}{\eta(\tau)}^{16}\,{\bar{\eta}(\bar{\tau})}^4}\nn
&\quad
\times
\left[
\paren{\bar{\vartheta}_{01}}^4\paren{\vartheta_{10}}^8\paren{\paren{\vartheta_{00}}^8-\paren{\vartheta_{01}}^8}
+\paren{\bar{\vartheta}_{10}}^4\paren{\vartheta_{01}}^8\paren{\paren{\vartheta_{00}}^8-\paren{\vartheta_{10}}^8}
\right],
	\label{partition function for non-SUSY}
}
where $F$ represents the fundamental region,
\al{
F	&:=	\Set{\paren{\tau_1,\tau_2}|-1/2\leq\tau_1\leq1/2,\ \ab{\tau}=\ab{\tau_1+i\tau_2}\geq 1},
}
and we have used the Jacobi's identity:
\al{
\paren{\bar{\vartheta}_{00}}^4-\paren{\bar{\vartheta}_{01}}^4-\paren{\bar{\vartheta}_{10}}^4=0.
}
We can see from this identity that the contributions between $\vec{w}_0$ and $\vec{w}_1$ cancel.
By the numerical calculation, we obtain~\cite{Dixon:1986iz,AlvarezGaume:1986jb}
\al{
\rho_{10}
	&=	-{Z_{T^2}\over V_{10}}\simeq \paren{3.9\times 10^{-6}} {1\over\alpha'^5}.
}
%%%%%%%%%%%%%%%%%%%%%%%%%%%%%%%%%%%%%%%%%%%%%%%%%%%%%%%%%%%%%%%%%%%%%%%5
\subsection{$S^1$ compactification with Wilson line}
\label{S1 with Wilson}
Now we compactify the $m=9$ direction on $S^1$ with radius $R$: $X^9\sim X^9+2\pi R$~\cite{Ginsparg:1986wr}.
Here we consider a large field limit of an extra $m=9$ dimensional component of the gauge field, $A_{m=9}^{I=1}$.
%We will show that the Wilson-line $A_9^1$ has a period $r$, which is a function of $A_9^1$ and $R$.
%We will show that $r$ gives the physical compactification radius.
We will find three possible large field limits discussed in the previous section.

The emission vertex for the gauge field with the polarization and momentum $\epsilon^m$ and $k$, respectively, is
\al{
\epsilon_m\left(i\bar{\partial}X^m+{\alpha'\over 2} \paren{k\cdot\psi_R} \psi^{m}_R\right)\partial X^I_L\,e^{ik\cdot X},
	\label{gauge vertex op}
}
where indices run such that $m=2,\dots,9$ and $I=1,\dots,16$.
We see by putting $k=0$ in Eq.~\eqref{gauge vertex op} that a constant background $A_m^I$ corresponds to adding
\al{
A_m^I\int \df^2z\,\bar{\partial}X^{m}\,\partial X^I_{L},
	\label{vertex op of A}
}
to the worldsheet action.\footnote{
In obtaining the constant background by putting $k=0$, it is again important that the $A$ is massless at the tree-level.
}
In particular, we switch on the component of $I=1$ and $m=9$, and write $A:=A_9^1$:
\al{
A\int \df^2z\,\bar{\partial}X^{9}\,\partial X^1_{L}.
	\label{vertex op of A comp}
}

Turning on the Wilson line background $A$ does not affect the oscillator modes since Eq.~\eqref{vertex op of A} is a total derivative in the worldsheet action; only the momentum lattice of the center-of-mass mode is changed by $A$.

Let $l_L$ be the momentum of $X^{I=1}_L$. After fermionization, we have
\al{
l_L&=\sqrt{{2\over\alpha'}}m,
}
where $m\in\mathbb{Z}$ and $\mathbb{Z}+1/2$ for the NS (anti-periodic) and R (periodic) boundary conditions, respectively.
Let $p_L$ and $p_R$ be the spacetime momenta of the $S^1$-compactified direction $X^{m=9}$ for the left and right movers, respectively:
\al{
p_L&={n\over R}+{R w\over\alpha'},\nn
p_R&={n\over R}-{R w\over\alpha'},
	\label{internal momentum}
}
where $n\in\mathbb Z$ and $w\in\mathbb Z$ are the KK and winding numbers, respectively.

Turning on the background $A$ corresponds to the boost on the momentum lattice~\cite{Narain:1986am}:
\al{
\bmat{
l_L'\\
p_R'
}
=
\bmat{
\cosh\eta&\sinh\eta\\
\sinh\eta&\cosh\eta
}
\bmat{
l_L\\
p_R
},\label{lL pR mixing}
}
since there appears only $l_L$ and $p_R$ in Eq.~\eqref{vertex op of A comp}.
This boost necessarily changes the radius of the compactification too.

we will see that the identification
\al{
A	&=	\sinh\eta,
	\label{A-id}
}
gives the correct answer below.
Let us define $r$ by
%\footnote{KK massをみて$r$を半径と呼んでいる. 
%T-dualityは$R\leftrightarrow \alpha'/R$. $r$で書くと $r\leftrightarrow \alpha'/\paren{1+A^2}r$.
%T-dualityはさらに$A\leftrightarrow A+\sqrt{2\alpha'}/r$というperiodicityを含む.
%$\tau_1=rA, \tau_2=r$と読み替えるとこれはまさにmodular変換 $SL(2,\mathbb{Z})$, $\tau\rightarrow \tau+1, \tau\rightarrow -1/\tau$.
%fundamental regionは$-1/2\leq rA\leq1/2, (rA)^2+r^2\geq 1$.
%partition functionは$\tau_1$方向にはperiodicityを持ち, $\tau_2$方向へはlinearに増えていく.
%See Fig.~\ref{Tduality}.
%}
\al{\label{r-id}
%A	&=	\sinh\eta,	&
r	&:=	{R\over\cosh\eta},
%{R\over r}
%	&=	\cosh\eta, %=\sqrt{1+A^2}.
}
which will turn out to be the compactification radius in the presence of $A$.
%We call $r$ physical radius because, as will shortly see, it gives the mass separation of the KK modes in the large $r$ limit.
Note that in the language of Sec.~\ref{momentum boost}, we have 17 left-moving and 1 right-moving internal dimensions ($p=17$ and $q=1$).
The non-trivial transformations on the compactified space are
\al{
{SO(17,1)\over SO(17)}.
	\label{tree moduli space}
}
Among them, we have chosen the boost between the left $I=1$ and right $m=9$ dimensions with the momenta $l_L$ and $p_R$, respectively.
The left momentum of the $m=9$ dimension, $p_L$, is untouched.
We will soon use the rotation between $l_L$ and $p_L$ that belongs to $SO(17)$.

We now show the validity of the identification~\eqref{A-id}.
%there is a periodicity in $A$ when we vary $A$ and $R$ in such a way that $r=R/\sqrt{1+A^2}$ is fixed. 
In terms of $A$ and $r$, we have
\al{\label{pR''}
p_R'	
	&=	p_R\cosh\eta +l_L\sinh\eta &
	&=	{n\over r}-{rw\over\alpha'}\paren{1+A^2}+\sqrt{{2\over\alpha'}}m A,\\
l_L'	
	&=	l_L\cosh\eta +p_R\sinh\eta	&
	&=	\sqrt{{2\over\alpha'}}m \sqrt{1+A^2}+{n\over r}{A\over\sqrt{1+A^2}}-{rw\over\alpha'}A\sqrt{1+A^2},\\
p_L'	
	&=	p_L	&
	&=	{n\over r}{1\over\sqrt{1+A^2}}+{r w\over\alpha'}\sqrt{1+A^2}.
}
We further rotate by a part of $SO(17)$ in Eq.~\eqref{tree moduli space},
\al{
\bmat{
l_L''\\
p_L''
}
&=	
\bmat{
\cos\theta&\sin\theta\\
-\sin\theta&\cos\theta
}
\bmat{
l_L'\\
p_L'
},	\label{lL pL mixing}\\
p_R''
	&=	p_R', \nonumber
}
with
\al{
\cos\theta	&=	{1\over\sqrt{1+A^2}},	&
\sin\theta	&=	-{A\over\sqrt{1+A^2}},
}
to get
\al{
l_L''&=\sqrt{{2\over\alpha'}}m-2{rw\over\alpha'}A,\\
p_L''&={n\over r}+{rw\over\alpha'}\paren{1-A^2}+\sqrt{{2\over\alpha'}}mA.
\label{pL''}
}
The spectrum becomes
\al{
\sum_\text{all modes}\paren{l_L''^2+p_L''^2+p_R'^2}
	&=	\sum_{m,n,w}\Bigg[
			\paren{\sqrt{{2\over\alpha'}}m-2{rw\over\alpha'}A}^2
			+\paren{{n\over r}+{rw\over\alpha'}\paren{1-A^2}+\sqrt{{2\over\alpha'}}mA}^2\nn
	&\phantom{\mbox{}=	\sum_{m,n,l}\Bigg[}
			+\paren{{n\over r}-{rw\over\alpha'}\paren{1+A^2}+\sqrt{{2\over\alpha'}}m A}^2
			\Bigg].
			\label{spectrum with Wilson line}
}
%The computation in terms of the momentum boost~\eqref{lL pR mixing} might look rather unconventional, but t
As promised, this result~\eqref{spectrum with Wilson line} correctly reproduces that in Ref.~\cite{Narain:1986am,Ginsparg:1986wr}, which is obtained from the quantization of the scalar field under constraints.
Furthermore, from Eq.~\eqref{spectrum with Wilson line}, we see 
\al{
\left.\paren{l_L''^2+p_L''^2}\right|_{m=w=0}
	=	\left.p_R''^2\right|_{m=w=0}
	=	{n^2\over r^2},
		\label{large r limit summation}
}
which indicates that $r$ is the physical radius of $S^1$.

Now let us discuss the T-dual transformations that can be read off from the above result.
\begin{itemize}
\item We can see that the shift
\al{
A	&\rightarrow
		A+{\sqrt{2\alpha'}\over r}
		\label{A shift}
}
leaves the spectrum~\eqref{spectrum with Wilson line} unchanged.\footnote{
After the shift of $A$, redefine the mode numbers by $n'=n+2m-2w$, $w'=w$, and $m'=m-2w$.
}
\item From Eq.~\eqref{internal momentum}, we see that the spectrum is invariant under the T-dual transformation~\cite{Kikkawa:1984cp,Sakai:1985cs,Maharana:1992my}
\al{
R	&\to {\alpha'\over R},
	\label{T-dual tf basic}
}
or in terms of $r$ and $A$, $r\to \alpha'/\paren{1+A^2}r$.
\end{itemize}
By defining
\al{\label{tautilde}
\tilde\tau
	&=	\tilde\tau_1+i\tilde\tau_2
	:=	{r A\over\sqrt{\alpha'}}+i{r\over\sqrt{\alpha'}},
}
we can write down the enlarged T-dual transformation:\footnote{
The $S$-transformation is the transformation~\eqref{T-dual tf basic} composed with $A\to-A$, while the $T$ is Eq.~\eqref{A shift}.
}
\al{
&S:\quad \tilde\tau\rightarrow-{1\over\tilde\tau}\nn
&T:\quad \tilde\tau\rightarrow\tilde\tau+\sqrt{2}.
	\label{T-duality transformation}
}
The general form of the T-dual transformation is
\al{\label{general transformation}
&\tilde\tau'={a\tilde\tau+b\over c\tilde\tau+d},
}
where $ad-bc=1$ and $a$, $b$, $c$, and $d$ are either
\al{
a	&\in	\mathbb Z,	&
b	&\in	\sqrt{2}\mathbb Z,	&
c	&\in	\sqrt{2}\mathbb Z,	&
d	&\in	\mathbb Z,	
}
or
\al{
a	&\in	\sqrt{2}\mathbb Z,	&
b	&\in	\mathbb Z,	&
c	&\in	\mathbb Z,	&
d	&\in	\sqrt{2}\mathbb Z.
}
The fundamental region is $-1/\sqrt{2}\leq \tilde\tau_1\leq1/\sqrt{2},\, \ab{\tilde\tau}\geq 1$.
More details can be found in Appendix~\ref{T-duality section}.

\subsection{Effective potential under Wilson line}
Let us write down the contribution from the momentum lattice after the bosonization
%fermionic CM modes 
in each sector $\alpha\vec w$; this time we include the momentum \eqref{internal momentum} of the $S^1$-compactified $X^{m=9}$ which is modified by the Wilson line $A$ as in Eqs.~\eqref{pR''} and \eqref{pL''}:
\al{
%\tilde Z_{T^2,\alpha\vec w}
%	&=	\Tr_{\text{momenta\,\&}\,X^9,\ \text{CM},\ \alpha\vec w} e^{2\pi i\tau_1\paren{L_0-\bar L_0}-2\pi\tau_2\paren{L_0+\bar L_0}}.
\tilde Z_{T^2,\alpha\vec w}
	&=	\Tr_{\alpha\vec w} e^{2\pi i\tau_1\paren{L_0-\bar L_0}-2\pi\tau_2\paren{L_0+\bar L_0}}\bigg|_{\text{momentum lattice}}.
}
Concretely,
\al{
\tilde Z_{T^2,\vec{0}}&=\phantom{-}{1\over8}\paren{\paren{\bar{\vartheta}_{00}}^4-\paren{\bar{\vartheta}_{01}}^4}\sum_{m\in\mathbb Z} g_m\fn{\eta,R}
\paren{\paren{\vartheta_{00}}^7+\paren{-1}^m\paren{\vartheta_{01}}^7}\paren{\paren{\vartheta_{00}}^8+\paren{\vartheta_{01}}^8},
\nn
\tilde Z_{T^2,\vec w_0}&=-{1\over8}
\paren{\bar{\vartheta}_{10}}^4
\sum_{m\in\mathbb Z+1/2}g_m\fn{\eta,R}
\paren{\vartheta_{10}}^{15},
\nn
\tilde Z_{T^2,\vec w_1}&=\phantom{-}{1\over8}\paren{\paren{\bar{\vartheta}_{00}}^4-\paren{\bar{\vartheta}_{01}}^4}
\sum_{m\in\mathbb Z+1/2}g_m\fn{\eta,R}
\paren{\vartheta_{10}}^{15},
\nn
\tilde Z_{T^2,\vec w_2}&=\phantom{-}{1\over8}\paren{\paren{\bar{\vartheta}_{00}}^4+\paren{\bar{\vartheta}_{01}}^4}
\sum_{m\in\mathbb Z+1/2}g_m\fn{\eta,R}
\paren{\vartheta_{10}}^7\paren{\paren{\vartheta_{00}}^8-\paren{\vartheta_{01}}^8},
\nn
\tilde Z_{T^2,\vec w_0+\vec w_1}&=-{1\over8}\paren{\bar{\vartheta}_{10}}^4
\sum_{m\in\mathbb Z}g_m\fn{\eta,R}
\paren{\paren{\vartheta_{00}}^7-\paren{-1}^m\paren{\vartheta_{01}}^7}
\paren{\paren{\vartheta_{00}}^8-\paren{\vartheta_{01}}^8},
\nn
\tilde Z_{T^2,\vec w_0+\vec w_2}&=-{1\over8}\paren{\bar{\vartheta}_{10}}^4
\sum_{m\in\mathbb Z}g_m\fn{\eta,R}
\paren{\paren{\vartheta_{00}}^7+\paren{-1}^m\paren{\vartheta_{01}}^7}
\paren{\vartheta_{10}}^8,
\nn
\tilde Z_{T^2,\vec w_1+\vec w_2}&=\phantom{-}{1\over8}\paren{\paren{\bar{\vartheta}_{00}}^4+\paren{\bar{\vartheta}_{01}}^4}
\sum_{m\in\mathbb Z}g_m\fn{\eta,R}
\paren{\paren{\vartheta_{00}}^7-\paren{-1}^m\paren{\vartheta_{01}}^7}\paren{\vartheta_{10}}^8,
\nn
\tilde Z_{T^2,\vec w_0+\vec w_1+\vec w_2}&=-{1\over8}\paren{\bar{\vartheta}_{10}}^4
\sum_{m\in\mathbb Z+1/2}g_m\fn{\eta,R}
\paren{\vartheta_{10}}^7
\paren{\paren{\vartheta_{00}}^8+\paren{\vartheta_{01}}^8},
	\label{partition function under A}
}
where
\al{
g_m\fn{\eta,R}
&=
\sum_{n,w=-\infty}^\infty \exp\left[
	\pi i \alpha'{\tau_1\over2}\paren{l_L''^2+p_L''^2-p_R''^2}
	-{\pi\over 2}\tau_2\alpha'\paren{l_L''^2+p_L''^2+p_R''^2}
\right]
\nn
&=\sum_{n,w=-\infty}^\infty \exp\left[
\pi i \tau_1\paren{m^2+2 n w}-{\pi\over 4}\tau_2\alpha'
	\paren{
	e^{2\eta}\paren{l_L+p_R}^2+e^{-2\eta}\paren{l_L-p_R}^2+2p_L^2
	}
\right]\label{Wilson-line-dependent part}
}
contains the information of the Wilson line.
We can check that the $\eta\to 0$ limit reduces Eq.~\eqref{partition function under A} to Eq.~\eqref{SO(16) partition function}, multiplied by the contribution from the compactified dimension shown in Appendix.~\ref{bosonic part}.

Including the oscillator modes and the spacetime coordinates $X^m$ ($m=2,\dots,9$), we get
\al{
Z_{T^2}
&={V_{9}\over\alpha'^{9/2}}{1\over2 (2\pi)^9}\int_F {d\tau_1 d\tau_2\over \tau_2^{11/2}}{1\over\ab{\eta(\tau)}^{16}{\eta(\tau)}^{16}{\bar{\eta}(\bar{\tau})}^4}
\sum_\text{sector $\alpha\vec w$}\tilde{Z}_{T^2,\alpha \vec{w}}\nn
&={V_{9}\over\alpha'^{9/2}}{1\over8 (2\pi)^9}\int_F {d\tau_1 d\tau_2\over \tau_2^{11/2}}{1\over\ab{\eta(\tau)}^{16}{\eta(\tau)}^{16}{\bar{\eta}(\bar{\tau})}^4}\nn
&\quad\times\Bigg(
\sum_{m\in\mathbb Z+1/2}g_m\fn{\eta,R}\paren{\vartheta_{10}}^7
	\paren{
		\paren{\bar{\vartheta}_{01}}^4\paren{\vartheta_{00}}^8
		-\paren{\bar{\vartheta}_{00}}^4\paren{\vartheta_{01}}^8
		}\nn
&\phantom{\quad\times\Bigg(}
	+\sum_{m\in\mathbb Z}g_m\fn{\eta,R}\bigg[
\paren{\vartheta_{00}}^7\paren{\paren{\bar{\vartheta}_{10}}^4\paren{\vartheta_{01}}^8+\paren{\bar{\vartheta}_{01}}^4\paren{\vartheta_{10}}^8}\nn
&\phantom{\quad\times\Bigg(+\sum_{m\in\mathbb Z}g_m\fn{\eta,R}\bigg[}
+\paren{-1}^m \paren{\vartheta_{01}}^7
\paren{\paren{\bar{\vartheta}_{10}}^4\paren{\vartheta_{00}}^8-\paren{\bar{\vartheta}_{00}}^4\paren{\vartheta_{10}}^8}
\bigg]
%&\left\{\paren{\bar{\vartheta}_{10}}^4(\vartheta_{00}^7\paren{\vartheta_{01}}^8+\paren{-1}^m\paren{\vartheta_{00}}^8\vartheta_{01}^7)
%+\paren{\vartheta_{10}}^8(\paren{\bar{\vartheta}_{01}}^4\vartheta_{00}^7
%+\paren{\bar{\vartheta}_{00}}^4\paren{-1}^{m+1}\vartheta_{01}^7)
%\right\}
\Bigg).\label{Wilsonline partition function}
}
The 9 dimensional energy density in the Jordan frame is given by
\al{
\rho_9
	&=	-{Z_{T^2}\over V_9};
}
see Eq.~\eqref{energy density}.

\begin{figure}[t]
\begin{center}
\hfill
\includegraphics[width=.4\textwidth]{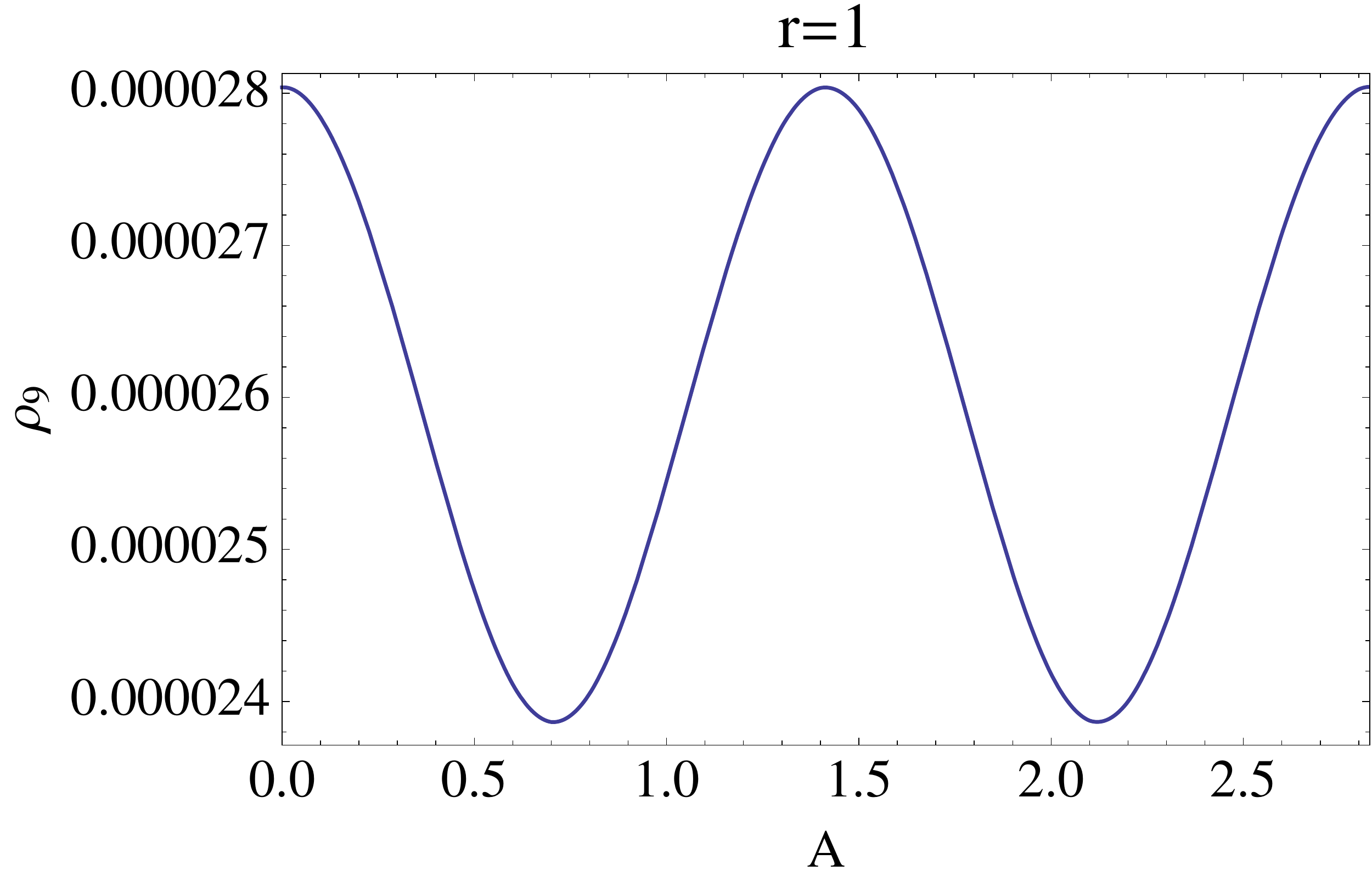}
\hfill
\includegraphics[width=.4\textwidth]{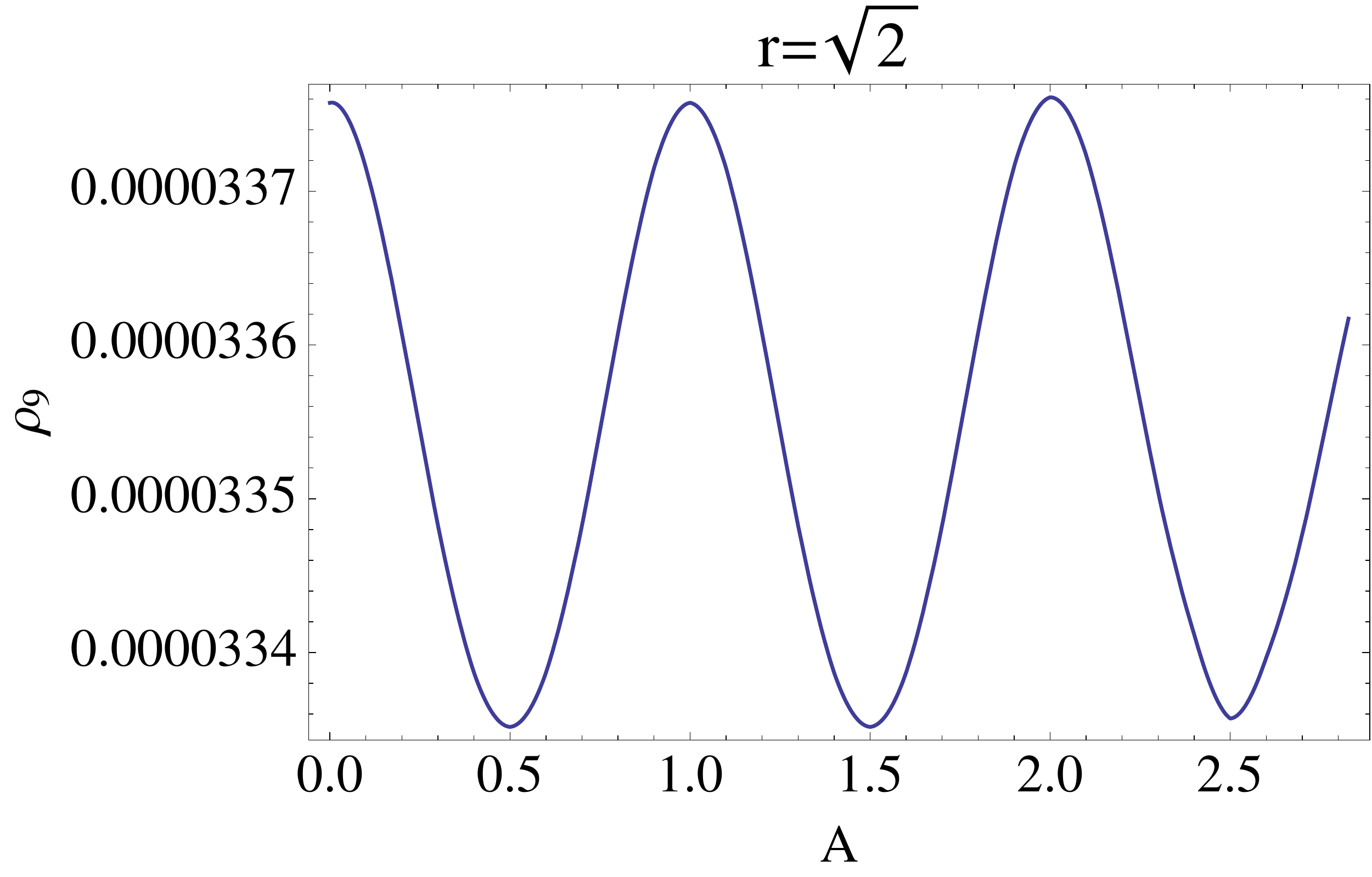}
\hfill\mbox{}
\caption{
The potential $\rho_9$ in Jordan frame as a function of $A$ with $r=\sqrt{\alpha'}$ (left) and $\sqrt{2\alpha'}$ (right), all in units of $\alpha'=1$.
We can see the periodicity $A\to A+\sqrt{2\alpha'}/r$, up to the distortions due to numerical errors.
}\label{periodicity}
\end{center}
\end{figure}
In Fig.~\ref{periodicity}, we plot $\rho_9$ as a function of $A$ for $r=\sqrt{\alpha'}$ (left) and $\sqrt{2\alpha'}$ (right), all in units of $\alpha'=1$.
The summation over $n$ and $m$ in Eqs.~\eqref{Wilson-line-dependent part} and \eqref{Wilsonline partition function} are truncated by $\ab{n},\ab{m}\leq10$ and the numerical integration is performed within $\tau_2\leq4$.
We can see the periodicity $A\to A+\sqrt{2\alpha'}/r$.

\begin{figure}[t]
\begin{center}
\hfill
\includegraphics[width=.35\textwidth]{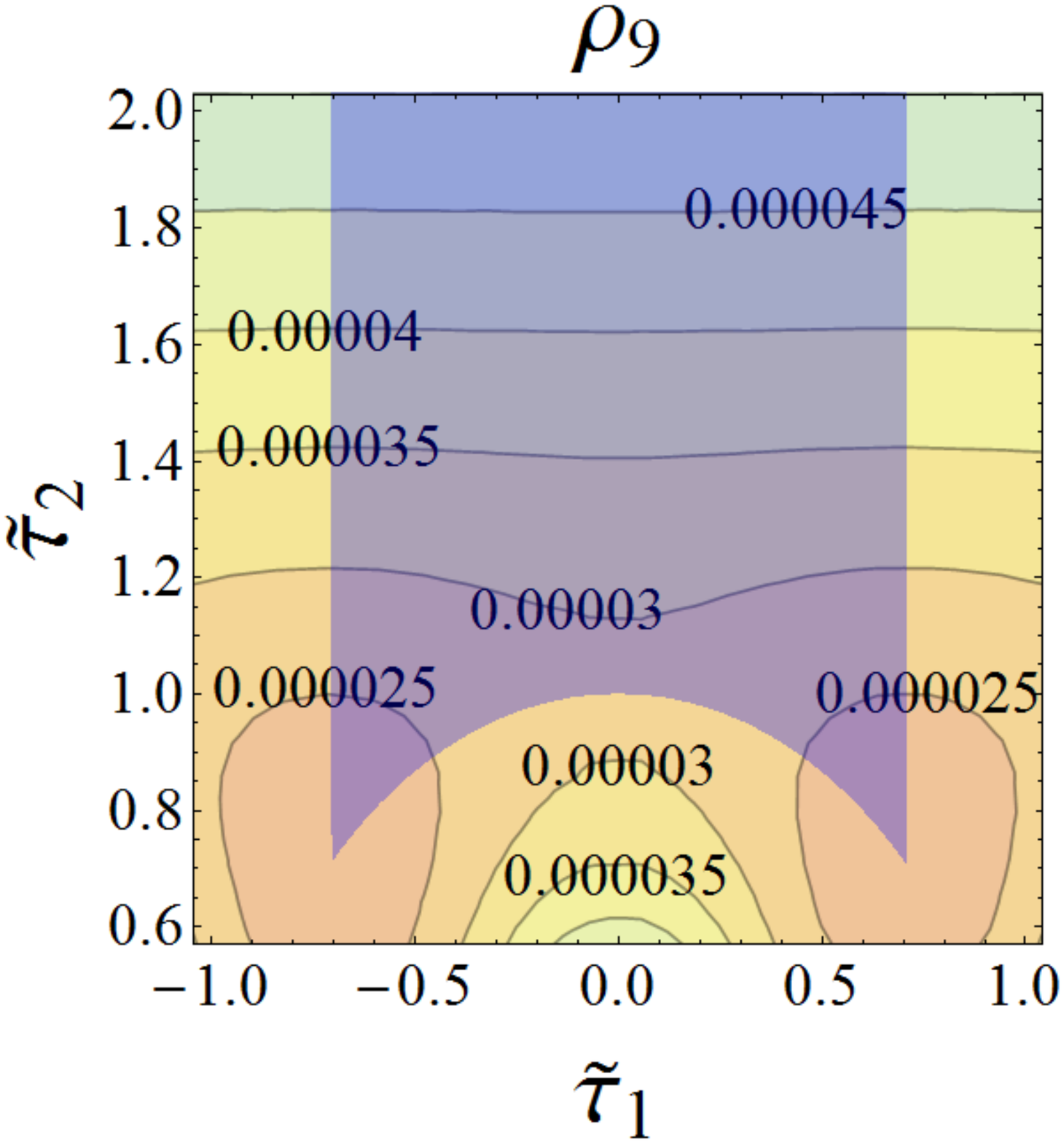}
\hfill
\includegraphics[width=.6\textwidth]{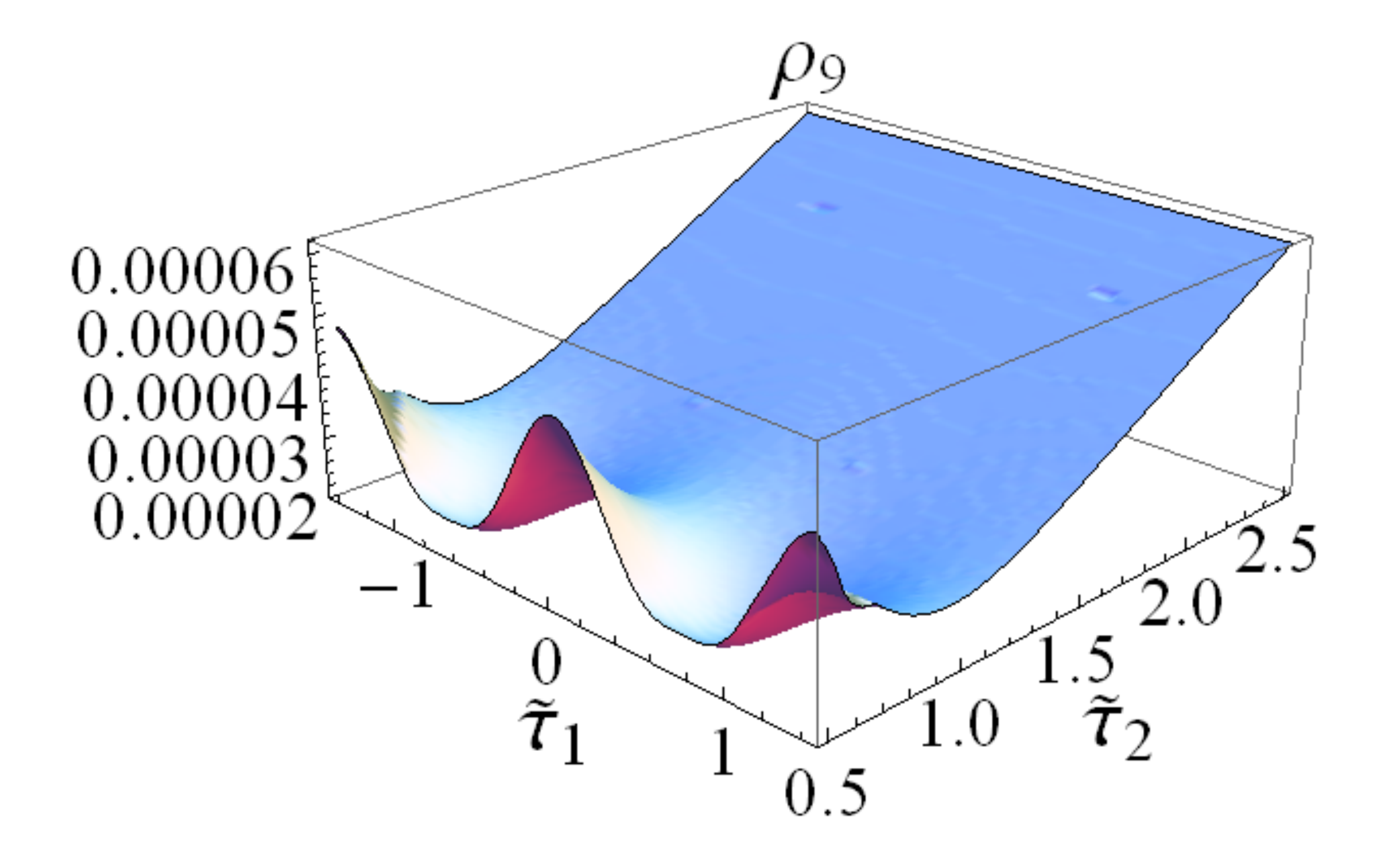}
\hfill\mbox{}
\caption{
Contour and 3D plots are shown in the left and right panels, respectively, for the energy density $\rho_9$ in the Jordan frame as a function of $\tilde\tau_1=rA/\sqrt{\alpha'}$ and $\tilde\tau_2=r/\sqrt{\alpha'}$, with all their values being given in $\alpha'=1$ units. In the left, we shade the fundamental region for the T-dual transformation: $\ab{\tau_1}\leq1/\sqrt{2}$, $\ab{\tau}\geq 1$. We can see the shift-symmetry $\tilde\tau_1\to\tilde\tau_1+\sqrt{2}$, up to distortions due to numerical errors.
}\label{Jordan potential}
\end{center}
\end{figure}
For varying $A$ and $r$, we plot $\rho_9$ as a function of $\tilde\tau_1=rA/\sqrt{\alpha'}$ and $\tilde\tau_2=r/\sqrt{\alpha'}$ in Fig.~\ref{Jordan potential}.
Note that in the large $r$ ($=\sqrt{\alpha'}\tilde\tau_2$) limit, the Jordan frame potential becomes proportional to $r$. This can also be seen analytically from the fact that in the large $r$ limit, the contributing modes are as in Eq.~\eqref{large r limit summation}, which results in the same expression as Eq.~\eqref{prop to R after Poisson}.
To repeat, we have obtained both numerically and analytically that the Jordan frame potential is proportional to $r$ at the one-loop level. For large $r$ limit, 
all the higher loop corrections have the same behavior since it comes from the fact that the energy is proportional to the volume of the compactified dimension.

%this behavior should not be altered by the higher loop corrections 

%\subsection{Switching to Einstein frame}

\begin{figure}[t]
\begin{center}
\hfill
\includegraphics[width=.35\textwidth]{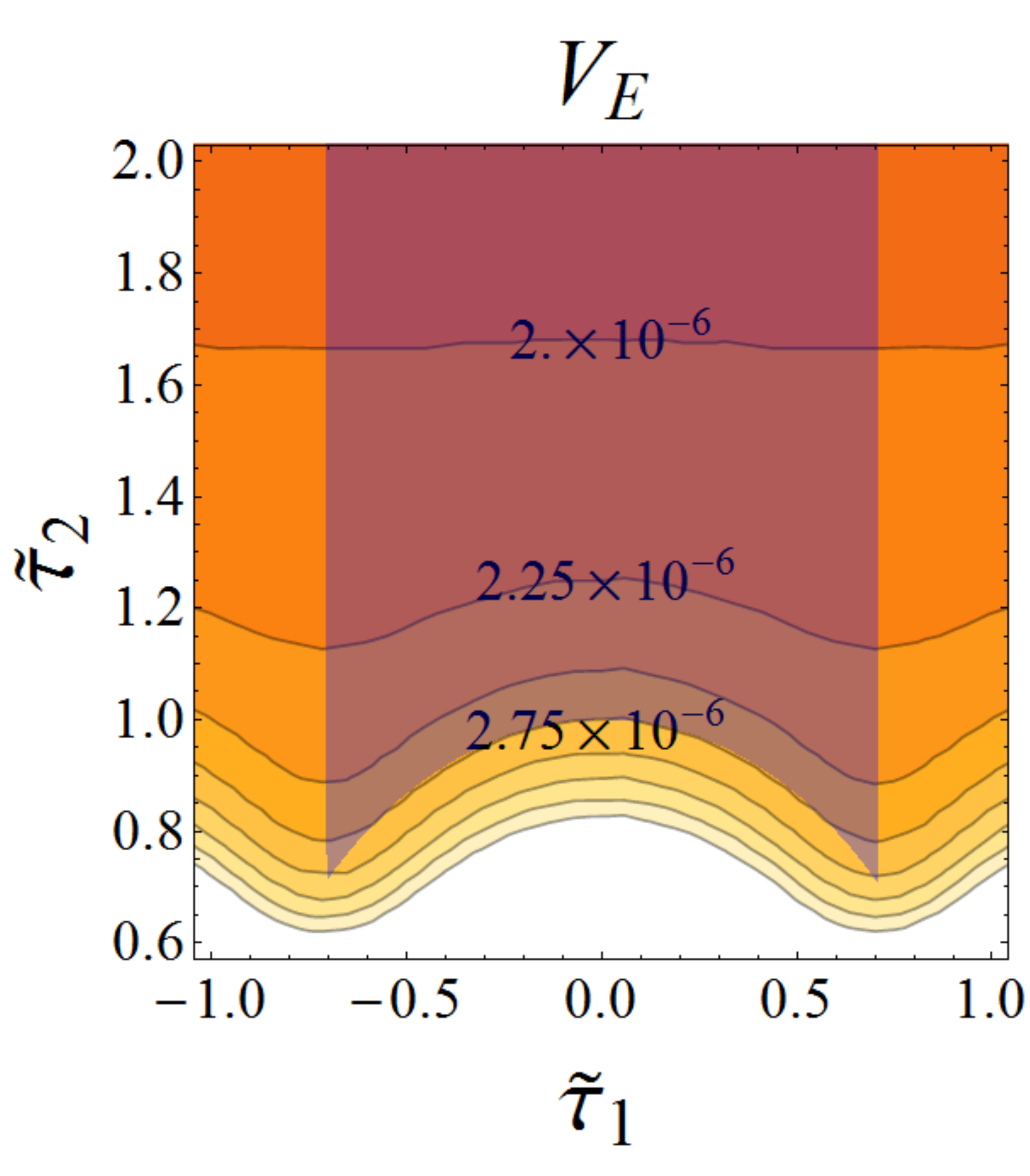}
\includegraphics[width=.6\textwidth]{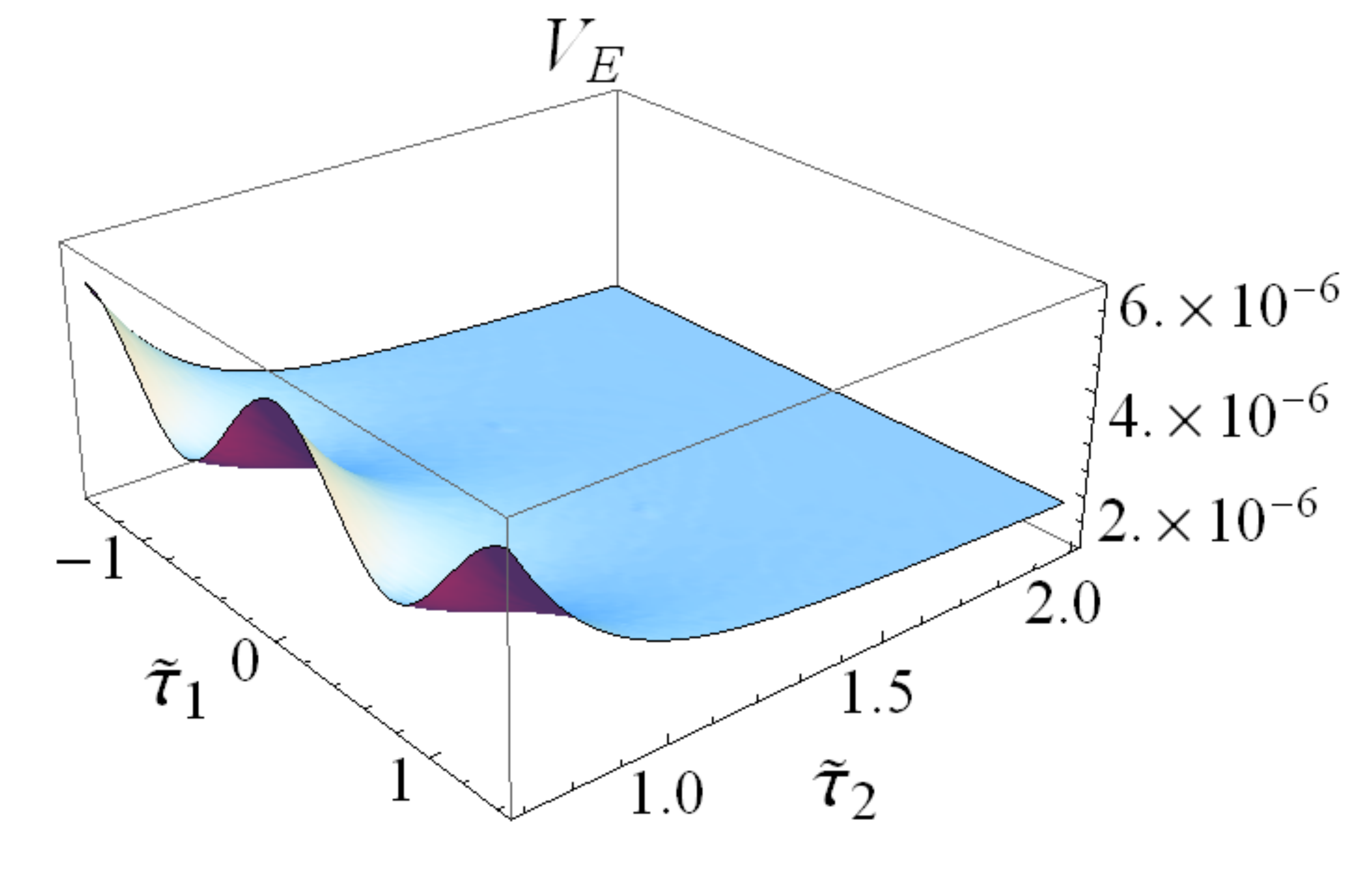}
\hfill\mbox{}
\caption{
Contour and 3D plots are shown in the left and right panels, respectively, for the energy density $V_\text{E}$ in the Einstein frame as a function of $\tilde\tau_1=rA/\sqrt{\alpha'}$ and $\tilde\tau_2=r/\sqrt{\alpha'}$, with all their values being given in $\alpha'=1$ units. 
The shaded fundamental region and the existence of the $\sqrt{2}$-shift are the same as in Fig.~\ref{Jordan potential}. We see that the potential becomes runaway for the large radius limit $r\gg\sqrt{\alpha'}$.
}\label{Einstein potential}
\end{center}
\end{figure}
Now let us turn to the Einstein frame:
\al{
V_E(r)&=-{1\over \paren{{2\pi r}}^{2/7}}{Z_{T^2}\over2\pi r V_9},\nn
&=-{1\over\alpha'^{9/2}}{1\over2 \paren{2\pi}^{72/7}}{1\over r^{9/7}}\int_F {d\tau_1 d\tau_2\over \tau_2^{11/2}}{1\over\ab{\eta(\tau)}^{16}{\eta(\tau)}^{16}{\bar{\eta}(\bar{\tau})}^4}
\sum_\text{sector $\alpha\vec w$}\tilde Z_{T^2,\alpha\vec w};
}
see Eq.~\eqref{to Einstein}.
We plot this potential in Fig.~\ref{Einstein potential}.
Important fact is that the potential in the Einstein frame becomes runaway for the large radius limit $r\gg\sqrt{\alpha'}$. As discussed above, this behavior should not be altered by the higher loop corrections.

Note that this effective potential in the Einstein frame is reliable only for large $r\gg\sqrt{\alpha'}$ since the treatment in terms of the effective field theory~\eqref{potential in lower dimensions} becomes valid only in this limit; furthermore, we can regard $r$ as the physical radius only in this limit; see also the argument around Eq.~\eqref{large r limit summation}.

\subsection{Large boost limit}
We want to examine the behavior of the Higgs potential in the large field limit.
However, in this nine dimensional toy model, there are two flat directions {\blue at this level}, namely, $A$ and $R$. %although we have only one direction in the SM.
If the Higgs comes from a similar mechanism to the gauge-Higgs unification,
the Higgs field should be identified with $A$.
Therefore, we check the large $A$ limit for a fixed $R$. %{\bf (Kawai-discussion needed for its physical meaning.)}
This limit is nothing but the large boost limit as is easily seen from Eq.~\eqref{A-id}: $\eta\rightarrow \infty$.
From Eqs.~\eqref{r-id} and \eqref{tautilde},
the trajectory in the $\tilde\tau_1$-$\tilde\tau_2$ plane is given by
\al{
&\tilde\tau_1={R\over\sqrt{\alpha'}}\tanh\eta,\nn
&\tilde\tau_2={R\over\sqrt{\alpha'}}{1\over\cosh\eta}.
	\label{trajectory of constant R}
}
Since $\tilde\tau_1^2+\tilde\tau_2^2=R^2/\alpha'$, this path starts from $\paren{0,\,R/\sqrt{\alpha'}}$ for $\eta=0$, and moves on the circle toward $\paren{R/\sqrt{\alpha'},\,0}$ as $\eta\to\infty$.
The question is what this trajectory is when mapped onto the fundamental region.
The large $\eta$ behavior depends on the value of $R/\sqrt{\alpha'}$:
\begin{itemize}
\item If $R/\sqrt{\alpha'}\in \sqrt{2}\mathbb Q$, then $\tilde\tau_2$ ($=r/\sqrt{\alpha'}$) goes to infinity in the large $\eta$ limit.
This can be seen as follows.
Since $\tilde\tau\to R/\sqrt{\alpha'}$ as $\eta\to\infty$, let us check to what point $R/\sqrt{\alpha'}$ is mapped in the fundamental region.
Let us write $R/\sqrt{\alpha'}=\sqrt{2}p/q$ with $p,q\in\mathbb Z$. 
By an appropriate times of $\sqrt{2}$-shifts ($T$-transfomation in \eqref{T-duality transformation}), we can always make $\ab{p}<\ab{q}$. 
Performing the inversion ($S$-transfomation in \eqref{T-duality transformation}), and again doing an appropriate times of $\sqrt{2}$-shifts, we can make the numerator $p$ smaller and smaller; eventually we get $p/q\to 0$.
This corresponds to the infinity $\tau_2\to\infty$ in the fundamental region. 

This behavior is expected from the discussion of the general momentum boost in Sec.~\ref{momentum boost}.
In fact, if and only if $R/\sqrt{\alpha'}\in \sqrt{2}\mathbb Q$, we can have a lattice point on the light cone in the momentum space, that is, there exist $n,m,w\in\mathbb Z$ such that\footnote{
%\begin{proof}
This can be proved as follows.
First we show that the conditions~\eqref{lL pR condition} and \eqref{pL condition} can be met for an arbitrary $R/\sqrt{\alpha'}\in\sqrt{2}\mathbb Q$ by an appropriate choice of $n,m,w$.
Let us write $R/\sqrt{\alpha'}=\sqrt{2}q_n/q_d$ with $q_n,q_d\in\mathbb Z$.
The condition~\eqref{pL condition} reads $nq_d/q_n+2wq_n/q_d=0$.
We can choose $n$ and $w$ such that $n=n'q_n$ and $w=w'q_d$ with $n',w'\in\mathbb Z$, resulting in the condition $n'q_d+2w'q_n=0$. This can be satisfied by setting $w'=q_d$ and $n'=-2q_n$.
Then the condition~\eqref{lL pR condition} reads
$0	\stackrel{!}{=}
%		m+{1\over\sqrt{2}}\paren{{nq_d\over\sqrt{2}q_n}-{\sqrt{2}q_nw\over q_d}}
%	=	m+{1\over2}\paren{{nq_d\over q_n}-{2q_nw\over q_d}}
		m+{1\over2}\paren{n'q_d-2q_nw'}
	=	m+n'q_d$, which can be satisfied by choosing $m=-n'q_d$.
	
Next we show that if $R/\sqrt{\alpha'}\slashed\in\sqrt{2}\mathbb Q$, there is no set of $n,m,w\in\mathbb Z$ that satisfies Eqs.~\eqref{lL pR condition} and \eqref{pL condition}.
%To satisfy Eq.~\eqref{pL condition}, it is necessary that $R^2/\alpha'\in\mathbb Q$.
By putting Eq.~\eqref{pL condition} into Eq.~\eqref{lL pR condition}, we get the condition
$m\pm \sqrt{2}n{\sqrt{\alpha'}\over R}=0$. Therefore, it is necessary that $R/\sqrt{\alpha'}=\mp\sqrt{2}n/m$ with $n,m\in\mathbb Z$.
%\end{proof}
}
\al{
l_L^2-p_R^2%={2\over\alpha'}m^2-\left({n\over R}-{Rw\over\alpha'}\right)^2
&={2\over\alpha'}
\sqbr{m+{1\over\sqrt{2}}\left({n\over R/\sqrt{\alpha'}}-{R\over\sqrt{\alpha'}}w\right)}
\sqbr{m-{1\over\sqrt{2}}\left({n\over R/\sqrt{\alpha'}}-{R\over\sqrt{\alpha'}}w\right)}
=0,	\label{lL pR condition}\\
p_L^2&={1\over\alpha'}\left({n\over R/\sqrt{\alpha'}}+{R\over\sqrt{\alpha'}}w\right)^2=0.
	\label{pL condition}
}
For $R/\sqrt{\alpha'}\in\sqrt{2}\mathbb Q$, there is a point on the light cone in the momentum space. Following the argument of Sec.~\ref{momentum boost}, the Lorentz boost between $l_L$ and $p_R$ opens up a new dimension. 

\item 
If $R/\sqrt{\alpha'}\slashed\in \sqrt{2}\mathbb Q$, the potential becomes either periodic or chaotic.
Let us check in what case we get the periodic potential.
\begin{itemize}
\item
The periodic case is realized if, starting from a point $\tilde{\tau}$~\eqref{trajectory of constant R} with the boost $\eta$, we get another point on the trajectory with the boost $\eta+\eta_c$,
\al{
\tilde\tau_1'	&=	{R\over\sqrt{\alpha'}}\tanh\fn{\eta+\eta_c}, \nn
\tilde\tau_2'	&=	{R\over\sqrt{\alpha'}}{1\over\cosh\fn{\eta+\eta_c}},
}
which can be mapped from $\tilde\tau$ by an appropriate T-dual transformation~\eqref{general transformation}.

In general, the transformation of $\tilde\tau_2$ is as shown in Eq.~\eqref{explicit T-dual}, and we get
\al{
\tilde\tau_2'
&={\tilde\tau_2\over\ab{c\tilde\tau+d}^2}={\tilde\tau_2\over c^2{R^2\over\alpha'}+d^2+2cd\tilde\tau_1}
={R\over\sqrt{\alpha'}}{1\over \paren{c^2{R^2\over\alpha'}+d^2}\cosh\eta+2cd{R\over\sqrt{\alpha'}}\sinh\eta}\nn
&={R\over\sqrt{\alpha'}}{1\over\cosh\fn{\eta-\eta_2}},
}
where we have defined $\eta_2$ by
\al{
\tanh\eta_2:=-{2cd{R\over\sqrt{\alpha'}}\over c^2{R^2\over\alpha'}+d^2}.
}
On the other hand,
the same transformation maps $\tilde{\tau}_1$ to
\al{
\tilde\tau_1'&={ac\ab{\tilde\tau}^2+\paren{ad+bc}\tilde\tau_1+bd\over \ab{c\tilde\tau+d}^2}
={ac{R^2\over\alpha'}+bd+\paren{ad+bc}\tau_1\over\cosh\fn{\eta-\eta_2}/\cosh\eta}\nn
&={\paren{ac{R^2\over\alpha'}+bd}\cosh\eta+{R\over\sqrt{\alpha'}}\paren{ad+bc}\sinh\eta\over\cosh\fn{\eta-\eta_2}}\nn
&=
{R\over\sqrt{\alpha'}}{\sinh\fn{\eta-\eta_1}\over\cosh\fn{\eta-\eta_2}},
}
where we have defined $\eta_1$ by
\al{
\tanh\eta_1=-{ac{R^2\over\alpha'}+bd\over {R\over\sqrt{\alpha'}}\paren{ad+bc}}.
}
The trajectory becomes periodic if and only if $\eta_1=\eta_2$, that is,
\al{
{2cd{R\over\sqrt{\alpha'}}\over c^2{R^2\over\alpha'}+d^2}
	&=	{ac{R^2\over\alpha'}+bd\over {R\over\sqrt{\alpha'}}\paren{ad+bc}},
}
or
\al{
\paren{d^2-c^2{R^2\over\alpha'}}\paren{bd-ac{R^2\over\alpha'}}
	&=	0.
}
Vanishing first factor means $\eta_2=\infty$, and the finite period is obtained when and only when the last factor becomes zero:
\al{
bd-ac{R^2\over\alpha'}
	&=	0.
}
Therefore, the partition function becomes periodic if and only if $R^2/\alpha'$ can be written as
\al{
&
{R\over \sqrt{2\alpha'}} \slashed\in \mathbb{Q}, &
&{R^2\over \alpha'}={bd\over ac},
		\label{periodicity condition}
}
where $ad-bc=1$ and either $a,d\in\sqrt{2}\mathbb Z$, $b,c\in\mathbb Z$ or $a,d\in\mathbb Z$, $b,c\in\sqrt{2}\mathbb Z$.

\item In particular, if $R^2/\alpha'$ is an irrational number then the condition~\eqref{periodicity condition} cannot be met (unless $ac=0$ that leads to the trivial $\eta_1=0$), and the partition function becomes non-periodic, namely chaotic.
\end{itemize}
%To summarize, for $R^2/\alpha'\in\mathbb Q$, the potential becomes periodic when Eq.~\eqref{periodicity condition} is met, and chaotic when otherwise.
\end{itemize}
\begin{figure}[t]
\begin{center}
\hfill
\includegraphics[width=.32\textwidth]{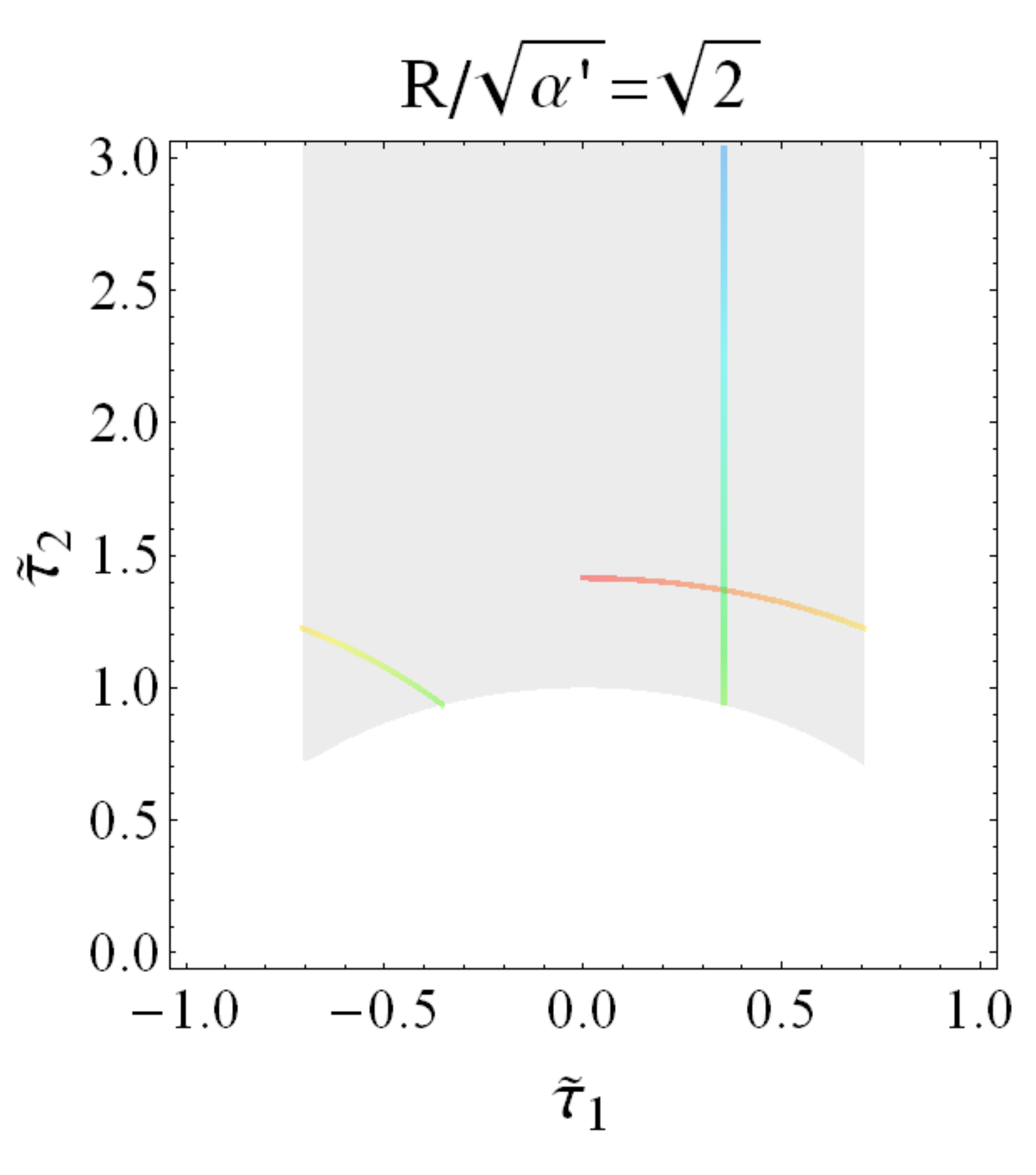}
\hfill
\includegraphics[width=.32\textwidth]{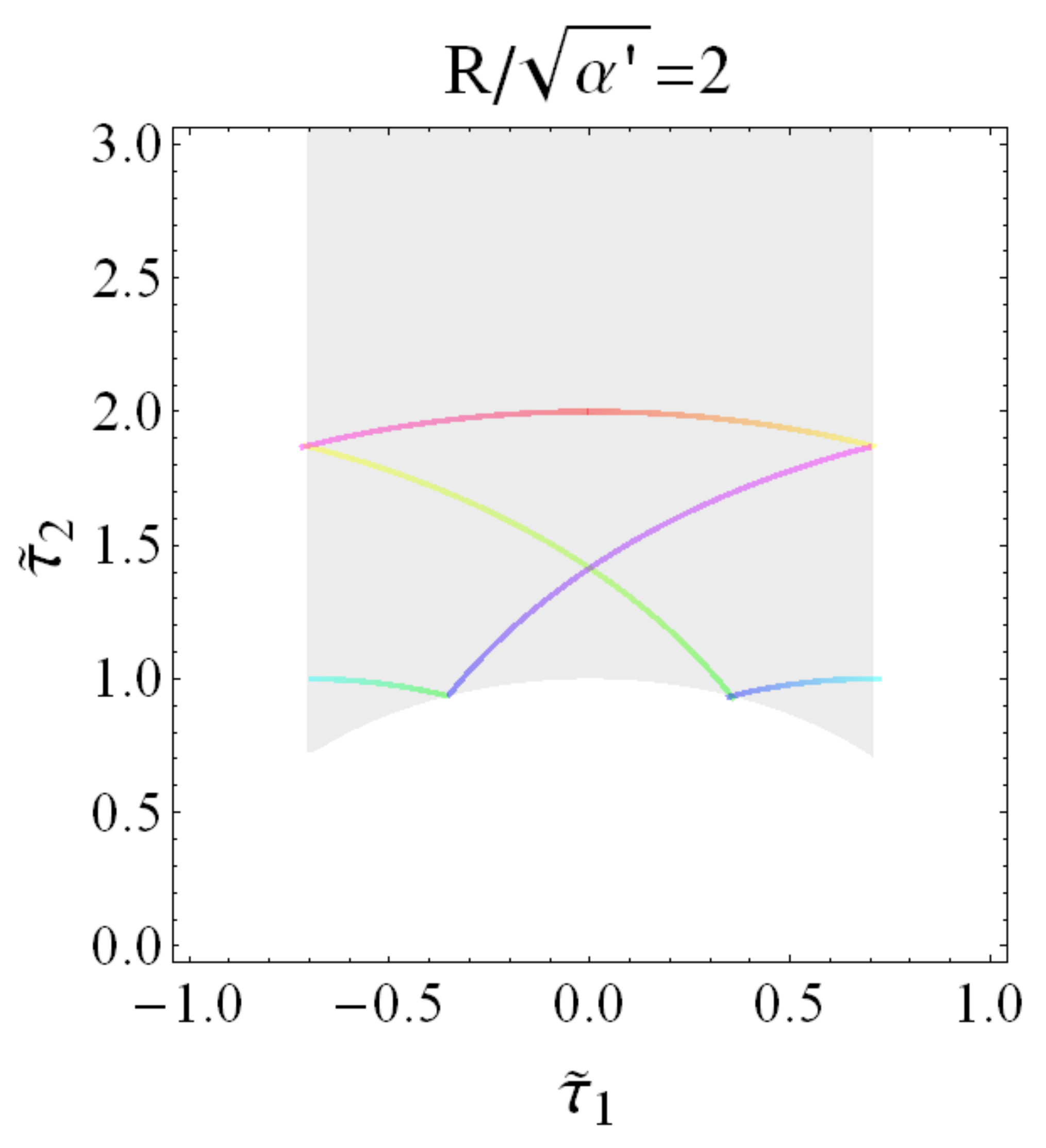}
\hfill
\includegraphics[width=.32\textwidth]{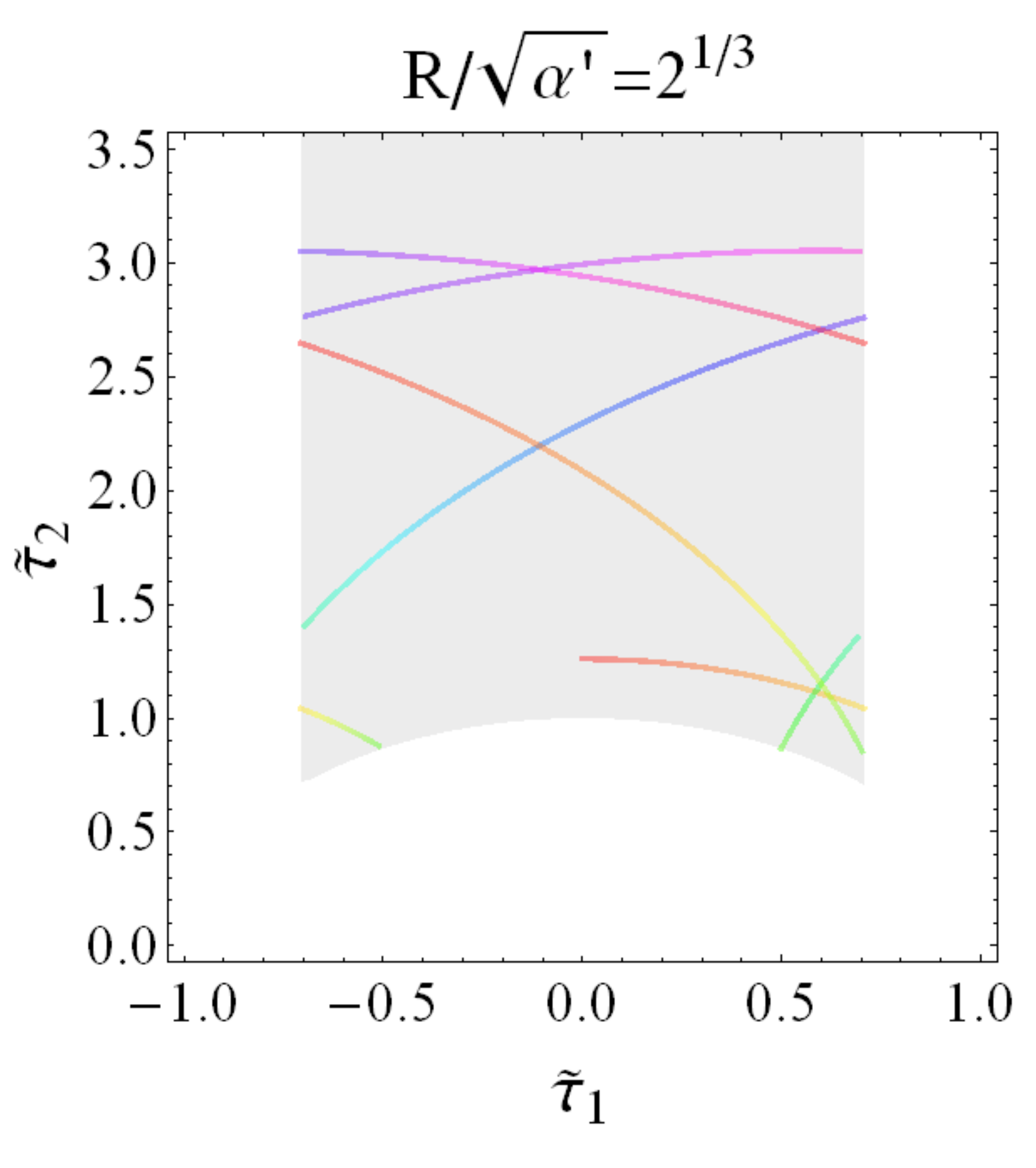}
\hfill\mbox{}
\caption{The trajectory that starts from $\eta=0$ at $\paren{\tilde\tau_1,\tilde\tau_2}=\paren{0,R/\sqrt{\alpha'}}$ for a fixed value of $R/\sqrt{\alpha'}$ being $\sqrt{2}$, 2, and $2^{1/3}$ in the left, center, and right panels, respectively, showing the runaway, periodic, and chaotic limits.
We have shaded the fundamental region for the T-dual transformations.
}\label{T-duality}
\end{center}
\end{figure}
As a check, we show the numerical results for $R/\sqrt{\alpha'}=\sqrt{2}$, $2$, and $2^{1/3}$ in Fig.~\ref{T-duality}.
We see that they show the runaway, periodic, and chaotic limits, respectively.
The result presented in this section provides a concrete example of the general argument presented in Sec.~\ref{classification}.
It is plausible that the large Higgs field limit in string theory fits into either one of these three.

%To summarize, We have computed the large boost limit for a fix $R$ in the $SO(16)\times SO(16)$ heterotic string theory in ten dimensions, one of which is compactified on $S^1$.

Note that our computation is based on the one-loop effective potential and that the higher order corrections are significant around the region $A,R^{-1}\sim M_\text{s}$ ($=1/\sqrt{\alpha'}$). Therefore, the result so far should be interpreted as an effort to guess what is the physical large field limit along a potential valley after including all the higher order corrections.
In Fig.~\ref{T-duality}, we have checked the large $A$ limit for a fixed $R$. Is this a physical limit, and if not, what should it be?
Comparing Figs.~\ref{Jordan potential} and \ref{Einstein potential}, we see that it is a generic feature that there is a runaway vacuum no matter what the structure is around $A,R^{-1}\sim M_\text{s}$. It seems plausible that if the physical large $A$ limit is not the one with fixed $\tilde\tau_2$, then large $A$ limit goes into the runaway vacuum after all. However, we consider all the three limits, runaway, periodic, and chaotic in order not to loose generality.

\begin{figure}[t]
\begin{center}
\hfill
\includegraphics[width=.5\textwidth]{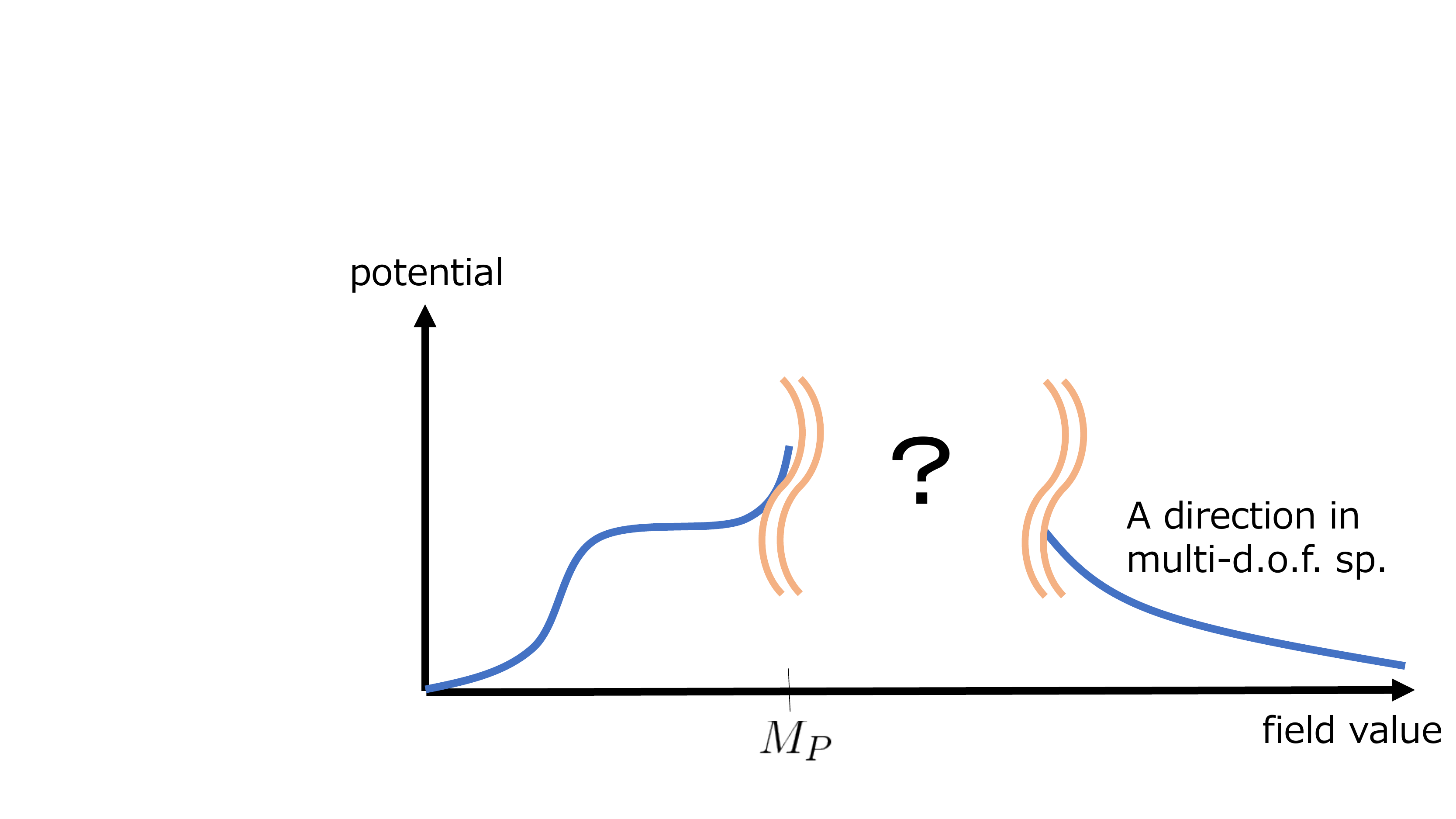}
\hfill\mbox{}
\caption{
Schematic figure for the Higgs potential. Low energy side is determined phenomenologically. High energy side represents a runaway direction in the multi degrees of freedom space.
}\label{potential}
\end{center}
\end{figure}

As said above, the extrapolation from the low energy data has revealed that there is the quasi-flat direction of the Higgs potential in the SM.
We are interested in the potential for the large field values.
Beyond the string or Planck scale, there opens up several quasi-flat directions in general. Therefore we need to consider a multi-dimensional field space. In the example examined in this section, it corresponds to the $A$-$R$ (or $\tilde\tau_1$-$\tilde\tau_2$) plane. As we have seen in this section, generally there is at least one runaway direction in this space that corresponds to opening up an extra dimension; see Fig.~\ref{potential}. We will discuss its physical implications in the subsequent sections.

\section{Eternal Higgs inflation}\label{eternal section}

%\begin{figure}[tn]
%\begin{center}
%\hfill
%\includegraphics[width=.7\textwidth]{low_potential.pdf}
%\hfill\mbox{}
%\caption{Schematic figure for the Higgs potential below the Planck scale in the SM at criticality; see e.g.\ Fig.~2 in Ref.~\cite{Hamada:2014wna} for more details.}\label{low scale potential}
%\end{center}
%\end{figure}

As shown in Introduction, the Higgs potential $V\sim\lambda_\text{eff}\ab{H}^4$ in the SM shows a quite peculiar behavior when extrapolated to very large field values: all of the $\lambda_\text{eff}$, its running, and the bare Higgs mass can be accidentally small.
%The quartic coupling $\lambda_\text{eff}$ depends on $\ab{H}$ logarithmically, and takes its minimum value around $\ab{H}\sim M_P$, regardless of the input values of the top quark mass, the strong coupling constant, and the Higgs mass at the weak scale within the experimental errors; see e.g.\ Ref.~\cite{Hamada:2014wna}. Furthermore, the potential $V$ can be zero at this minimum within the current experimental bounds.
In Ref.~\cite{Hamada:2014iga}, we have proposed a possibility that this behavior, so to say the criticality, is a consequence of the Planck scale physics and that the criticality is closely related to the cosmic inflation.
%{\blue
%Furthermore, the height of the phenomenological Higgs potential at the critical point is very close to that of massless state potential calculated in Sec.~\ref{SO(16)} if $M_\text{s}\sim10^{17}\GeV$ and the compactification scale is the order of $M_\text{s}$.
%This leads us to identifying the SM Higgs as the massless state of string theory.  
%}

\begin{figure}[tn]
\begin{center}
\hfill
\includegraphics[width=.7\textwidth]{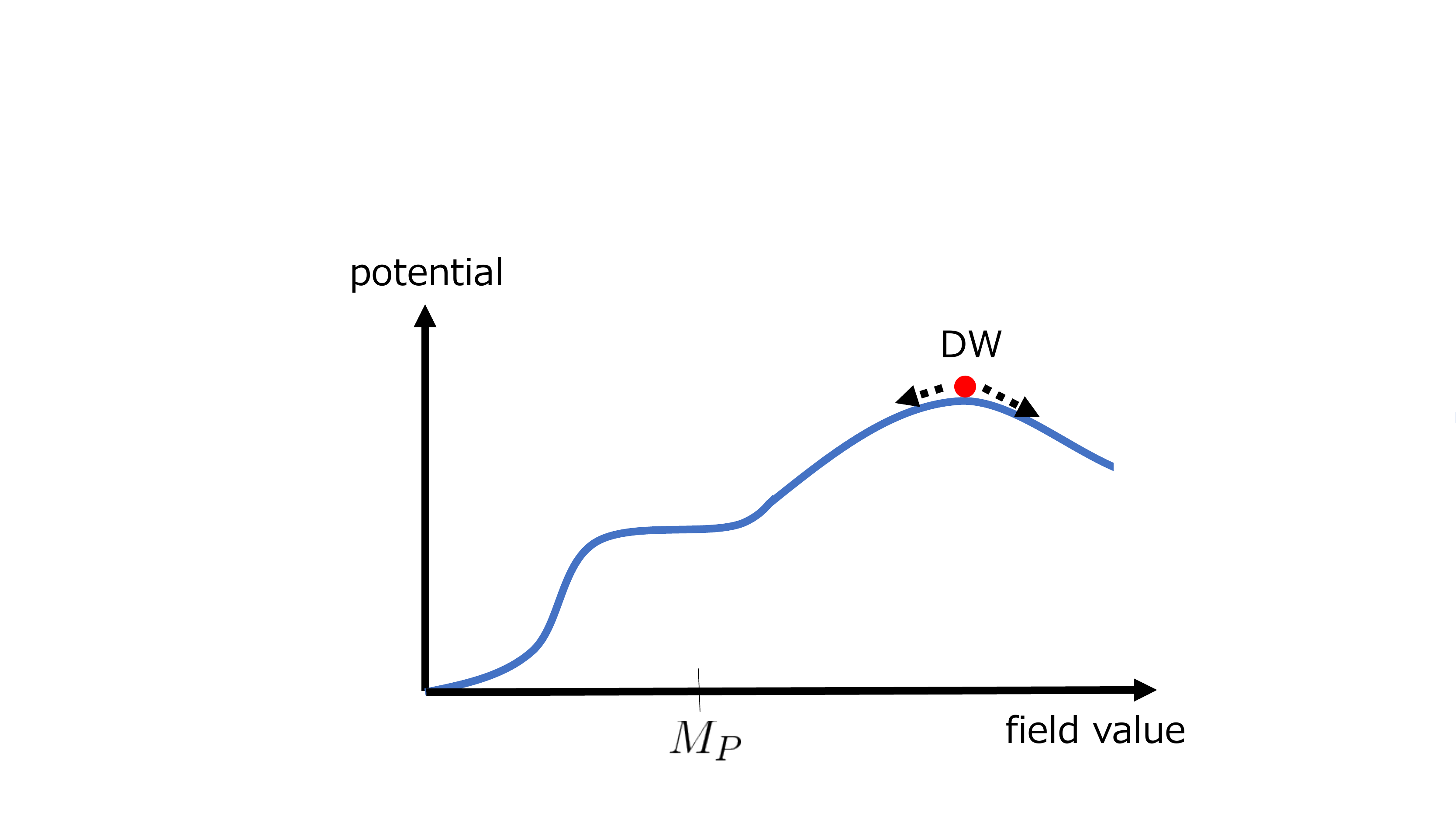}
\hfill\mbox{}
\caption{Schematic figure for the maximum that yields the domain wall, which becomes the source for the eternal inflation.
}\label{domainwall}
\end{center}
\end{figure}
We have seen that the large field limit goes down to a runaway direction, which corresponds to opening up an extra dimension, in the multi degrees of freedom space, as shown in Fig.~\ref{potential}.
Therefore, there is at least one maximum of the potential around the Planck scale; see Fig.~\ref{domainwall}.
This maximum can be a source of an eternal inflation at the core of the domain wall~\cite{Hamada:2014raa} between the electroweak vacuum and the runaway vacuum, in which the fifth dimension is opened up. In order for this to work, the curvature of the potential at the maximum must be sufficiently small~\cite{Sakai:1995nh}:
\al{
\left.M_P^2{V_{\varphi\varphi}\over V}\right|_\text{maximum}\lesssim 1.4.
	\label{DW condition}
}
In our scenario, this can be naturally satisfied as follows.
The potential for the fifth dimension can be seen by putting $D=5$ in Eq.~\eqref{to Einstein}. In stringy language, the action for the fifth dimension $R'\gg M_\text{s}^{-1}$ is coming from the one-loop potential: In the Einstein frame, we get
\al{
S_\text{eff}
	&\sim	{M_\text{s}^2\over g_\text{s}^2}\int \df^4x\sqrt{-g}
			\paren{
				\mathcal R-{\paren{\p R'}^2\over R'^2}-g_\text{s}^2M_\text{s}^2{1\over R'}
				}.
}
Switching to the canonical field $R'=e^{g_\text{s}\chi/M_\text{s}}$, we get
\al{
S_\text{eff}
	&\sim	\int \df^4x\sqrt{-g}\paren{
				M_\text{P}^2\mathcal R-\paren{\p\chi}^2-e^{-\chi/M_\text{P}}M_\text{s}^4
				},
}
where $M_\text{P}=M_\text{s}/g_\text{s}$.
Therefore, the stringy potential also gives
\al{
V_{\chi\chi}
	&\sim	{V\over M_\text{P}^2}.
}
It is remarkable that the potential changes of order unity when we vary $\chi$ by $M_\text{P}$, not by $M_\text{s}$, for large $\chi$. On the other hand at low energies, the SM potential in the Einstein frame exhibits the same behavior if the non-minimal coupling $\xi$ is of order ten~\cite{Hamada:2014wna}.
Therefore, it is natural to conclude that the condition~\eqref{DW condition} is also met around the maximum.

We note that in the original version of the topological Higgs inflation~\cite{Hamada:2014raa}, $\xi$ is used to make the maximum of the potential, and hence that it cannot account for the observed fluctuation of the cosmic microwave background (CMB). On the other hand, the scenario proposed in this paper allows the Higgs to be the source for both the eternal topological inflation and for the one that accounts for the CMB fluctuation, simultaneously.

%Here we discuss the implication of the three large field limits, runaway, periodic, and chaotic, on this Higgs inflation scenario at the criticality.
%In Fig.~\ref{low scale potential}, we show a schematic Higgs potential in this scenario below the Planck scale.

%\begin{figure}[tn]
%\begin{center}
%\hfill
%\includegraphics[width=.7\textwidth]{}
%\hfill\mbox{}
%\caption{
%}\label{}
%\end{center}
%\end{figure}
%
\begin{figure}[tn]
\begin{center}
\hfill
\includegraphics[width=.5\textwidth]{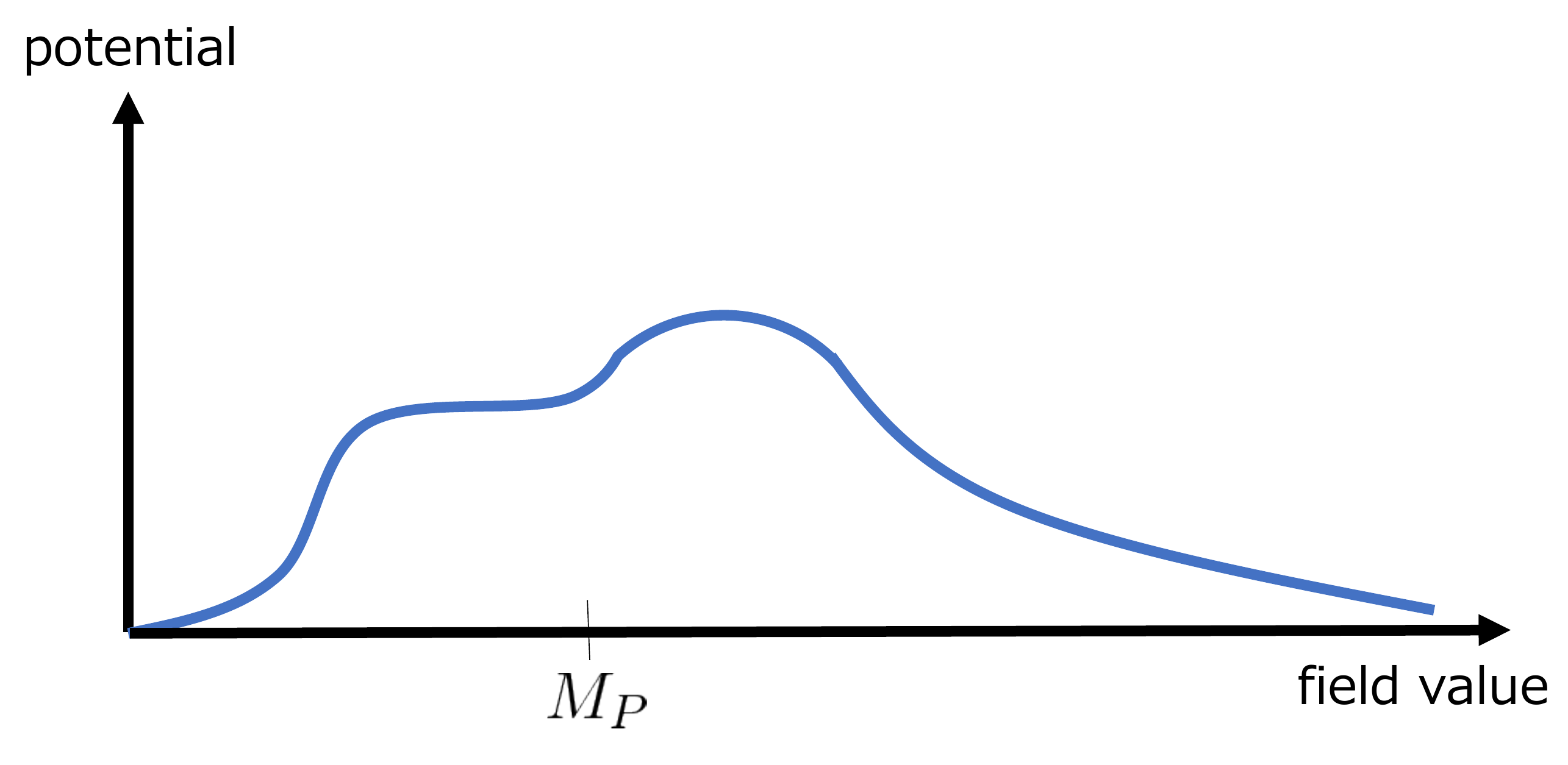}
\hfill\mbox{}
\caption{
Schematic figure for the Higgs potential smoothly connected to the runaway direction.
}\label{runaway figure}
\end{center}
\end{figure}
\begin{figure}[tn]
\begin{center}
\hfill
\includegraphics[width=.4\textwidth]{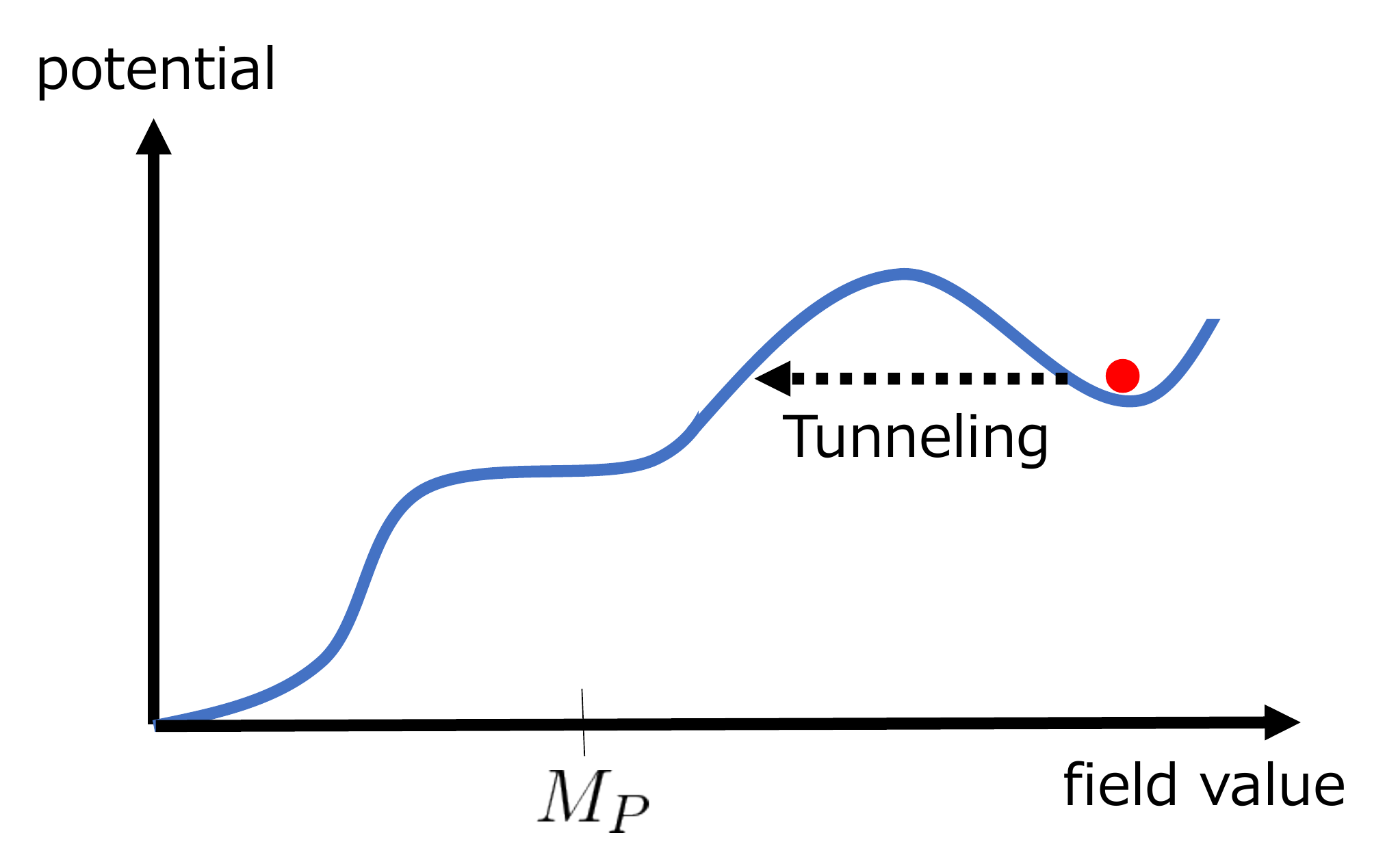}
\hfill
\includegraphics[width=.5\textwidth]{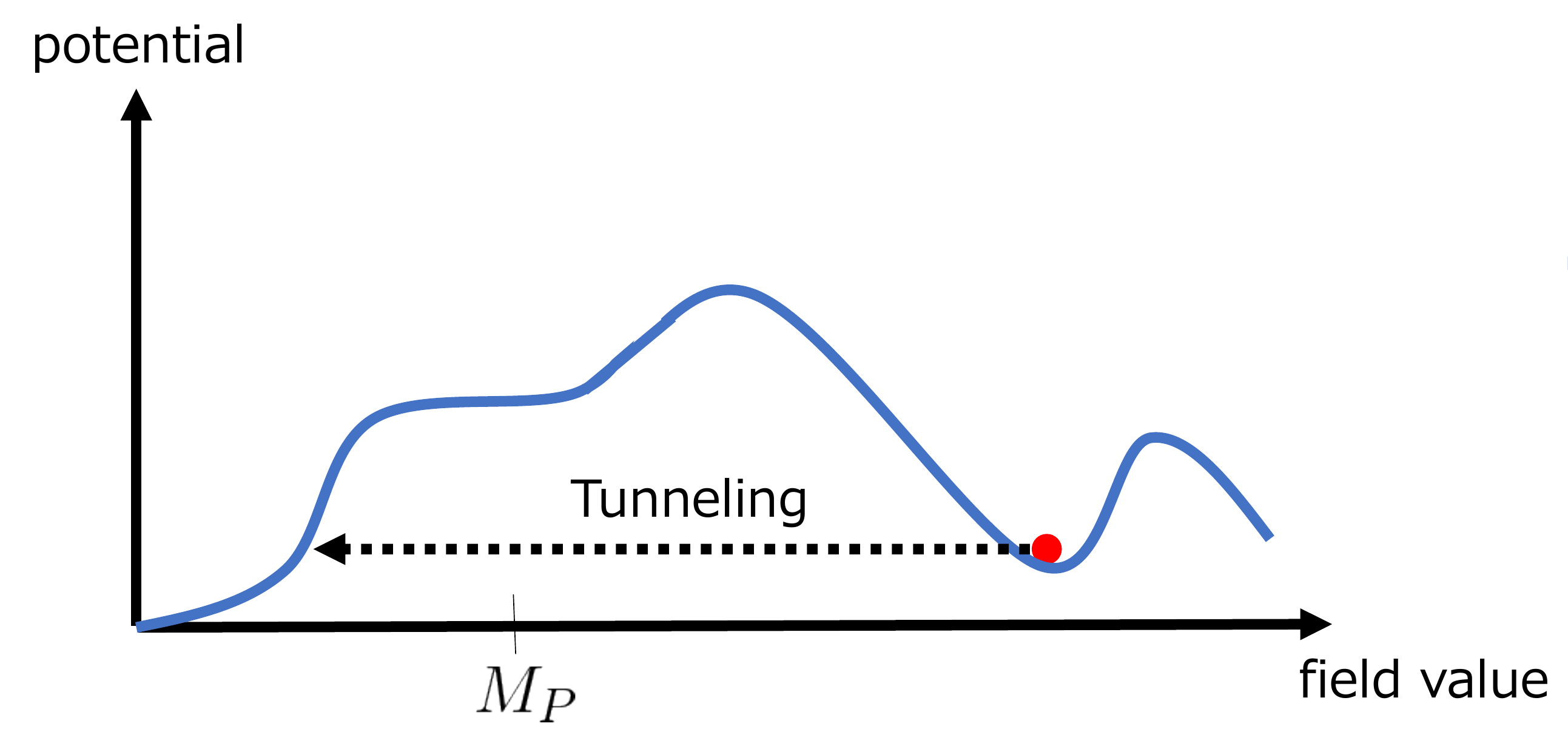}
\hfill\mbox{}
\caption{
Schematic figure for the Higgs potential. On the left, the false vacuum has higher energy than the quasi-flat potential in the SM, while on the right, it has lower energy.
}\label{false vacuum}
\end{center}
\end{figure}
There are two possibilities for the potential beyond the maximum:
\begin{itemize}
\item The potential smoothly becomes runaway as in Fig.~\ref{runaway figure}.
\item The potential has another local minimum as in Fig.~\ref{false vacuum}.
\end{itemize}
In the latter case, the false vacuum gives another mechanism of eternal inflation.
This situation is similar to some of the originators' idea of the inflation using a first order phase transition~\cite{Sato:1980yn,Guth:1980zm}. In the medium of the false vacuum, which is indicated by the (red) dot in Fig.~\ref{false vacuum}, there appears a bubble of the electroweak vacuum due to the tunneling, which is indicated by the dotted arrow.
This eternal inflation in the false vacuum had caused the so-called the graceful exit problem in the old inflation scenario~\cite{Linde:1981mu,Hawking:1982ga,Guth:1982pn}. 
However in the left case in Fig.~\ref{false vacuum}, the space inside the bubble experiences the second stage of inflation~\cite{Hamada:2014iga,Hamada:2014wna}, after the dotted arrow in the figure, and hence this problem is ameliorated as we do not need bubbles to collide.
In the right case in Fig.~\ref{false vacuum}, we need another inflation to account for the observed CMB fluctuation such as the $B-L$ Higgs inflation.

\section{Cosmological constant}\label{solution to cosmological constant}
As is reviewed in detail in Appendix~\ref{MPP review}, the MPP requires  degenerate vacua at the field value of the order of the Planck scale~\cite{Froggatt:1995rt,Froggatt:2001pa,Nielsen:2012pu}.
The cosmological constant of the runaway vacuum is exactly zero.
%; see footnote~\ref{limit is free}.
Then the MPP tells us that our electroweak vacuum must have the zero cosmological constant too.
This is a new solution to the cosmological constant problem in terms of the MPP.\footnote{
See also Ref.~\cite{Nielsen:2012pu} in which the cosmological constant problem is discussed in a different perspective.
}

%Note that this solution applies regardless of the existence of the eternal  inflation at the domain wall.
On the other hand,
%the observed value of the cosmological constant is~\cite{Ade:2013zuv}
the current universe is being dominated by the cosmological constant%~\eqref{cosmological constant observed}
~\cite{Ade:2013zuv} 
\al{
\rho_\Lambda^\text{obs}
	%&=	\paren{0.686\pm0.020}{3H_0^2M_P^2}
	\simeq
		\paren{2.2\,\text{meV}}^4,
		\label{cosmological constant observed}
}
and is entering the second inflationary stage.
This will eventually lead to the de Sitter space $dS_4$ with the length scale $H^{-1}$, where
\al{
H^2	=	{\rho_\Lambda^\text{obs}\over 3M_P^2}.
	%\sim	\paren{10^{-33}\,\text{eV}}^2,
}
%where $M_P:={1/\sqrt{8\pi G}}=2.4\times10^{18}\GeV$ and
%\al{
%H_0	&=	\paren{67.4\pm1.4}{\text{km}/\text{s}\over\text{Mpc}}.
%	\label{observed Hubble}
%}
%Then how can we explain the observed value of the cosmological constant?
%This is also explained in the same framework as follows.
%\simeq10^{27}\,\text{m}$
 %In Planck units, this is
%\al{
%{r_U\over l_P}
%	&=	{M_P\over H}.
	%\sim
	%	10^{60}.
%}
%
%We will show that the existence of the finite cosmological constant can be explained by a consideration of the statistical fluctuation.
We will discuss the possibility that the existence of the finite cosmological constant is understood as a statistical fluctuation.

\begin{figure}[tn]
\begin{center}
\hfill
\includegraphics[width=.4\textwidth]{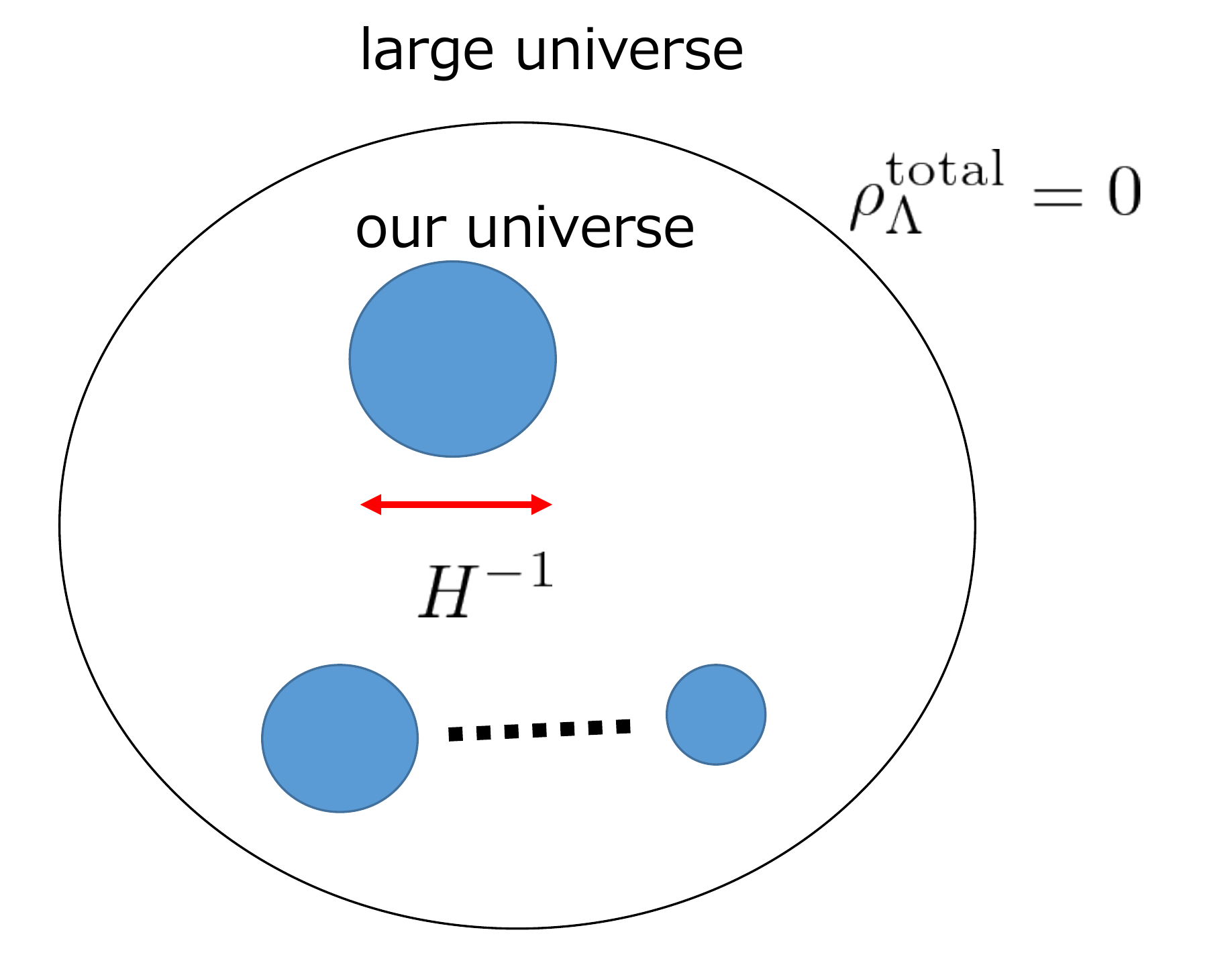}
\hfill\mbox{}
\caption{
Universe is divided into parts that will eventually become causally disconnected to each other in the end of their histories.
}\label{universe}
\end{center}
\end{figure}
First we point out that our universe is a part of a large universe 
whose cosmological constant is fixed to zero by the MPP.\footnote{
The argument in this section may also apply for the multiverse~\cite{Kawai:2011qb,Kawai:2013wwa,Hamada:2014ofa,Kawana:2014vra,Hamada:2014xra}.
}
The large universe can be divided into parts that will eventually become causally disconnected de Sitter spaces %to each other 
in the end of their histories, as in Fig.~\ref{universe}.
After the Euclideanization, each de Sitter space becomes $S^4$ with radius $r_U=1/H$.
%As we will see below, the effective cosmological constant of our universe can have a finite value.
%on which MPP claims that 
%The current universe is being dominated by the cosmological constant~\eqref{cosmological constant observed} and is entering the second inflationary stage with the Hubble constant
%\al{
%H^2	\sim	{\rho_\Lambda^\text{obs}\over 3M_P^2}
%	\sim	\paren{10^{-33}\,\text{eV}}^2,
%}
%which becomes the de Sitter space $dS_4$.
%
%Let us recall that in the ordinary statistical mechanics, the partition function
%\al{\label{partition function}
%Z\fn{T}=\sum_n e^{-E_n/T}
%\simeq
%\int dE\,e^{-\paren{E-TS}/T}
%}
%is dominated by the point where $E-TS$ takes its minimum value:
%\al{
%{\df S\over \df E}={1\over T}.
%}
%Let $E_*$ denote the value satisfying this condition. 
%By expanding the integrand of \eqref{partition function}, we obtain   
%denote the thermal fluctuation of $E$ around this value by $\delta E$:
%\al{
%Z\fn{T}\simeq \int \df\paren{\delta E}\,\exp\left[-{E_*-TS_*\over T}-{1\over2 {\df %E\over \df T} T^2}\paren{\delta E}^2
%\right],
%}
%where $\delta E$ stands for the fluctuation of $E$ around $E_*$.
%We see that the typical thermal fluctuation of the energy is $\sqrt{{\df E\over \df T} T^2}$.
%
%Let us turn to the QFT version.

We consider one of the $S^4$'s and latticize it by the lattice spacing of the order of $\l_P=1/M_\text{P}$, and 
let $S_i$ be the action on each site labeled by $i$.
%put the scalar field $\varphi_i$ on each site labeled by $i$.
The total action for the $S^4$ becomes the sum over positions:
\al{
S	&=	\sum_{i=1}^{r_U^4/l_P^4} S_i.
}
%For simplicity we neglect the contribution from the kinetic term: $S_i= S_i(\varphi_i)$.\footnote{
%The kinetic term would contribute with the same order as $S_i(\varphi_i)$ at each site, namely of order unity in Planck units, and should not alter the order-of-magnitude argument shown here.
%}
%The partition function becomes
%\al{
%Z&=\int \left[\prod_i \df\varphi_i\right] e^{-S}\nn
%&=\prod_i \left(\int  \df  S_i {\df\varphi_i \over \df  S_i}e^{-S_i}\right)\nn
%&= \prod_i \left(\int  \df  S_i\,e^{-f_i\fn{ S_i}}\right),
%}
%where $f_i:= S_i-\ln{\df\varphi_i\over\df S_i}$.
%Suppose $f_i$ takes its minimum $f_{i*}$ at $S_i= S_{i*}$.
%Then we have %expand $f_i$ around its minimum $f_{i*}$ at the minimum $S_i= S_{i*}$:
%\al{
%f_i\simeq f_{i*}+{A_i\over2}\paren{\delta  S_i}^2,
%}
%where $A_i$ is the inverse-square of the variance of $ S_i$ and $\delta S_i$ stands for the fluctuation of $S_i$. 
%
%Let us recall what was the cosmological constant problem.
%In ordinary QFT, natural values of $S_{i*}$ would be order of unity, and $S$ will take a typical value
%\al{
%S_*=\sum_{i=1}^{r_U^4/l_P^4}S_{i*}\sim {r_U^4\over l_P^4}.
%	\label{typical action}
%}
%The action $S$ can be written by energy density $\rho_\Lambda$ as
%\al{\label{S4 action}
%S	&\sim \int_{S^4} \df^4x\rho_\Lambda \sim	r_U^4\rho_\Lambda.
%}
%If we simply identified \eqref{typical action} and \eqref{S4 action}, we would obtain
%Eq.~\eqref{typical action} implies
%\al{
%\rho_\Lambda
%	\stackrel{?}{\sim} {1\over l_P^4}=M_P^4,
%}
%which is of the 120 orders of magnitude larger than the observation~\eqref{cosmological constant observed}.
Assuming that $S_i$ are independent of each other, 
the vanishing cosmological constant for the large universe leads to $\Braket{S_i}=0$ for each $i$ and in particular to $\Braket{S}=0$ for this part.
%let us assume that $S_*$ is tuned to be zero e.g.\ by the mechanism proposed in this paper, namely the MPP with runaway vacuum.
%
Therefore the value of $S$ fluctuates around zero and its variance can be evaluated as
\al{\label{fluctuation}
\Braket{S^2}\sim
	N:={r_U^4\over l_P^4}, %\sim 10^{120}.
}
where we have assumed that the variance of each $S_i$ is of order unity.
%Then the value of the cosmological constant that we observe today will be the root mean square of the fluctuations from zero. For each site $i$, the natural value of the variance of $S_i$ would be of order unity. There are totally $N:=r_U^4/l_P^4$ sites, and the total%average of the 
%fluctuation is
%\al{\label{fluctuation}
%\Braket{\delta S}\sim
%	\sqrt{N}={r_U^2\over l_P^2}\sim 10^{120}.
%}

We interpret Eq.~\eqref{fluctuation} as the variance of the actions of the $S^4$'s in the large universe.
%of 
%that is integrated over such a part of the large universe that will eventually becomes $dS_4$ with length scale $H^{-1}$ in the end of its history. 
Then the typical amount of the energy density of one $S^4$ is estimated as
%For each site $i$ with volume $l_P^4$, the fluctuation of the cosmological constant $\delta\rho_{\Lambda i}$ for a given $\delta S_i$ is $\delta \rho_{\Lambda i}=\delta S_i/l_P^4$.
%Therefore, the root mean square value of the cosmological constant becomes
\al{
\rho_\Lambda
	\sim	{\sqrt{\Braket{S^2}}\over r_U^4}
	\sim	{1\over l_P^2 r_U^2}\sim\paren{\meV}^4.
}
Thus, we have obtained the right amount of the cosmological constant as %the root mean square of the 
the fluctuation from zero.
{\blue This result has been obtained in Ref.~\cite{Sorkin:2007bd} in the context of causal set theory.}

We note that the value of $H$ is not really a prediction in this argument. We have rather provided a consistent explanation of having a finite amount of the cosmological constant, even though it is fixed to be zero for the large universe.

%%%%%%%%%%%%%%%%%%%%%%%%%%%%%%%%%%%%%%%%%%%%%%%%%%%%%%%%%%%%
\section{Summary \red and discussions}\label{summary}
We have studied possible large field limits of the SM Higgs, assuming that it is coming from a massless state at the tree level in heterotic string theory with its supersymmetry broken at the string scale.
In the toroidal compactification, putting a background for such a massless state corresponds to a boost in the momentum lattice. We have classified the large boost limits with fixed radius into three categories: runaway, periodic, and chaotic.

As a concrete toy model, we have examined the ten-dimensional $SO(16)\times SO(16)$ non-supersymmetric heterotic string, with a dimension being compactified on $S^1$ with the radius $R$.
We have considered the large field limit of a Wilson line on the $S^1$ with fixed $R$, and reproduced these three limits.
We have argued that this behavior is universal if the zero momentum limit of the emission vertex of the Higgs is written as a product of holomorphic (1,0) and anti-holomorphic (0,1) operators, not only in the case of toroidal compactification.
In the known models of fermionic construction and of orbifolding, the emission vertex tends to be written as such a product, and our argument applies for these wide class of models.

Physically several degrees of freedom appears when the Higgs field value becomes larger than the Planck scale. %\footnote{
%\red Such a direction typically acquires a one-loop suppressed mass compared to $M_\text{s}$ in this non-supersymmetric setup, and is different from the moduli in the ordinary supersymmetric context, which have the mass of the order of the supersymmetry breaking scale that is far smaller than $M_\text{s}$. The moduli problem in the supersymmetric vacua does not exist in the non-supersymmetric ones.}
We have argued that there exists an runaway direction in this multi degrees of freedom space. This runaway vacuum corresponds to opening up an extra dimension.

It is noteworthy that this potential fits into the criteria of the MPP proposed by Froggatt and Nielsen. 
The MPP requires that the electroweak vacuum is degenerate with this runaway vacuum, and hence that the cosmological constant of the electroweak vacuum is tuned to be zero in the large universe.
%This may give a solution to the cosmological constant problem.
We have speculated that the observed amount of the cosmological constant can be understood as a fluctuation from zero in the framework of the MPP.

We may get the eternal inflation from this potential. 
It is realized either as a topological inflation at the domain wall between the two vacua or as a decay from the false vacuum that traps the Higgs field.
In both cases, the Higgs field, which is rolling down the potential, may cause the succeeding inflation, which accounts for the observed CMB fluctuations, along the quasi-flat potential around the critical point.

%without spoiling the criticality of the SM Higgs potential which is necessary for the secondary inflation that accounts for the observed CMB fluctuation.

It would be interesting to study the limit in more realistic SM-like model with the orbifolding, fermionic constructions, etc; see e.g.\ Ref.~\cite{Blaszczyk:2014qoa}.

{\red
Finally we comment on the dilaton potential. Though we consider the general compactifications which may not even have a geometric interpretation, let us illustrate the situation starting from a conventional ten dimensional string theory. The low energy effective action in ten dimensions reads
\al{
S	&=	{M_\text{s}^8\over g_\text{s}^2}\int\df^{10}x\sqrt{-g}\,e^{-2\Phi}
			\paren{\mathcal R+4\p_\mu\Phi\,\p^\mu\Phi+\cdots}\nn
	&\quad	+M_\text{s}^{10}\int\df^{10}x\sqrt{-g}\paren{-C+\cdots}\nn
	&\quad	+\mathcal O\fn{g_s^2e^{2\Phi}},
}
where $\Phi$ is the dilaton field and $C$ is the dimensionless cosmological constant induced at the one-loop level. We note that in this string frame, $g_\text{s}$ and $\Phi$ always appear in the combination $g_\text{s}e^\Phi$. After the compactification,
\al{
S	&=	{M_\text{s}^2\over g_\text{s}^2}\paren{M_\text{s}^6V_6}\int\df^4x\sqrt{-g_4}\,e^{-2\Phi}
			\paren{\mathcal R_4+4\p_\mu\Phi\,\p^\mu\Phi+\cdots}\nn
	&\quad	+M_\text{s}^4\paren{M_\text{s}^6V_6}\int\df^4x\sqrt{-g_4}\paren{-C+\cdots}\nn
	&\quad	+\mathcal O\fn{g_s^2e^{2\Phi}},
		\label{string frame}
}
where $V_6$ is the compactification volume. Switching to the Einstein frame, we get
\al{
S	&=	{M_\text{s}^2\over g_\text{s}^2}\paren{M_\text{s}^6V_6}\int\df^4x\sqrt{-g_E}
			\paren{\mathcal R_E-2\p_\mu\Phi\,\p^\mu\Phi+\cdots}\nn
	&\quad	+M_\text{s}^4\paren{M_\text{s}^6V_6}\int\df^4x\sqrt{-g_E}\paren{-C\,e^{4\Phi}+\cdots}\nn
	&\quad	+\cdots.
}
We see from the second line that the dilaton has the runaway potential $e^{4\Phi}$ for $\Phi\to-\infty$ if the cosmological constant $C$ is positive. In this limit, the expansion parameter $g_se^\Phi$ in Eq.~\eqref{string frame} becomes small, and the theory is weakly coupled. Since all the higher-loop corrections come with this combination as well, the runaway behavior is not altered by taking them into account. Therefore, this direction $\Phi\to-\infty$ necessarily comprises one of the runaway directions~\cite{Dine:1985he} in Fig.~\ref{potential}, and hence the arguments in Sections~\ref{eternal section} and \ref{solution to cosmological constant} apply quite generally.
}

\subsection*{Acknowledgement}
We thank Michael Blaszczyk, Stefan Groot Nibbelink, Orestis Loukas, S\'aul Ramos-S\'anchez, and Toshifumi Yamashita for useful comments.
This work is in part supported by the Grant-in-Aid for Scientific Research Nos.\ 22540277 (HK), 23104009, 20244028, and 23740192 (KO). The work of Y. H. is supported by a Grant-in-Aid for Japan Society for the Promotion of Science (JSPS) Fellows No.25$\cdot$1107. 
\appendix
%%%%%%%%%%%%%%%%%%%%%%%%%%%%%%%%%%%%%%%%%%%%%%%%%%%%%%%%%%%%%%%
\section{Theta functions}\label{notation}
We list the notations for the functions that we use in the computation of the partition function. (The notations are the same as in Polchinski's textbook but we list them anyway for convenience.) The Dedekind eta function is
\al{
\eta(\tau)=q^{1/24}\prod_{n=1}^\infty\paren{1-q^n},
}
where $q=e^{2\pi\tau}$.
We write theta function with characteristics as
\al{
\vartheta
\begin{bmatrix}
a\\
b
\end{bmatrix}
(\nu,\tau)&=
\sum_{n=-\infty}^\infty \exp\fn{\pi i\paren{n+a}^2\tau+2\pi i\paren{n+a}\paren{\nu+b}},
}
and introduce the following shorthand notations:
\al{
\vartheta_{00}(\tau)&=\vartheta
\begin{bmatrix}
0\\
0
\end{bmatrix}
(0,\tau),\\
\vartheta_{01}(\tau)&=\vartheta
\begin{bmatrix}
0\\
1/2
\end{bmatrix}
(0,\tau),\\
\vartheta_{10}(\tau)&=\vartheta
\begin{bmatrix}
1/2\\
0
\end{bmatrix}
(0,\tau),\\
\vartheta_{11}(\tau)&=\vartheta
\begin{bmatrix}
1/2\\
1/2
\end{bmatrix}
(0,\tau)=0.
}
The Jacobi's identity reads
\al{
\paren{\vartheta_{00}}^4-\paren{\vartheta_{01}}^4-\paren{\vartheta_{10}}^4=0.
}

\section{Fermionic construction manual for ten dimensions}\label{fermionic construction}
We review the $SO(16)\times SO(16)$ heterotic string theory in terms of the fermionic construction, and show the computation of its one-loop partition function.
In heterotic string theory, the right-moving modes are the same as the superstring in 10 dimensions $X^\mu$ ($\mu=0,\dots,9$), while the left-movers as the bosonic string in 26 dimensions with their ``internal'' $X_L^I$ ($I=1,\dots,16$) being compactified.

\subsection{Generalized GSO projection}\label{formalism section}
We work in the light-cone gauge, where $X^+$ is identified with the time direction and $X^-$ is written in terms of the transverse modes, where
\al{
X_\pm={1\over\sqrt{2}}(X^0\pm X^1).
}
We express the left-moving extra degrees of freedom $X_L^I$ by 16 complex fermions, while we form 4 complex fermions by pairing the right-moving fermions $\psi_R^m$ ($m=2,\dots,9$).
Hereafter, $\psi^a$ ($a=1,\dots,4$) denote the 4 complex fermions representing $\psi_R^m$, and $\psi^a$ ($a=5,\dots,20$) denote the 16 complex fermions representing $X_L^I$.

\begin{figure}[tn]
\begin{center}
\hfill
\includegraphics[width=.8\textwidth]{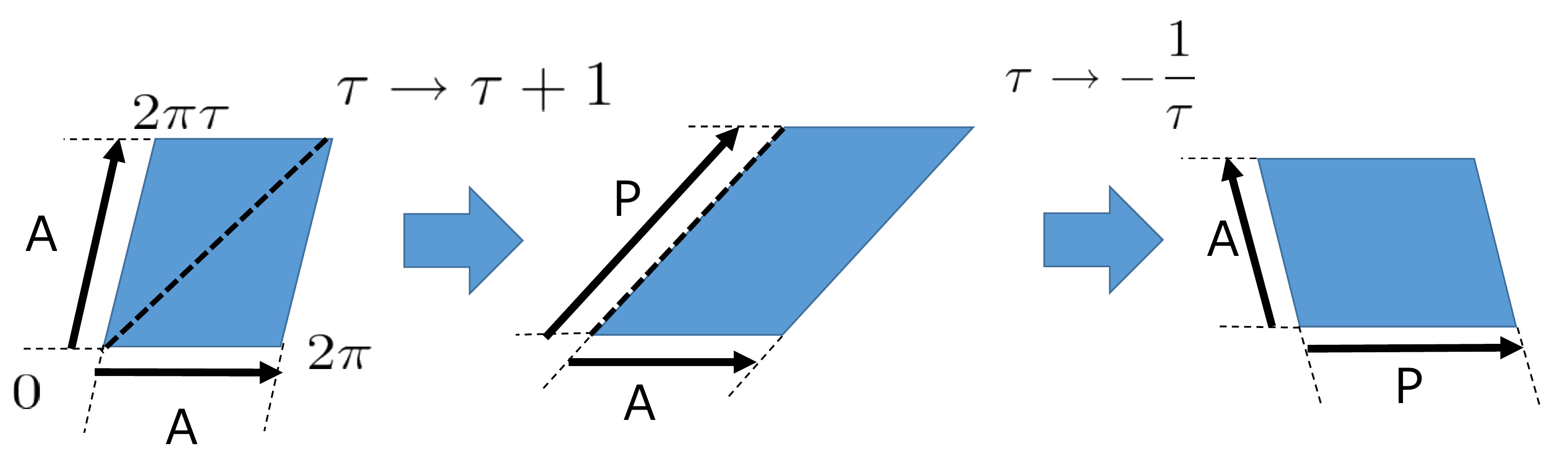}
%\hfill\mbox{}
%\\
%\hfill
%\includegraphics[width=.25\textwidth]{interaction.pdf}
\hfill\mbox{}
\caption{
A and P denote the anti-periodic and periodic boundary conditions, respectively.
The horizontal direction is the spatial one, $\sigma$.
%In the horizontal (namely spatial) direction, A and P correspond to the NS and R boundary conditions, respectively.
Given the all-A boundary condition for both the time and spatial directions, the modular invariance necessitates the all-P boundary condition for the time or spatial direction.
}\label{modular invariance}
\end{center}
\end{figure}

\begin{figure}[tn]
\begin{center}
\hfill
\includegraphics[width=.25\textwidth]{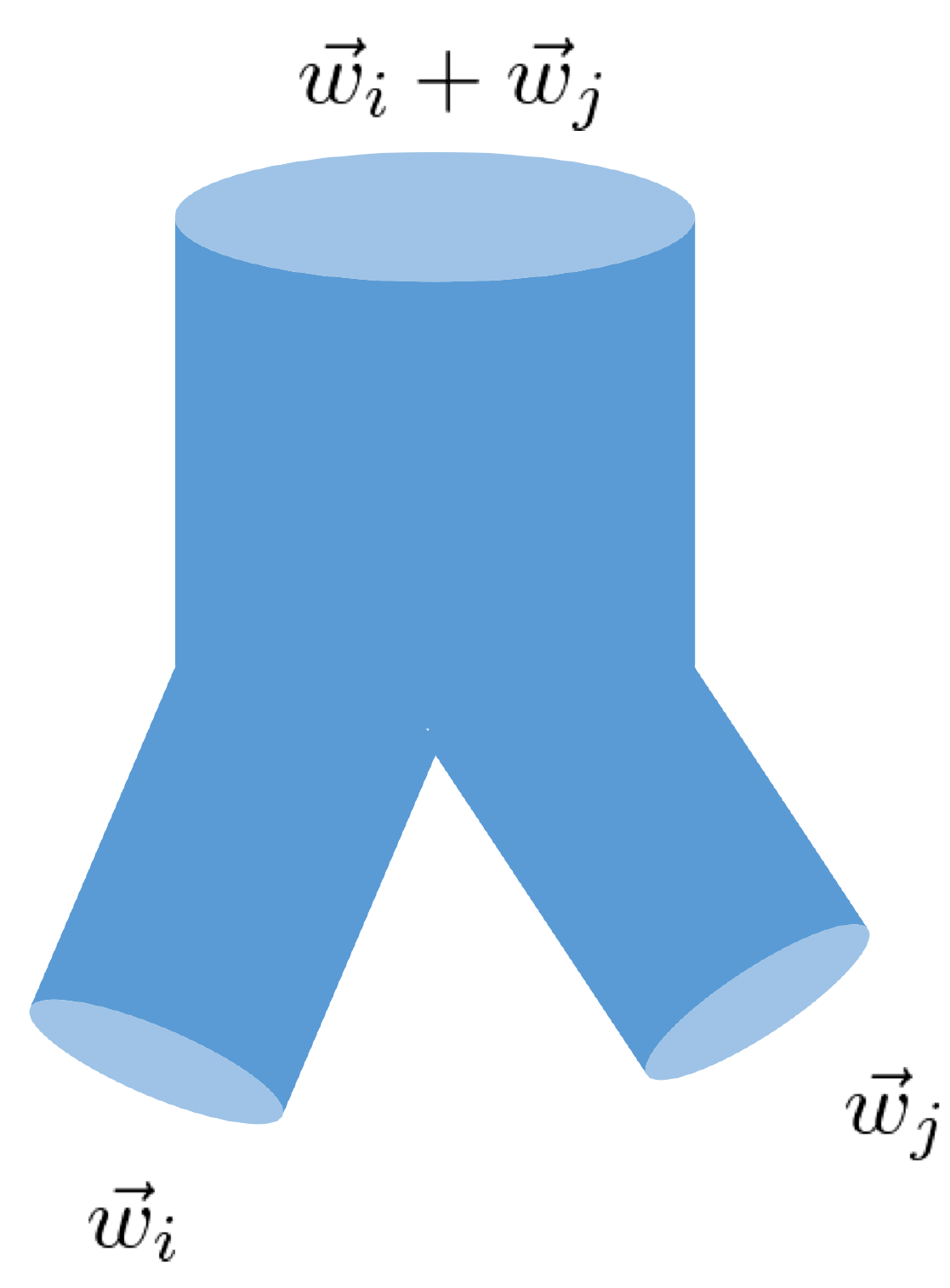}
\hfill\mbox{}
\caption{
Schematic picture for the string interaction joining $\vec{w}_i$ and $\vec{w}_j$ strings to make $\vec w_i+\vec w_j$ string.
If there exist the sets of boundary conditions $\vec w_i$ and $\vec w_j$, then there must be $\vec w_i+\vec w_j$ in $W$.
}\label{interaction}
\end{center}
\end{figure}
Let us review the procedure to retain the modular invariance.
Here we assume that each of the above-listed complex fermions $\psi^a$ ($a=1,\dots,20$) either has the NS (anti-periodic) or R (periodic) boundary condition on the worldsheet $\sigma\sim \sigma+2\pi$:\footnote{
One can consider a more general boundary condition such as $\psi^a\fn{\sigma=2\pi}=\pm\ol{\psi^a}\fn{\sigma=0}$ and $\psi^a\fn{\sigma=2\pi}=-e^{2\pi iw_a}\psi^a\fn{\sigma=0}$ for arbitrary rational $w^a$~\cite{Kawai:1986vd}.
}
\al{
\text{NS: }\quad
\psi^a\fn{\sigma=2\pi}
	&=	-\psi^a\fn{\sigma=0},\nn
\text{R: }\quad
\psi^a\fn{\sigma=2\pi}
	&=	\psi^a\fn{\sigma=0}.
}
We can write them collectively
\al{
\psi^a\fn{\sigma=2\pi}
	&=	-e^{2\pi i w^a}\psi^a\fn{\sigma=0},
}
where the vector $\vec w=\paren{w^a}_{a=1,\dots,20}$ consists of either 0 (NS, anti-periodic) or $1/2$ (R, periodic) modulo 1.

We classify the possible boundary conditions by the following procedure.
Let $W$ be a vector space over $Z_2$ spanned by the bases $\Set{\vec w_i}_{i=0,\dots,l}$.
$W$ is the set of boundary conditions, appearing in a theory, that are required by the string interaction and the modular invariance:
\begin{itemize}
\item There must be all anti-periodic boundary condition
\al{
\vec 0=\paren{0}_{a=1,\dots,20}=\paren{0,\dots,0}
}
in $W$: When considering a partition function on the torus, there must be the sector in which all the fermions are anti-periodic in the time direction in order to get the identity operator in the trace, which is needed to form a projection operator; then the $S$-transformation $\tau\to-1/\tau$ maps this condition to the space direction.
\item Then the modular invariance necessitates the all-periodic boundary condition
\al{
\vec w_0:=\paren{1/2}_{a=1,\dots,20}=\paren{1/2,\dots,1/2}
}
in $W$; see Fig.~\ref{modular invariance}. 
\item If $\vec w_i$ and $\vec w_j$ exist in $W$, then $\vec{w}_i+\vec{w}_j$ must also be in $W$; see Fig.~\ref{interaction}.
\end{itemize} 
A boundary condition belonging to $W$ can be written as
\al{
\psi^a\fn{\sigma=2\pi}
	&=	-e^{2\pi i\paren{\alpha\vec w}^a}\psi^a\fn{\sigma=0},
}
where $\alpha\vec w:=\sum_{i=0}^l\alpha_i\vec w_i$, namely $\paren{\alpha\vec w}^a:=\sum_{i=0}^l\alpha_i w_i^a$, with $\alpha_i$ ($i=0,\dots, l$) being either 0 or 1 and the vector $\vec w_i=\paren{w_i^a}_{a=1,\dots,20}$  consisting of either 0 or $1/2$ again.
%We split $W$ into sectors: $\alpha\vec w$ and $\beta\vec w$ belong to the same sector if ${\alpha_i/2}\modeq{\beta_i/2}$ for all $i=0,\dots,l$.

The partition function becomes modular invariant if and only if we impose the generalized GSO projection, under which the surviving states satisfy the following condition~\cite{Kawai:1986vd}:
\al{
e^{2\pi i \vec{w}_i\cdot \vec{N}_{\alpha \vec{w}}}
	&=	e^{2\pi i\paren{\sum_{j=0}^l k_{ij}\alpha_j-\vec{w}_i\cdot \ol{\ol{\alpha \vec w}}+s_i}},
			\qquad\text{for each $i=0,\dots,l$,}
}
that is,
\al{
\vec{w}_i\cdot \vec{N}_{\alpha \vec{w}}
	&\modeq
		\sum_{j=0}^l k_{ij}\alpha_j-\vec{w}_i\cdot \ol{\ol{\alpha \vec w}}+s_i,
			\qquad\text{for each $i=0,\dots,l$,}
			\label{GSO projection}
}
where $\modeq$ stands for the equality modulo 1; $\vec N_{\alpha\vec w}$ is the vector consisting of the worldsheet fermion numbers for the $\alpha\vec w$ sector; the inner product is Lorentzian such that $+$ and $-$ are respectively assigned for right and left movers,
\al{
\vec{w}_i\cdot \vec{w}_j
	&:=	\sum_{a=1}^4 {w}_i^a{w}_j^a-\sum_{b=5}^{20} {w}_i^b{w}_j^b;
}
$s_i$ denotes the value of the right-moving components of $w_i^a$ ($a=1,\dots,4$),
\al{
s_i:=w_i^1=w_i^2=w_i^3=w_i^4
	\label{s_i defined}
}
($\sum_{i=0}^l\alpha_is_i=0$ and $1/2$ respectively indicate that the $\alpha\vec w$ sector is a spacetime boson and fermion);\footnote{
All the right-moving components must take the same value as in Eq.~\eqref{s_i defined} since we assume the 10 dimensional Lorentz invariance.
}
$\ol{\ol{\alpha \vec w}}$ is the vector, each of its component being the fractional part of the corresponding component of $\alpha \vec w$, that is, $\ol{\ol{\alpha \vec{w}}}$ is the fractional vector in the decomposition\footnote{
The fractional vector is chosen in such a way that all its components are within $[-1/2,1/2)$.
In our application, $\vec w_i\cdot\ol{\ol{\alpha \vec{w}}}$ turns out to be zero.
}
\al{
\alpha \vec{w}=\sum_{i=0}^l\alpha_i\vec{w}_i=\text{(integer vector)}+\text{(fractional vector)};
}
and $k_{ij}$ is the solution to the following conditions
\al{
k_{ij}+k_{ji}&\modeq \vec{w}_i\cdot \vec{w}_j\nn
k_{ij}m_j&\modeq 0,	&&\text{all the indices $i,j=0,\dots,l$ unsummed,}\nn
k_{ii}+k_{i0}+s_i&\modeq {1\over 2}\vec{w}_i\cdot\vec{w}_i,\label{kequation}
}
with $m_i$ being the smallest integer that satisfies $m_i\vec{w}_i\modeq \vec{0}$ for each $i$ unsummed (In our case $m_i=2$).
%($s_i$ is fixed for a given $W$, whereas $\vec N_{\alpha\vec w}$ depends on the integers $\alpha_i$.)

For a given $W$, the condition~\eqref{kequation} may have several solutions for $k_{ij}$.
Each solution $k_{ij}$ gives a 10 dimensional string theory that is in general physically distinct from the others.
These solutions are believed to complete all the possible consistent string theories in 10 dimensions~\cite{Kawai:1986vd}.
To summarize, once a set of basis vectors $\Set{\vec w_i}_{i=0,\dots,l}$ is given, then one can construct consistent string theories according to the above procedure.
A concrete example is shown below.

\subsection{$E_8\times E_8$ and $SO(16)\times SO(16)$ string theories}
\label{concrete heterotic models}
We can obtain the $E_8\times E_8$ superstring theory and the $SO(16)\times SO(16)$ non-supersymmetric string theory by the following choice of basis:\footnote{
$SO(32)$ supersymmetric string corresponds to the bases $\Set{\vec{w}_0,\vec{w}_1}$.
}
\al{
%\vec{0}&=
%\begin{pmatrix}
%0^4&|&0^8 & 0^8
%\end{pmatrix}
%,\nn
\vec w_0&=
\paren{
\left({1\over 2}\right)^4\,\bigg|\,\left({1\over 2}\right)^8\,\left({1\over 2}\right)^8
},\nn
\vec w_1&=
\paren{
0^4\,\bigg|\,\left({1\over 2}\right)^8\,\left({1\over 2}\right)^8
}
,\nn
\vec w_2&=
\paren{
0^4\,\bigg|\,\left({1\over 2}\right)^8\,0^8
},\label{bases w}
}
where $0$ and $1/2$ represent the anti-periodic and periodic boundary conditions, respectively, as explained above;
those on the left (right) of $|$ are the boundary conditions for the right (left) moving fermions;
and e.g.\ $0^4$ denote that there are four $0$s in the slots.

Let us obtain $k_{ij}$ for the basis~\eqref{bases w}. We express $k_{ij}$ by a three-by-three matrix:
\al{
k	=	\bmat{
		k_{00}&k_{01}&k_{02}\\
		k_{10}&k_{11}&k_{12}\\
		k_{20}&k_{21}&k_{22}
		}
	=	\bmat{
		a&b&c\\
		d&e&f\\
		g&h&i
		}.
}
Noting that
$m_0=m_1=m_2=2$, 
$s_0=1/2$, 
$s_1=s_2=0$, 
$\vec{w}_0\cdot\vec{w}_0=-3$, 
$\vec w_1\cdot\vec w_1=\vec w_0\cdot\vec w_1=-4$, and
$\vec w_2\cdot\vec w_2=\vec w_0\cdot\vec w_2=\vec w_1\cdot\vec w_2=-2$, 
we obtain from Eq.~\eqref{kequation}
\al{
k	&\modeq
\bmat{
a&b&c\\
b&b&f\\
c&f&c
}, &
a,b,c,f
	&\modeq	0\text{ or }{1\over2}.
}
From $2\vec w_i\modeq\vec 0$, we have the eight sectors shown in the table below.
\als{
\begin{array}{|c|cccccccc|}
\hline
\alpha&
\paren{0,0,0}&
\paren{1,0,0}&
\paren{0,1,0}&
\paren{0,0,1}&
\paren{1,1,0}&
\paren{1,0,1}&
\paren{0,1,1}&
\paren{1,1,1}\\
\hline
\alpha\vec w&
\vec{0} &
\vec{w}_0 &
\vec{w}_1 &
\vec{w}_2 &
\vec{w}_0+\vec{w}_1 &
\vec{w}_0+\vec{w}_2 &
\vec{w}_1+\vec{w}_2 &
\vec{w}_0+\vec{w}_1+\vec{w}_2\\
\hline
\end{array}
}
Note that $\vec w_i\cdot\ol{\ol{\alpha\vec w}}\modeq0$ for all sectors in this case.

Let us see the massless spectrum of each sector.
The ground state energies of the $\vec 0$ sector are $-M_\text{s}/2$ and $-M_\text{s}$ for the right and left movers, respectively; recall that we have taken $M_\text{s}:=\sqrt{1/\alpha'}$.
Changing the boundary condition of each slot ($a=1,\dots,20$) from NS (anti-periodic) to R (periodic) raises the vacuum energy by $M_\text{s}/8$.
The lowest bosonic and fermionic modes raise the energy by $M_\text{s}$ and $M_\text{s}/2$, respectively.
The level matching condition says that the left and right levels should be the same.  
We see that the possible problem of having tachyonic modes resides only in the $\vec 0$ sector; we will check that they are safely projected out.

Let $N_R$ be the number of right-moving complex fermions in the first 4 slots, where $\alpha\vec w$-dependence is made implicit for simplicity.
Similarly, the subsequent 8 slots for the left-movers are numbered as $N_{L1}$ and the last 8 slots $N_{L2}$.
We can write
\al{
\vec w_i\cdot\vec N_{\alpha\vec w}
	&=	\sum_{a=1}^4w_i^a N_R^a
		-\sum_{a=5}^{12}w_i^a N_{L1}^a
		-\sum_{a=13}^{20}w_i^a N_{L2}^a,
}
where
\al{
N_R
	&=	\sum_{a=1}^4N_R^a,	&
N_{L1}
	&=	\sum_{a=5}^{12}N_{L1}^a,	&
N_{L2}
	&=	\sum_{a=13}^{20}N_{L2}^a.
}
In our case~\eqref{bases w}, we get
\al{
\vec{w}_0\cdot \vec{N}_{\alpha\vec w}
	&=	{N_R\over2}-{N_{L1}\over2}-{N_{L2}\over2},	\nn
\vec{w}_1\cdot \vec{N}_{\alpha\vec w}
	&=	-{N_{L1}\over2}-{N_{L2}\over2},	\nn
\vec{w}_2\cdot \vec{N}_{\alpha\vec w}
	&=	-{N_{L1}\over2}.
}
For the fermions with R (periodic) boundary condition, it is convenient to use
\al{
\Gamma_R
	&:=	\paren{-1}^{N_R},	&
\Gamma_{L1}
	&:=	\paren{-1}^{N_{L1}},	&
\Gamma_{L2}
	&:=	\paren{-1}^{N_{L2}},	
}
since $\Gamma_R$ gives the chirality of the 10 dimensional spinor for the right-moving fermions and $\Gamma_{L1}, \Gamma_{L2}$ give the chirality of the $SO(16)$ spinor for the left-moving fermions.

We look for the surviving states under the three projections $i=0,1,2$ in Eq.~\eqref{GSO projection}.
\begin{itemize}
\item $\vec{0}$ sector: All the fermions have the NS (anti-periodic) boundary condition.
The projection~\eqref{GSO projection} reads
\al{
\vec{w}_0\cdot \vec{N}_{\vec{0}}
	=	{N_R\over2}-{N_{L1}\over2}-{N_{L2}\over2}
	&\modeq	{1\over 2},	\nn
\vec{w}_1\cdot \vec{N}_{\vec{0}}
	=	-{N_{L1}\over2}-{N_{L2}\over2}
	&\modeq	0,	\nn
\vec{w}_2\cdot \vec{N}_{\vec{0}}
	=	-{N_{L1}\over2}
	&\modeq	0.
}
When exponentiated, it results in
\al{
\paren{-1}^{N_R}&=-1,	&
\paren{-1}^{N_{L1}}&=1,	&
\paren{-1}^{N_{L2}}&=1.
}

We see that we need at least one mode of the right-moving fermion $\psi_R^m$, which raises the mass level from $-M_\text{s}/2$ at least to 0. Then the level matching condition tells that the left levels start from 0 too.
Therefore, there remains no tachyonic mode.

The massless states in this sector are
\al{
\psi_{R,-1/2}^mX_{L,-1}^n\ket{0}_{\vec 0}
}
($m,n=2,\dots,9$) that becomes a graviton, an antisymmetric tensor, and a dilation in ten dimensions and
\al{
\psi_{R,-1/2}^m\psi_{L,-1/2}^a\psi_{L,-1/2}^b\ket{0}_{\vec 0}
}
($a,b=5,\dots,12$ or $a,b=13,\dots,20$) that becomes $SO(16)\times SO(16)$ gauge boson.
To summarize, the massless states are $(\bs{35},\bs1,\bs1)+(\bs{28},\bs1,\bs1)+(\bs1,\bs1,\bs1)$ and $(\bs8_v,\bs{120},\bs1)+(\bs8_v,\bs1,\bs{120})$ in terms of $SO(8)\times SO(16) \times SO(16)$, where $\bs8_v$ is the vector representation.
This sector is common for the $E_8\times E_8$ superstring and the $SO(16)\times SO(16)$ non-supersymmetric string.

\item $\vec{w}_0=\paren{\paren{1\over2}^4\,\Big|\,\paren{1\over2}^8\,\paren{1\over2}^8}$ sector:
All the fermions have the R (periodic) boundary condition. The projection~\eqref{GSO projection} is
\al{
\vec{w}_0\cdot \vec{N}_{\vec{w}_0}
	=	{N_R\over2}-{N_{L1}\over2}-{N_{L2}\over2}
	&\modeq a+{1\over2},\nn
\vec{w}_1\cdot \vec{N}_{\vec{w}_0}
	=	-{N_{L1}\over2}-{N_{L2}\over2}
	&\modeq b,\nn
\vec{w}_2\cdot \vec{N}_{\vec{w}_0}
	=	-{N_{L1}\over2}
	&\modeq c.
}
That is,
\al{
\Gamma_{R}&=\paren{-1}^{2\paren{a+b}+1},\nn
\Gamma_{L1}&=\paren{-1}^{2c},\nn
\Gamma_{L2}&=\paren{-1}^{2(b+c)}.
}
The left ground state is raised by $16\times {M_\text{s}\over8}$ from $-M_\text{s}$ due to the R (periodic) boundary conditions. The lightest left states start from $M_\text{s}$. So do the right states due to the level matching condition.
There is no massless state in this sector.
\item $\vec{w}_1=\paren{0^4\,\Big|\,\paren{1\over2}^8\,\paren{1\over2}^8}$ sector:
The right and left movers have the NS (anti-periodic) and R (periodic) boundary conditions, respectively.
The projection~\eqref{GSO projection} is
\al{
\vec{w}_0\cdot \vec{N}_{\vec{w}_1}
	=	{N_R\over2}-{N_{L1}\over2}-{N_{L2}\over2}
	&\modeq
		b+{1\over 2},\nn
\vec{w}_1\cdot \vec{N}_{\vec{w}_1}
	=	-{N_{L1}\over2}-{N_{L2}\over2}
	&\modeq
		b,\nn
\vec{w}_2\cdot \vec{N}_{\vec{w}_1}
	=	-{N_{L1}\over2}
	&\modeq
		f,
}
that is,
\al{
\paren{-1}^{N_R}&=-1,\nn
\Gamma_{L1}&=\paren{-1}^{2f},\nn
\Gamma_{L2}&=\paren{-1}^{2\paren{b+f}}.
}
Following the same reasoning as the $\vec w_0$ sector, there is no massless state in this sector.
\item $\vec{w}_2=\paren{0^4\,\Big|\,\paren{1\over2}^8\,0^8}$ sector:
The projection~\eqref{GSO projection} is
\al{
\vec{w}_0\cdot \vec N_{\vec{w}_2}
	=	{N_R\over2}-{N_{L1}\over2}-{N_{L2}\over2}
	&\modeq c+{1\over 2},\nn
\vec{w}_1\cdot \vec N_{\vec{w}_2}
	=	-{N_{L1}\over2}-{N_{L2}\over2}
	&\modeq f,\nn
\vec{w}_2\cdot \vec N_{\vec{w}_2}
	=	-{N_{L1}\over2}
	&\modeq c,
}
that is,
\al{
\paren{-1}^{N_R}&=\paren{-1}^{2\paren{c+f}+1},\nn
\Gamma_{L1}&=\paren{-1}^{2c},\nn
\paren{-1}^{N_{L2}}&=\paren{-1}^{2\paren{c+f}}.
}
The massless spectrum depends on the value of $c+f$.
If $c+f\modeq 0$, the massless states form a spacetime vector: $\paren{\bs8_v,\bs{128},\bs1}$ of $SO(8)\times SO(16) \times SO(16)$, which is a part of the $E_8\times E_8$ gauge boson in the superstring theory.
If $c+f\modeq 1/2$, there is no massless state. 

\item $\vec{w}_0+\vec{w}_1=\paren{\paren{1\over2}^4\,\Big|\,0^8\,0^8}$ sector:
The projection~\eqref{GSO projection} is
\al{
\vec{w}_0\cdot \vec N_{\vec{w}_0+\vec{w}_1}
	=	{N_R\over2}-{N_{L1}\over2}-{N_{L2}\over2}
	&\modeq a+b+{1\over2},\nn
\vec{w}_1\cdot \vec N_{\vec{w}_0+\vec{w}_1}
	=	-{N_{L1}\over2}-{N_{L2}\over2}
	&\modeq 0,\nn
\vec{w}_2\cdot \vec N_{\vec{w}_0+\vec{w}_1}
	=	-{N_{L1}\over2}
	&\modeq c+f,
}
that is,
\al{
\Gamma_{R}&=\paren{-1}^{2\paren{a+b}+1},\nn
\paren{-1}^{N_{L1}}&=\paren{-1}^{2\paren{c+f}},\nn
\paren{-1}^{N_{L2}}&=\paren{-1}^{2\paren{c+f}}.
}
The massless spectrum depends on the value of $c+f$.
If $c+f\modeq 0$, the massless state becomes the gravitino and dilatino $\paren{\bs{56},\bs1,\bs1}+\paren{\bs8',\bs1,\bs1}$ and the gaugino $\paren{\bs8,\bs{120},\bs1}+\paren{\bs8,\bs1,\bs{120}}$ in terms of $SO(8)\times SO(16) \times SO(16)$, where $\bs8$ and $\bs8'$ are two spacetime spinor representations with different chiralities. We see that a spacetime supersymmetry remains.
If $c+f\modeq 1/2$, the massless state becomes a spacetime spinor $\paren{\bs8,\bs{16},\bs{16}}$ which belongs to the bi-fundamental representation of the gauge group.
This theory does not have a gravitino nor a gaugino, and hence the supersymmetry is not left.

\item $\vec{w}_0+\vec{w}_2=\paren{\paren{1\over2}^4\,\Big|\,0^8\,\paren{1\over2}^8}$ sector:
The projection~\eqref{GSO projection} is
\al{
\vec{w}_0\cdot \vec N_{\vec{w}_0+\vec{w}_2}
	=	{N_R\over2}-{N_{L1}\over2}-{N_{L2}\over2}
	&\modeq a+c+{1\over2},\nn
\vec{w}_1\cdot \vec N_{\vec{w}_0+\vec{w}_2}
	=	-{N_{L1}\over2}-{N_{L2}\over2}
	&\modeq b+f,\nn
\vec{w}_2\cdot \vec N_{\vec{w}_0+\vec{w}_2}
	=	-{N_{L1}\over2}
	&\modeq 0,
}
that is,
\al{
\Gamma_R&=\paren{-1}^{2\paren{a+b+c+f}+1},\nn
\paren{-1}^{N_{L1}}&=1,\nn
\Gamma_{L2}&=\paren{-1}^{2\paren{b+f}}.
}
The massless states form a spacetime spinor that is $\paren{\bs8,\bs1,\bs{128}}$ representation of $SO(8)\times SO(16) \times SO(16)$.

\item $\vec{w}_1+\vec{w}_2=\paren{0^4\,\Big|\,0^8\,\paren{1\over2}^8}$ sector:
The projection~\eqref{GSO projection} reads
\al{
\vec{w}_0\cdot \vec N_{\vec{w}_1+\vec{w}_2}
	=	{N_R\over2}-{N_{L1}\over2}-{N_{L2}\over2}
	&\modeq b+c+{1\over2},\nn
\vec{w}_1\cdot \vec N_{\vec{w}_1+\vec{w}_2}
	=	-{N_{L1}\over2}-{N_{L2}\over2}
	&\modeq b+f,\nn
\vec{w}_2\cdot \vec N_{\vec{w}_1+\vec{w}_2}
	=	-{N_{L1}\over2}
	&\modeq c+f,
}
that is,
\al{
\paren{-1}^{N_R}&=\paren{-1}^{2\paren{c+f}+1},\nn
\paren{-1}^{N_{L1}}&=\paren{-1}^{2\paren{c+f}},\nn
\Gamma_{L2}&=\paren{-1}^{2(b+c)}.
}
The massless spectrum depends on the value of $c+f$.
If $c+f\modeq 0$, the massless states form a spacetime vector: $\paren{\bs8_v,\bs1,\bs{128}}$ of $SO(8)\times SO(16) \times SO(16)$. This becomes a part of the $E_8\times E_8$ gauge boson.
If $c+f\modeq 1/2$, there is no massless state.

\item $\vec{w}_0+\vec{w}_1+\vec{w}_2=\paren{\paren{1\over2}^4\,\Big|\,\paren{1\over2}^8\,0^8}$ sector:
The projection~\eqref{GSO projection} is
\al{
\vec{w}_0\cdot \vec N_{\vec{w}_0+\vec{w}_1+\vec{w}_2}
	=	{N_R\over2}-{N_{L1}\over2}-{N_{L2}\over2}
	&\modeq a+b+c+{1\over 2},\nn
\vec{w}_1\cdot \vec N_{\vec{w}_0+\vec{w}_1+\vec{w}_2}
	=	-{N_{L1}\over2}-{N_{L2}\over2}
	&\modeq f,\nn
\vec{w}_2\cdot \vec N_{\vec{w}_0+\vec{w}_1+\vec{w}_2}
	=	-{N_{L1}\over2}
	&\modeq f,
}
that is,
\al{
\Gamma_{R}&=\paren{-1}^{2\paren{a+b+c+f}+1},\nn
\Gamma_{L1}&=\paren{-1}^{2f},\nn
\paren{-1}^{N_{L2}}&=1.
}
The massless states form a spacetime fermion: $\paren{\bs8,\bs{128},\bs1}$ of $SO(8)\times SO(16) \times SO(16)$.
\end{itemize}
To summarize, if $c+f\modeq0$, the theory has a supersymmetry, and the massless states form the supergravity multiplet and the $E_8\times E_8$ vector multiplet in 10 dimensions.
If $c+f\modeq 1/2$, the theory is non-supersymmetric, and the massless states are 
 \al{
\paren{\bs{56},\bs1,\bs1}+\paren{\bs{28},\bs1,\bs1}+\paren{\bs1,\bs1,\bs1}\nn
+\paren{\bs8_v,\bs{120},\bs1}+\paren{\bs8_v,\bs1,\bs{120}}\nn
+\paren{\bs8,\bs{128},\bs1}+\paren{\bs8,\bs1,\bs{128}}+\paren{\bs8',\bs{16},\bs{16}},
 }
of $SO(8)\times SO(16) \times SO(16)$.
In this paper, we consider the latter.

We comment on the choice of chirality.
%The choice of the sign of $\Gamma$ is not necessarily physical, but its relative sign is physical.
Since the chirality of each sector takes either value of
\al{
\Gamma_R
	&=	\paren{-1}^{2\paren{a+b}+1}
	&\text{or}&
	&&	\paren{-1}^{2\paren{a+b+c+f}+1},\nn
\Gamma_{L1}
	&=	\paren{-1}^{2c}
	&	\text{or}&
	&&	\paren{-1}^{2f},\nn
\Gamma_{L2}
	&=	\paren{-1}^{2\paren{b+c}}
	&	\text{or}&
	&&	\paren{-1}^{2\paren{b+f}},
}
the relative difference of the chirality depends only on the combination $c+f$.
Therefore, it suffices to determine $c+f$ in order to classify the theories.

\subsection{Contributions from worldsheet fermions to one-loop partition function}\label{fermion one-loop}
Let us compute the contribution from worldsheet fermions to the one-loop partition function $Z_{T^2}$ in the fermionic construction of the $SO(16)\times SO(16)$ heterotic string theory. (We treat the contributions from the spacetime coordinates in the next section.)

We use the bosonization technique that replaces each worldsheet complex fermion by a worldsheet boson.
The contribution from the oscillator modes of the bosons is the same as in the free boson case, resulting in the factor $1/\bar{\eta}^4\eta^{16}$.
As we will see below, the contribution from the boson zero modes that are constant along $\sigma$ is computed as follows:
The momentum of the boson zero mode is equal to the fermion number of the corresponding fermion; %including the vacuum charge;
therefore for the momentum lattice of the bosons is the same as the charge lattice of the fermions; from NS (anti-periodic) fermions, we replace $N_R$, $N_{L1}$ and $N_{L2}$ in the partition function by the corresponding momentum lattice of the boson zero mode;
for the R (periodic) fermion, we shift the momentum lattice by half of the lattice spacing in order to take the vacuum charge into account.

Let us check the contribution from each sector of fermions.
As we have explained above, $c+f\modeq 1/2$ in the non-supersymmetric heterotic string; we take $a\modeq f\modeq 1/2$ and $b\modeq c\modeq 0$ without loss of generality. 
\begin{itemize}
\item $\vec{0}$ sector:
The momentum lattice is
\al{
\Gamma_{\vec 0}
	&=	\Set{
		\paren{n_1,...,n_4\mid m_1,...,m_8,\,l_1,...,l_8}|
		N\in\textsl{odd},\,
		M\in\textsl{even},\,
		L\in\textsl{even}
		},
}
where $\textsl{even}=2\mathbb Z$, $\textsl{odd}=2\mathbb Z+1$, and we define
\al{
N	&:=	\sum_{i=1}^4 n_i,	&
M	&:=	\sum_{i=1}^8 m_i,	&
L	&:=	\sum_{i=1}^8 l_i.
}
The summation over the momenta of the boson zero modes becomes
\al{
\Zhat_{\vec 0}
&=	\sum_{\br{p_R,\,p_L}\in\Gamma_{\vec 0}} \bar{q}^{p_R^2/2} q^{p_L^2/2}
	\nn
&:=
\sum_{\br{n_1,\dots,n_4,\,m_1,\dots,m_8,\,l_1,\dots,l_8}\in\Gamma_{\vec 0}} 
e^{-\pi i\bar{\tau}\sum_{i=1}^4 n_i^2}
e^{\pi i \tau \sum_{i=1}^8\paren{m_i^2+l_i^2}}\nn
&=
\sum_{\br{n_1,\dots,n_4,\,m_1,\dots,m_8,\,l_1,\dots,l_8}\in\mathbb{Z}^{20}}
	{1-\paren{-1}^{N}\over2}{1+\paren{-1}^{M}\over2}{1+\paren{-1}^{L}\over2}\nn
&\phantom{=\sum_{\br{n_1,\dots,n_4,\,m_1,\dots,m_8,\,l_1,\dots,l_8}\in\mathbb{Z}^{20}}}
	\times e^{-\pi i\bar{\tau}\sum_i n_i^2}
	e^{\pi i \tau \sum_i\paren{m_i^2+l_i^2}}
\nn
&=
{1\over 8}
\paren{\paren{\bar{\vartheta}_{00}}^4-\paren{\bar{\vartheta}_{01}}^4}
\paren{\paren{\vartheta_{00}}^8+\paren{\vartheta_{01}}^8}
\paren{\paren{\vartheta_{00}}^8+\paren{\vartheta_{01}}^8},
}
where the theta functions are listed in Appendix~\ref{notation}.
%$p_R^2:=n_1^2+\cdots+n_4^2$, $p_L^2:=m_1^2+\cdots+m_8^2+l_1^2+\cdots+l_8^2$, and

\item $\vec{w}_0$ sector: The momentum lattice is
\al{
\Gamma_{\vec w_0}
	&=	\bigg\{\paren{n_1+{1\over2},\,...,\,n_4+{1\over2}\,\Big|\, 
			m_1+{1\over2},\,...,\,m_8+{1\over2},\,l_1+{1\over2},\,...,\,l_8+{1\over2}}\bigg|\nn
	&\qquad
			N\in\textsl{even}, \, M\in\textsl{even}, \, L\in\textsl{even}
			\bigg\}.
}
The summation is
\al{
\Zhat_{\vec w_0}
=	-\sum_{\br{p_R,\,p_L}\in\Gamma_{\vec w_0}} \bar{q}^{p_R^2/2} q^{p_L^2/2}
=	-{1\over8}
\paren{\bar{\vartheta}_{10}}^4
\paren{\vartheta_{10}}^8
\paren{\vartheta_{10}}^8.
}
Note that the extra minus sign is put for spacetime fermions.

\item $\vec{w}_1$ sector: The momentum lattice is
\al{
\Gamma_{\vec w_1}
	&=	\Set{\paren{n_1,...,n_4\mid m_1+{1\over2},...,m_8+{1\over2},l_1+{1\over2},...,l_8+{1\over2}}|
		N\in\textsl{odd},\,M\in\textsl{odd},\,L\in\textsl{odd}}.
}
The summation is
\al{
\Zhat_{\vec w_1}=
\sum_{\br{p_R,\,p_L}\in\Gamma_{\vec w_1}} \bar{q}^{p_R^2/2} q^{p_L^2/2}
=
{1\over8}
\paren{\paren{\bar{\vartheta}_{00}}^4-\paren{\bar{\vartheta}_{01}}^4}
\paren{\vartheta_{10}}^8
\paren{\vartheta_{10}}^8.
}

\item $\vec{w}_2$ sector: The momentum lattice is
\al{
\Gamma_{\vec w_2}
	&=	\Set{\paren{n_1,...,n_4\mid m_1+{1\over2},\,...,\,m_8+{1\over2},\,l_1,...,l_8}|
		N\in\textsl{even},\,M\in\textsl{even},\,L=\textsl{odd}}.
}
The summation is
\al{
\Zhat_{\vec w_2}
=	
\sum_{\br{p_R,\,p_L}\in\Gamma_{\vec w_2}} \bar{q}^{p_R^2/2} q^{p_L^2/2}&=
{1\over8}
\paren{\paren{\bar{\vartheta}_{00}}^4+\paren{\bar{\vartheta}_{01}}^4}
\paren{\vartheta_{10}}^8
\paren{\paren{\vartheta_{00}}^8-\paren{\vartheta_{01}}^8}.
}

\item $\vec{w}_0+\vec{w}_1$ sector: The momentum lattice is
\al{
\Gamma_{\vec{w}_0+\vec{w}_1}
	&=	\Set{\paren{n_1+{1\over2},...,n_4+{1\over2}\mid m_1,...,m_8,l_1,...,l_8}|
		N\in\textsl{even},\,M\in\textsl{odd},\,L\in\textsl{odd}}.}
The summation is
\al{
\Zhat_{\vec w_0+\vec w_1}=
-\sum_{\br{p_R,\,p_L}\in\Gamma_{\vec w_0+\vec w_1}} \bar{q}^{p_R^2/2} q^{p_L^2/2}&=
-{1\over8}
\paren{\bar{\vartheta}_{10}}^4
\paren{\paren{\vartheta_{00}}^8-\paren{\vartheta_{01}}^8}
\paren{\paren{\vartheta_{00}}^8-\paren{\vartheta_{01}}^8}.
}

\item $\vec{w}_0+\vec{w}_2$ sector: The momentum lattice is
\al{
\Gamma_{\vec{w}_0+\vec{w}_2}
	&=	\bigg\{\paren{n_1+{1\over2},\,...,\,n_4+{1\over2}\,\Big|\, 
			m_1,\,...,\,m_8,\,l_1+{1\over2},\,...,\,l_8+{1\over2}}\bigg|\nn
	&\qquad
			N\in\textsl{odd}, \, M\in\textsl{even}, \, L\in\textsl{odd}
			\bigg\}.
}
The summation is
\al{
\Zhat_{\vec w_0+\vec w_2}=
-\sum_{\br{p_R,\,p_L}\in\Gamma_{\vec w_0+\vec w_2}} \bar{q}^{p_R^2/2} q^{p_L^2/2}&=
-{1\over8}
\paren{\bar{\vartheta}_{10}}^4
\paren{\paren{\vartheta_{00}}^8+\paren{\vartheta_{01}}^8}
\paren{\vartheta_{10}}^8.
}

\item $\vec{w}_1+\vec{w}_2$ sector: The momentum lattice is
\al{
\Gamma_{\vec{w}_1+\vec{w}_2}
	&=	\Set{\paren{n_1,...,n_4\,\Big|\, m_1,...,m_8,l_1+{1\over2},...,l_8+{1\over2}}|
		N\in\textsl{even},\,M\in\textsl{odd},\,L\in\textsl{even}}.
}
The summation is
\al{
\Zhat_{\vec w_1+\vec w_2}=
\sum_{\br{p_R,\,p_L}\in\Gamma_{\vec w_1+\vec w_2}} \bar{q}^{p_R^2/2} q^{p_L^2/2}&=
{1\over8}
\paren{\paren{\bar{\vartheta}_{00}}^4+(\paren{\bar{\vartheta}_{01}}^4}
\paren{\paren{\vartheta_{00}}^8-\paren{\vartheta_{01}}^8}
\paren{\vartheta_{10}}^8.
}

\item $\vec{w}_0+\vec{w}_1+\vec{w}_2$ sector: The momentum lattice is
\al{
\Gamma_{\vec{w}_0+\vec{w}_1+\vec{w}_2}
	&=	\bigg\{\paren{n_1+{1\over2},\,...,\,n_4+{1\over2}\,\Big|\, 
			m_1+{1\over2},\,...,\,m_8+{1\over2},\,l_1,\,...,\,l_8}\bigg|\nn
	&\qquad
			N\in\textsl{odd}, \, M\in\textsl{odd}, \, L\in\textsl{even}
			\bigg\}.
}
The summation is
\al{
\Zhat_{\vec w_0+\vec w_1+\vec w_2}=
-\sum_{\br{p_R,\,p_L}\in\Gamma_{\vec w_0+\vec w_1+\vec w_2}} \bar{q}^{p_R^2/2} q^{p_L^2/2}&=
-{1\over8}
\paren{\bar{\vartheta}_{10}}^4
\paren{\vartheta_{10}}^8
\paren{\paren{\vartheta_{00}}^8+\paren{\vartheta_{01}}^8}.
}
\end{itemize}
Summing up the contributions from all the sectors, and including the trivial contribution from the spacetime bosons shown in Sec.~\ref{bosonic part}, we get Eq.~\eqref{partition function for non-SUSY}. 

Note that in Eq.~\eqref{partition function for non-SUSY}, the overall normalization is chosen to match the field theoretical computation as follows:
Summing up loops of a point particle with length $\alpha$, we get
\al{\label{point particle}
Z_{S^1}=V_d\int{d^dp\over(2\pi)^d}\int_0^\infty {d\alpha\over 2\alpha}e^{-\alpha{(p^2+m^2)/2}},
}
where the factor $2\alpha$ comes from the redundancy to choose the initial point of the loop and its direction. In string theory, we want to fix the normalization $A$ in
\al{
Z_{T^2}=A V_d\int {d^dp\over(2\pi)^d} \int {d\tau_1d\tau_2\over \tau_2} \exp\fn{
2\pi i\tau_1\paren{L_0-\bar{L}_0}-2\pi \tau_2\paren{L_0+\bar{L}_0-{1\over24}(c+\bar{c})}}.
}
The $\tau_1$ integral gives the level matching condition $L_0=\bar{L}_0$.
To compare with the point particle computation, we concentrate on the spacetime momentum: $L_0+\bar{L}_0=p^2\alpha'/2+(\text{neglected oscillators})$.
After the $\tau_1$ integral, we get
\al{
Z_{T^2}=
A V_d\int {d^dp\over(2\pi)^d} \int {d\tau_2\over \tau_2} \exp\paren{
-\pi \tau_2 p^2 \alpha'}
	+(\text{contribution from oscillators}).
}
Comparing this expression with Eq.~\eqref{point particle}, we see
\al{
A={1\over 2}.
}

%%%%%%%%%%%%%%%%%%%%%%%%%%%%%%%%%%%%%%%%%%%%%%%%%%%%%%%%%%%%%%%%%%%
\subsection{Contributions from spacetime coordinates to one-loop partition function}\label{bosonic part}
Let us briefly recall the basic computation of the remaining contributions from the spacetime coordinates.

We start from the $D=10$ dimensional free bosonic string:
\al{
H_X&=L_0+\bar{L}_0,\\
L_0&={\alpha'\over4}p^2+\sum_{n=1}^\infty\sum_{m=2}^{D-1} \alpha_{-n}^m\alpha_n^m-{D-2\over 24},\\
\bar{L}_0&={\alpha'\over4}p^2+\sum_{\tilde{n}=1}^\infty\sum_{m=2}^{D-1} \tilde{\alpha}_{-\tilde{n}}^m\tilde{\alpha}_{\tilde{n}}^m-{D-2\over 24},
}
where $p^2:=\sum_{\mu=0}^Dp^\mu p_\mu$.
Its contribution reads
\al{
\Tr\paren{q^{L_0}\bar{q}^{\bar{L}_0}}
&=V_D\,q^{-{D-2\over 24}}\bar{q}^{-{D-2\over 24}}%e^{2\pi i \tau_1\paren{L_0-\bar{L}_0}}
\int {d^Dp\over(2\pi)^D} 
\exp\fn{-\pi \tau_2 p^2 \alpha'}
\prod_{i,n,\tilde{n}} 
\sum_{N_{i,n},\tilde{N}_{i,n}=1}^{\infty}
q^{n N_{i,n}}\bar{q}^{\tilde{n}\tilde N_{i,n}}\nn
&=i {V_D\over(2\pi)^D} q^{-{D-2\over 24}}\bar{q}^{-{D-2\over 24}} %e^{2\pi i \tau_1\paren{L_0-\bar{L}_0}} 
\left({1\over\tau_2\alpha'}\right)^{D/2}\prod_{i,n,\tilde{n}}\paren{1-q^n}^{-1}\paren{1-\bar{q}^{\tilde{n}}}^{-1}\nn
&=
i {V_D\over(2\pi)^D} %e^{2\pi i \tau_1\paren{L_0-\bar{L}_0}}
 \left({1\over\tau_2\alpha'}\right)^{D/2} {1\over\eta(\tau)^{D-2}\,\bar{\eta}(\bar{\tau})^{D-2}},
}
where $N$ and $\tilde{N}$ are the occupation numbers.

Next we compactify the $(D-1)$th direction on $S^1$:  $X^{D-1}\sim X^{D-1}+2\pi R$.
The $(D-1)$th momentum becomes discrete, which we replace in $L_0$ and $\bar{L}_0$ as
\al{
{\alpha'\paren{p^{D-1}}^2\over 4}
	&\rightarrow {\alpha'\over 4}\paren{p_L^2+p_R^2},
}
where
\al{
p_L	&=	{n\over R}+{wR\over \alpha'},	&
p_R	&=	{n\over R}-{wR\over \alpha'}.
}
We then obtain
\al{
\Tr\paren{q^{L_0}\bar{q}^{\bar{L}_0}}
&\rightarrow
i {V_{D-1}\over(2\pi)^{D-1}} \left({1\over\tau_2\alpha'}\right)^{\paren{D-1}/2}
 	{1\over {\eta\fn{\tau}}^{D-2} {\bar{\eta}\fn{\bar{\tau}}}^{D-2}}\nn
&\qquad
\times\sum_{n,w}
 e^{2\pi i \tau_1{\alpha'\over4}\paren{p_L^2-p_R^2}}
 \exp\fn{-\pi \tau_2\alpha'\left(
 {n^2\over R^2}+{w^2R^2\over\alpha'^2}
 \right)
 }.
}
\section{T-duality}\label{T-duality section}
In this Appendix, we show that successive $S$ and $T$ transformations~\eqref{T-duality transformation} yield the Eq.~\eqref{general transformation}.
%\al{\label{general transformation}
%&\tilde\tau'={a\tilde\tau+b\over c\tilde\tau+d},
%}
%where $ad-bc=1$ and $a$, $b$, $c$, and $d$ are either
%\al{
%a	&\in	\mathbb Z,	&
%b	&\in	\sqrt{2}\mathbb Z,	&
%c	&\in	\sqrt{2}\mathbb Z,	&
%d	&\in	\mathbb Z,	
%}
%or
%\al{
%a	&\in	\sqrt{2}\mathbb Z,	&
%b	&\in	\mathbb Z,	&
%c	&\in	\mathbb Z,	&
%d	&\in	\sqrt{2}\mathbb Z.
%}
More explicitly,
\al{
\tilde\tau_1	&\rightarrow {ac\ab{\tilde\tau}^2+\paren{ad+bc}\tilde\tau_1+bd\over\ab{c\tilde\tau+d}^2},&
\tilde\tau_2	&\rightarrow {\tilde\tau_2\over\ab{c\tilde\tau+d}^2}.
	\label{explicit T-dual}
}
Using this duality, we will check in Sec.~\ref{current T-dual tf} if $\tilde\tau_2=r/\sqrt{\alpha'}$ stays finite or goes to infinity in the large boost limit $\eta\to\infty$.

\subsection{Review on ordinary modular transformation}\label{ordinary modular tf}
Let us first recall how we have shown that the general form of the transformation generated by $\tau\rightarrow \tau+1$ and $\tau\rightarrow -1/\tau$ is given by
\al{	
\tau	&\rightarrow \tau'={a\tau+b\over c\tau+d},	&
ad-bc	&=	1.	\label{SL(2,Z)}
}
%are generated by
%from those on the worldsheet torus: 
%$\tau\rightarrow \tau+1$ and $\tau\rightarrow -1/\tau$. 
First we point out that the set of transformations Eq.~\eqref{SL(2,Z)} forms the $SL(2,\mathbb{Z})$ group, from which the closure of the transformation is obvious.
%by .
%Therefore, 
In fact, if we identify the transformation with the matrix $\bmat{a&b\\ c&d}$, the composition of two transformations
\al{
\tau''
	=	{a'{a\tau+b\over c\tau+d}+b'\over c'{a\tau+b\over c\tau+d}+d'}
	=	{\paren{a'a+b'c}\tau+\paren{a'b+b'd}\over \paren{c'a+d'c}\tau+\paren{c'b+d'd}}
}
is equivalent to the multiplication of the corresponding matrices $\bmat{a'&b'\\ c'&d'}\bmat{a&b\\ c&d}$.
Moreover, the inverse of $\tau\to\tau'$
\al{
\tau	&=	{-d\tau'+b\over c\tau'-a}
}
is equivalent to the inverse matrix $\bmat{a&b\\ c&d}^{-1}$.

Since $\tau\to\tau+1$ and $\tau\to-1/\tau$ are special cases of Eq.~\eqref{SL(2,Z)}, the transformation generated by them also has the form Eq.~\eqref{SL(2,Z)}.
On the other hand, any transformation Eq.~\eqref{SL(2,Z)} can be obtained as
successive applications of %the modular transformation on the worldsheet torus
$\tau\rightarrow \tau+1$ and/or $\tau\rightarrow -1/\tau$. %result in .
\begin{proof}
We start from the general form
\al{
{a\tau+b\over c\tau+d}
}
of the transformation, and show that it reduces to $\tau$ by applying $\tau\to-1/\tau$ and $\tau\to\tau+1$.
By the $n$ times of shift, we get
\al{
\tau'
	&=	{a\tau+b\over c\tau+d}+n
	=	{\paren{a+nc}\tau+\paren{b+nd}\over c\tau+d}.
}
Choosing $n\in\mathbb Z$ appropriately, we can make $a'=a+nc$ satisfy $\ab{a'}<\ab{c}$.
The inversion $\tau'\to-1/\tau'$ gives
\al{
\tau'	&\to	\tau''={-c\tau-d\over a'\tau+b'}.
}
Now $a''$ ($=-c$) and $c''$ ($=a'$) satisfy $\ab{a''}>\ab{c''}$.
By doing this cycle of shift and inversion successively, we can always reduce the value of $a$ to eventually get
\al{
{b\over c\tau+d}.
}
From the condition for the determinant to be unity, we get $bc=-1$, which reads $b=\pm1$ for $b$ and $c$ are integers. Finally by the inversion, we get
\al{
{b\over c\tau+d}	&\to	-{c\over b}\tau-{d\over b}=\tau-{d\over b},
}
from which we obtain $\tau$ by the integer shift.%is a shift by an integer $\mp d$ from $\tau$.
\end{proof}
%%%%%%%%%%%%%%%%%%%%%%%%%%%%%%%%%%%%%%%%%%%%%%%%%%%%%%%%%%%%%%%
\subsection{T-dual transformation}\label{current T-dual tf}
We follow the argument above to show that we can get the general form~\eqref{general transformation} from the $\sqrt{2}$-shift and inversion in Eq.~\eqref{T-duality transformation}.
Let us start from
\al{
&{a\tilde\tau+b\sqrt{2}\over c\sqrt{2}\tilde\tau+d},	&
&ad-2bc=1, &
&a,b,c,d\in\mathbb{Z}.
	\label{general form of T-dual}
}
The closure and the existence of the inverse can be shown in the same way as above.

By the $n$ times of $\sqrt{2}$-shift, we get
\al{
{a\tilde\tau+b\sqrt{2}\over c\sqrt{2}\tilde\tau+d}
	&\to	\tilde\tau'
	=		{a\tilde\tau+b\sqrt{2}\over c\sqrt{2}\tilde\tau+d}+n\sqrt{2}
	=		{\paren{a+2nc}\tilde\tau+\paren{b+nd}\sqrt{2}\over c\sqrt{2}\tilde\tau+d}
	=		{a'\tilde\tau+b'\sqrt{2}\over c'\sqrt{2}\tilde\tau+d'}.
}
Choosing appropriate $n\in\mathbb Z$, we can always make $\ab{a'}\leq\ab{c}$. Further performing the inversion and the $n'$ times of $\sqrt{2}$-shift, we get
\al{
\tilde\tau'
	&\to	\tilde\tau''
	=		-{c'\sqrt{2}\tilde\tau+d'\over a'\tilde\tau+b'\sqrt{2}}+n'\sqrt{2}
	=		{\paren{-c'+n'a'}\sqrt{2}\tilde\tau+\paren{-d'+2n'b'}\over a'\tilde\tau+b'\sqrt{2}}.
}
Again inverting, we get
\al{
\tilde\tau''
&\to	\tilde\tau'''
=		-{  a'\tilde\tau+b'\sqrt{2}\over \paren{-c'+n'a'}\sqrt{2}\tilde\tau+\paren{-d'+2n'b'}}
=		{a'''\tilde\tau+\sqrt{2}b'''\over c'''\sqrt{2}\tilde\tau+d'''}.
}
Choosing $n'\in\mathbb Z$ appropriately, we can always make $\ab{c'''}\leq |a'''/2|=|a'/2|\leq |c/2|$.

By repeating this cycle, we can make the absolute value of the coefficient $c$ in Eq.~\eqref{general form of T-dual} smaller and smaller to get $c=0$ eventually:
\al{
{a\tilde\tau+b\sqrt{2}\over d}={a\over d}\tilde\tau+{b\over d}\sqrt{2}.
	\label{sqrt2-shift}
}
Since $ad=1$ due to the condition for the determinant to be unity. From Eq.~\eqref{sqrt2-shift}, we obtain $\tilde{\tau}$ by the $\sqrt{2}$-shifts.

The case
\al{
&{a\sqrt{2}\tilde\tau+b\over c\tilde\tau+d\sqrt{2}},	&
&2ab-bc=1,&
&a,b,c,d\in\mathbb{Z},
}
is an inversion of Eq.~\eqref{general form of T-dual}.

%%%%%%%%%%%%%%%%%%%%%%%%%%%%%%%%%%%%%%%%%%%%%%%%%%%%%%%%%%%%%%%

\section{Multiple point principle}\label{MPP review}
We review the original argument for the MPP that says that the SM parameters should be tuned so that our SM vacuum is degenerate with another one whose vacuum expectation value of the Higgs field is around the Planck scale~\cite{Froggatt:1995rt,Froggatt:2001pa,Nielsen:2012pu}. 

The quantum field theory (QFT) is formulated by the path integral
\al{
Z\fn{\br{\lambda}}	&=	\int[\df\varphi]\,e^{-S\fn{\br{\lambda}}\sqbr{\varphi}},
	\label{partition function in QFT}
}
where $\br{\lambda}$ denotes the dependence on the coupling constants (and mass) collectively. 
The partition function~\eqref{partition function in QFT} is analogous to the one in the canonical ensemble in the statistical mechanics:
\al{
Z(\beta)
	&=	\sum_n e^{-\beta H_n}.	\label{partition function}
}
However in the statistical mechanics, the most fundamental concept is the micro-canonical ensemble:
\al{
\Omega\fn{E}	&=	\sum_n \delta\fn{H_n-E}.
}
Froggatt and Nielsen argue that more fundamental formulation of the QFT may be analogous to the micro-canonical ensemble, in which rather the average field value is fixed while the coupling constants are determined dynamically. Let us review their argument step by step.

The canonical ensemble becomes equivalent to the micro-canonical one in the thermodynamic (large volume) limit: Given the partition function~\eqref{partition function},
we can compute the multiplicity

\al{
\ol\Omega(E)
	:=	\int\df\beta\,e^{\beta E}Z(\beta)
	&=	\int\df\beta\int\df\E\paren{\sum_n\delta\fn{H_n-\E}}e^{-\beta\paren{\E-E}}\nn
	&=	\int\df\beta\int\df \E\,\Omega\fn{\E}e^{-\beta\paren{\E-E}}\nn
	&=	\int\df\beta\int\df \E\,e^{\mathcal S(\E)-\beta\paren{\E-E}},
		\label{multiplicity}
}
where we used the entropy $\mathcal S(\E):=\ln\Omega(\E)$;
noting that $\mathcal S(\E)$, $\E$, and $E$ are extensive variables, in the thermodynamic limit, the integral over $\beta$ and $\E$ is dominated by the strong peak at their stationary values; by taking variations of $\E$ and $\beta$, we get $\df \mathcal S/\df E=\beta$ and $\E=E$:
\al{
\ol\Omega(E)
	&\to	e^{\mathcal S(E)}
	=	\Omega(E).
}
The energy is fixed first, and then the temperature $T:=1/\beta$ is determined dynamically.
Later we will see, in the QFT language, that the inverse-temperature $\beta$ corresponds to the coupling constants, that the energy $E,\E$ to the spatial integral over field values $\int\df^Dx\,\ab{\varphi}^n$, and that the summation over the states $\sum_n$ to the path integration $\int[\df\varphi]$.

As an illustration, let us consider a system of co-existing water and vapor with a fixed pressure in a piston, placed in a room temperature. We add heat into the piston. The temperature $\beta^{-1}$ in the piston rises to the boiling point. Even if we further continue to add the heat, it is used to make the water into the vapor, without changing the temperature. This way, for a large range of energy, the temperature is tuned to be the boiling point due to the two co-existing phases. In QFT language, this will be translated to the statement that even if Nature changes the field value in the micro-canonical version of the QFT, the coupling constant (mass) is tuned to the value that allows two co-existing  vacua.\footnote{
The effective potential must be convex, which is realized as a spatially inhomogeneous configuration with $\varphi=\varphi_1$ in some regions and $\varphi=\varphi_2$ in other places, where $\varphi_1$ and $\varphi_2$ are local minima of the potential; see e.g.\ Ref.~\cite{Weinberg:1987vp}.
}

The ordinary QFT starts from the path integral~\eqref{partition function in QFT}.
Let us illustrate the situation by a simple toy model:
\al{
S\fn{\Lambda,m^2,\lambda,\dots}\sqbr{\varphi}
	&=	\int\df^Dx\,\paren{\ab{\partial\varphi}^2+\Lambda+m^2\ab{\varphi}^2+\lambda\ab{\varphi}^4+\cdots}.
}
The partition function reads
\al{
Z\fn{\Lambda,m^2,\lambda,\dots}
	&=	\int[\df\varphi]\,e^{-S\fn{\Lambda,m^2,\lambda,\dots}\sqbr{\varphi}}.
}
The counterpart of Eq.~\eqref{multiplicity} should be the following:
\al{
\ol\Omega\fn{I_0,I_2,I_4,\dots}
	&=	\paren{\int\df\Lambda\int\df m^2\int\df\lambda\cdots}\,
			e^{\Lambda I_0+m^2I_2+\lambda I_4+\cdots} Z\fn{\Lambda,m^2,\lambda,\dots}\nn
	&=	\paren{\int\df\Lambda\int\df m^2\int\df\lambda\cdots}\,
			e^{\Lambda I_0+m^2I_2+\lambda I_4+\cdots}
			\int[\df\varphi]\,e^{-S\fn{\Lambda,m^2,\lambda,\dots}\sqbr{\varphi}}\nn
	&=	\paren{\int\df\Lambda\int\df m^2\int\df\lambda\cdots}\,
		\paren{\int\df\I_0\int\df \I_2\int\df\I_4\cdots}\nn
	&\qquad\times
			e^{-\Lambda\paren{\I_0-I_0}-m^2\paren{\I_2-I_2}-\lambda\paren{\I_4-I_4}+\cdots}\nn
	&\qquad\times\Bigg[\int[\df\varphi]\,
			e^{-\int\df^Dx\,\paren{\partial\varphi}^2}\nn
	&\phantom{\qquad\times\Bigg[\,}
			\delta\fn{\int\df^Dx-\I_0}
			\delta\fn{\int\df^Dx\,\ab{\varphi}^2-\I_2}
			\delta\fn{\int\df^Dx\,\ab{\varphi}^4-\I_4}
			\cdots
		\Bigg],
		\label{FN integral}
}
where the dimensionality is
\al{
\sqbr{\varphi}
	&=	{D-2\over2},	&
%\sqbr{\varphi^2}
%	&=	{D-2},	&
\sqbr{\I_0}
	&=	-D,	&
\sqbr{\I_2}	&=	-2,	&
\sqbr{\I_4}
	&=	D-4,	
}
etc.

From the observation, we know that the volume of the universe $\mathcal V$ is much larger than the Planck volume: $\mathcal V:=\int \df^Dx\ggg M_P^{-D}$.
%Therefore, the integral~\eqref{FN integral} must be dominated by the configuration of large $\I_0$ ($=I_0$). This is nothing but 
In the thermodynamic limit $\mathcal V\to\infty$, we will recover the multiplicity in the micro-canonical ensemble:
\footnote{
Here we leave the kinetic term as is. one might apply the same argument for the kinetic term as well.
}
\al{
\ol\Omega\fn{I_0,I_2,I_4,\dots}
	&\to\int[\df\varphi]\,e^{-\int\df^Dx\,\paren{\partial\varphi}^2}
				\delta\fn{\int\df^Dx-I_0}
				\delta\fn{\int\df^Dx\,\ab{\varphi}^2-I_2}
				\delta\fn{\int\df^Dx\,\ab{\varphi}^4-I_4}
				\cdots\nn
	&=:	\Omega\fn{I_0,I_2,I_4,\dots}.
}
The ``entropy'' is given by
\al{
\mathcal S\fn{I_0,I_2,I_4,\dots}
	&=	\ln\Omega\fn{I_0,I_2,I_4,\dots}.
}

In the micro-canonical version of the QFT, Nature chooses a set of extensive variables $\Set{I_0,I_2,\dots}$. Natural choice would be the values of order unity in Planck units, multiplied by the volume $\mathcal V$:
\al{
I_0	&\sim	\mathcal V,	&
I_2	&\sim	\mathcal V M_P^{D-2},	&
I_4	&\sim	\mathcal VM_P^{2D-4},	&
\cdots.
	\label{generic extensive variables}
}
Suppose that such a generic set of extensive variables are given in the micro-canonical picture. Then the integral over the intensive variables $\Lambda, m^2, \lambda, \dots$ in Eq.~\eqref{FN integral} must be dominated by such values that allow the co-existing vacua, whose mixture can reproduce the values~\eqref{FN integral} as their mean value. This is just as in the heuristic example shown above.
The field values in such vacua other than ours must be around the Planck scale.

We comment that the effective potential can be approximated by the quartic term because the running Higgs mass is almost zero in Planck units in a mass independent renormalization scheme. Therefore both the quartic coupling and its beta function must be zero at the Planck scale in order to allow the other vacuum. This has led to the predictions of the top mass $173\pm5\GeV$ and the Higgs mass $135\pm9\GeV$~\cite{Froggatt:1995rt}, nearly twenty years before the Higgs discovery.

We note that the bare Higgs mass becomes accidentally small for a Planck scale cutoff, given the low energy data at the electroweak scale~\cite{Hamada:2012bp,Jegerlehner:2013cta,Jegerlehner:2013nna,Hamada:2013cta,Masina:2013wja,Alsarhi:1991ji,Jones:2013aua}.
This smallness of the bare mass can be accounted for by the above argument if we employ a regularization scheme in which the bare Higgs mass appears in the effective potential near the cutoff; see e.g.\ Appendix~B in Ref.~\cite{Hamada:2013mya}.

In Ref.~\cite{Froggatt:2001pa}, this argument has been extended to the meta-stable vacua.
In Ref.~\cite{Nielsen:2012pu}, the delta function in this argument has been promoted to an arbitrary function having appropriate peaks.

%%%%%%%%%%%%%%%%%%%%%%%%%%%%%%%%%%%%%%%%%%%%%%%%%%%%%%%%%%%%%%%

\bibliographystyle{TitleAndArxiv}
\bibliography{HKO}

\end{document}